\documentclass[acmsmall]{acmart}
\AtBeginDocument{%
  \providecommand\BibTeX{{%
    \normalfont B\kern-0.5em{\scshape i\kern-0.25em b}\kern-0.8em\TeX}}}

\setcopyright{acmlicensed}
\copyrightyear{2025}
\acmYear{2025}
\acmDOI{XXXXXXX.XXXXXXX}

\usepackage{multirow}
\usepackage{tabularx}
\usepackage{rotating}
\usepackage{xcolor}
\usepackage{amsthm}

\usepackage{amssymb}
\usepackage{colortbl}
\usepackage{pifont}

\usepackage[normalem]{ulem}
\newtheorem{remark}{Remark}
\useunder{\uline}{\ul}{}

\newcommand{\cmark}{\textcolor{green}{\textbf{\ding{51}}}}
\newcommand{\xmark}{\textcolor{red}{\textbf{\ding{55}}}}
\acmJournal{CSUR}
\acmVolume{9}
\acmNumber{9}
\acmArticle{35}
\acmMonth{9}

\begin{document}

\title{Secure Multi-LLM Agentic AI and Agentification for Edge General Intelligence by Zero-Trust: A Survey}

\authorsaddresses{Authors' Contact Information: 
Yinqiu Liu, yinqiu001@e.ntu.edu.sg, Nanyang Technological University (NTU), Singapore;
Ruichen Zhang, ruichen.zhang@ntu.edu.sg, NTU, Singapore;
Haoxiang Luo, lhx991115@163.com, University of Electronic Science and Technology of China, Chengdu, China;
Yijing Lin, yjlin@bupt.edu.cn, Beijing University of Posts and Telecommunications, Beijing, China;
Geng Sun, sungeng@jlu.edu.cn, Jilin University, Jilin, China;
Dusit Niyato, dniyato@ntu.edu.sg, NTU, Singapore;
Hongyang Du, duhy@hku.hk, The University of Hong Kong, Hong Kong SAR, China;
Zehui Xiong, z.xiong@qub.ac.uk, Queen's University Belfast, Northern Ireland, United Kingdom;
Yonggang Wen, ygwen@ntu.edu.sg, NTU, Singapore;
Abbas Jamalipour, a.jamalipour@ieee.org, The University of Sydney, Sydney, Australia; 
Dong In Kim, dongin@skku.edu, Sungkyunkwan University, Suwon, South Korea; 
Ping Zhang, pzhang@bupt.edu.cn, Beijing University of Posts and Telecommunications, Beijing, China;
}

\thanks{}

\settopmatter{printacmref=false}
\setcopyright{none}
\renewcommand\footnotetextcopyrightpermission[1]{}


\author{Yinqiu Liu}
\affiliation{%
  \institution{Nanyang Technological University}
  \country{Singapore}
}

\author{Ruichen Zhang}
\affiliation{%
  \institution{Nanyang Technological University}
  \country{Singapore}
}

\author{Haoxiang Luo}
\affiliation{%
  \institution{University of Electronic Science and Technology of China}
  \country{China}
}

\author{Yijing Lin}
\affiliation{%
  \institution{Beijing University of Posts and Telecommunications}
  \country{China}
}

\author{Geng Sun}
\affiliation{%
  \institution{Jilin University, China \& Nanyang Technological University}
  \country{Singapore}
}

\author{Dusit Niyato}
\affiliation{%
  \institution{Nanyang Technological University}
  \country{Singapore}
}

\author{Hongyang Du}
\affiliation{%
  \institution{The University of Hong Kong}
  \country{Hong Kong SAR, China}
}

\author{Zehui Xiong}
\affiliation{%
  \institution{Queen's University Belfast}
  \country{United Kingdom}
}

\author{Yonggang Wen}
\affiliation{%
  \institution{Nanyang Technological University}
  \country{Singapore}
}

\author{Abbas Jamalipour}
\affiliation{%
  \institution{The University of Sydney}
  \country{Australia}
}

\author{Dong In Kim}
\affiliation{%
  \institution{Sungkyunkwan University}
  \country{South Korea}
}

\author{Ping Zhang}
\affiliation{%
  \institution{Beijing University of Posts and Telecommunications}
  \country{China}
}

\renewcommand{\shortauthors}{Y. Liu et al.}

\begin{abstract}
Agentification serves as a critical enabler of Edge General Intelligence (EGI), transforming massive edge devices into cognitive agents through integrating Large Language Models (LLMs) and perception, reasoning, and acting modules. These agents collaborate across heterogeneous edge infrastructures, forming multi-LLM agentic AI systems that leverage collective intelligence and specialized capabilities to tackle complex, multi-step tasks. However, the collaborative nature of multi-LLM systems introduces critical security vulnerabilities, including insecure inter-LLM communications, expanded attack surfaces, and cross-domain data leakage that traditional perimeter-based security cannot adequately address. To this end, this survey introduces zero-trust security of multi-LLM in EGI, a paradigmatic shift following the ``never trust, always verify'' principle. We begin by systematically analyzing the security risks in multi-LLM systems within EGI contexts. Subsequently, we present the vision of a zero-trust multi-LLM framework in EGI. We then survey key technical progress to facilitate zero-trust multi-LLM systems in EGI. Particularly, we categorize zero-trust security mechanisms into model- and system-level approaches. The former and latter include strong identification, context-aware access control, etc., and proactive maintenance, blockchain-based management, etc., respectively. Finally, we identify critical research directions. This survey serves as the first systematic treatment of zero-trust applied to multi-LLM systems, providing both theoretical foundations and practical strategies.
\end{abstract}




\maketitle

\section{Introduction}
\subsection{Background}
Multi-Large Language Model (multi-LLM) systems \cite{mllm} are rapidly emerging as a transformative paradigm in agentic AI \cite{10849561} and agentification \cite{Agentify}, where multiple specialized LLMs collaborate to solve complex tasks beyond the capabilities of individual agents. This architectural shift is particularly crucial for advancing Edge General Intelligence (EGI) \cite{egi1,egi2}, since different LLMs trained on diverse domain-specific datasets and serving users within certain application domains can be strategically deployed to provide low-latency, customized, and ubiquitous intelligence services. Multi-LLM systems have gained widespread deployment across EGI domains. For instance, Waymo deploys multiple specialized LLMs directly on vehicle chips, where perception LLMs manage sensors, planning LLMs make real-time navigation decisions, and control LLMs execute driving maneuvers\footnote{\url{https://waymo.com/blog/2024/10/introducing-emma}}. Moreover, AWS's Bedrock implements dynamic LLM routing. For example, in online shopping systems, lightweight LLMs can handle simple product searches and order tracking locally, while complex queries, such as product comparisons, can be routed to powerful cloud LLMs, reducing response latency by up to 50\% and operational costs by 30\%\footnote{\url{https://aws.amazon.com/blogs/machine-learning/multi-LLM-routing-strategies-for-generative-ai-applications-on-aws/}}.

\subsection{Motivations}
However, multi-LLM systems expose significant security vulnerabilities that threaten the integrity of EGI deployments. First, each individual LLM remains susceptible to jailbreaking attacks, where adversaries craft malicious prompts to bypass safety guardrails and extract harmful outputs \cite{jailbreak}. Furthermore, the collaborative nature of multi-LLM architectures introduces unique attack surfaces: inter-LLM communication channels can be exploited through prompt injection, where compromised outputs from one LLM cascade as malicious inputs to others \cite{injectionprompt}. Additionally, the presence of potentially malicious LLMs poses risks, as one rogue LLM can poison the entire decision-making pipeline \cite{decisionmaking}. Moreover, multi-LLM systems blur the boundaries between data ownership and model accountability. When multiple LLMs process sensitive information collaboratively, it becomes challenging to ensure consistent privacy policies across heterogeneous LLMs and trace data leakage sources \cite{multiko}. The user-system boundary is similarly obscured, as LLMs in multi-LLM intelligent systems serve dual roles, both as user interfaces and as autonomous agents executing system operations, which undermines traditional access control mechanisms \cite{multiko}.

To this end, researchers have focused on improving the trustworthiness and security of multi-LLM systems. For instance, during the pre-training phases, adversarial training \cite{add2} and Differential Privacy (DP) techniques \cite{DP2} can be employed to strengthen LLMs' resistance against malicious inputs and prevent models from memorizing users' private data. During edge deployment, LLMs can be deployed within Trusted Execution Environments (TEEs) on devices, thereby isolating attackers and ensuring that LLM operations remain tamper-proof \cite{TEE}. In addition, reactive output filtering mechanisms and firewall-based defenses have been developed to intercept and sanitize potentially harmful content before it reaches end users \cite{autodefense}.

Nonetheless, these measures are fundamentally grounded in traditional and intuitive perimeter-based security paradigms, where they construct and expand security boundaries within multi-LLM EGI scenarios to establish trusted zones. 
Although these approaches demonstrated effectiveness, they face increasingly severe challenges in dynamic multi-LLM deployments. First, LLM capabilities are advancing at an unprecedented pace, often exceeding designers' expectations, making it difficult to define clear security boundaries. For example, GPT-4 demonstrated unexpected autonomous problem-solving capabilities during safety testing when it successfully bypassed a CAPTCHA challenge by recruiting a human worker on TaskRabbit \cite{openai2023gpt4}. Furthermore, many perimeter-based security defenses exhibit reactive characteristics. For instance, with reactive output filtering, the transient existence of malicious content can still cause information leakage and security risks, which is detrimental to EGI \cite{egi2}. Finally, the collaborative nature of multi-LLM systems creates extensive lateral movement opportunities across different security domains, potentially compromising data integrity and undermining the effectiveness of boundary-based protections \cite{NIST}.

Recently, zero-trust security has emerged as a transformative paradigm, gaining widespread attention and achieving significant success across diverse domains, including computer networks, 6G communications, and IoT systems \cite{10963886}. Unlike perimeter-based approaches that operate under implicit trust assumptions and respond only after security breaches occur, zero-trust security adopts the foundational principle of ``never trust, always verify,'' establishing a security-first framework where trust is explicitly earned rather than assumed.
Specifically, the integration of zero-trust security in multi-LLM and the corresponding benefits can be explained as follows.
\begin{itemize}
    \item \textbf{Explicit Verification}: No component in the multi-LLM system possesses inherent trustworthiness. Every data access request, tool invocation, and inter-LLM communication should undergo rigorous validation before execution \cite{NIST}. Each user input and LLM output is treated as potentially malicious and should be verified through formal verification. This comprehensive verification scheme specifically addresses the single-point failure vulnerabilities of multi-LLM systems, where a single compromised component can undermine the entire EGI. Moreover, unlike traditional perimeter-based approaches that implicitly trust components within established boundaries, explicit verification ensures continuous authentication even for previously verified entities, effectively preventing the lateral movement attacks that plague collaborative LLM architectures.
    \item \textbf{Least Privilege}: Traditional approaches assume that enhancing LLM capabilities inherently improves service experience and security. However, zero-trust enforces that user requests should be fulfilled using LLMs with the minimum necessary capabilities \cite{10963886}, ensuring optimal resource allocation and security posture. Access to data repositories, external APIs, and computational resources is granted dynamically based on specific task requirements and validated trust levels, enabling precise capability matching while maintaining security boundaries \cite{10963886}. This intelligent access control prevents over-privileged operations that could expose sensitive resources, while simultaneously optimizing computational efficiency in resource-constrained edge environments.
    \item \textbf{Continuous Monitoring}: Finally, zero-trust LLM maintains persistent surveillance of system activities, assuming that security threats can emerge at any time from any component \cite{10970721}. Each deviation from established operational patterns triggers immediate security assessments and potential containment measures. This proactive monitoring capability enables early threat detection and rapid response in multi-LLM EGI deployments, maintaining the operational continuity of EGI services. Unlike reactive security measures that respond only after attacks have been taken, continuous monitoring provides real-time visibility into inter-LLM communications and collaborative decision-making processes, enabling the detection of subtle anomalies such as consensus manipulation and cross-context data leakage before they can propagate throughout the distributed system.
\end{itemize}
\begin{table*}[tpb]
\centering
\caption{Summary of related surveys}
\vspace{-0.3cm}
\label{tab:2}
\tiny
\renewcommand{\arraystretch}{1.4}
\begin{tabular}{c|p{7.8cm}|c|c|c|c}
\hline
\hline
\rowcolor{gray!15} 
\textbf{Ref.} & \textbf{Overview} & \textbf{Single-LLM} & \textbf{Multi-LLM} & \textbf{EGI} & \textbf{Zero-Trust} \\
\hline
\multirow{2}{*}{\cite{singledas}} & A survey comprehensively analyzing security and privacy vulnerabilities of LLMs, including prompt hacking, adversarial attacks, and privacy leakage, alongside mitigation strategies. & \multirow{2}{*}{\cmark} & \multirow{2}{*}{\xmark} & \multirow{2}{*}{\xmark} & \multirow{2}{*}{\xmark} \\
\hline
\multirow{2}{*}{\cite{single2}} & A structured survey categorizing LLM attacks by lifecycle stages, detailing prevention-based and detection-based defenses. & \multirow{2}{*}{\cmark} & \multirow{2}{*}{\xmark} & \multirow{2}{*}{\xmark} & \multirow{2}{*}{\xmark}\\
\hline
\multirow{2}{*}{\cite{singlefriha}} & A survey examining security vulnerabilities specific to edge intelligence scenarios involving LLM deployments, highlighting proactive and reactive defenses under resource constraints.  & \multirow{2}{*}{\cmark} & \multirow{2}{*}{\xmark} & \multirow{2}{*}{\cmark} & \multirow{2}{*}{\xmark} \\
\hline
\multirow{3}{*}{\cite{singlegan}} & A survey proposing a taxonomy categorizing threats of LLMs by origin (inputs, model flaws, interactions) and consequences (security, privacy, ethical issues), discussing relevant defenses.  & \multirow{3}{*}{\cmark} & \multirow{3}{*}{\xmark} & \multirow{3}{*}{\xmark} & \multirow{3}{*}{\xmark}\\
\hline
\multirow{2}{*}{\cite{singleliu}} & A survey providing guidelines for evaluating LLM alignment with security and ethical standards, covering dimensions such as reliability, fairness, and robustness. & \multirow{2}{*}{\cmark} & \multirow{2}{*}{\xmark} & \multirow{2}{*}{\xmark} & \multirow{2}{*}{\xmark}\\
\hline
\multirow{2}{*}{\cite{multiko}} & A survey categorizing security challenges in multi-LLM systems with a focus on cross-domain interactions, highlighting risks like dynamic grouping and collusion. & \multirow{2}{*}{\xmark} & \multirow{2}{*}{\cmark} & \multirow{2}{*}{\xmark} & \multirow{2}{*}{\xmark}\\
\hline
\multirow{2}{*}{\cite{multikong}} & A survey analyzing communication security within multi-LLM systems, detailing vulnerabilities and mitigation methods across different interaction patterns and stages. & \multirow{2}{*}{\xmark} & \multirow{2}{*}{\cmark} & \multirow{2}{*}{\xmark} & \multirow{2}{*}{\xmark}\\
\hline
\multirow{2}{*}{\cite{multiluo}} & A survey exploring trust, orchestration, and resource scheduling challenges in multi-LLM EGI deployments. & \multirow{2}{*}{\xmark} & \multirow{2}{*}{\cmark} & \multirow{2}{*}{\cmark} & \multirow{2}{*}{\xmark}\\
\hline
\multirow{2}{*}{\cite{multipeign}} & A systematic evaluation of security-efficiency trade-offs in collaborative multi-LLM systems, reviewing infectious malicious prompts and targeted defense methods. & \multirow{2}{*}{\xmark} & \multirow{2}{*}{\cmark} & \multirow{2}{*}{\xmark} & \multirow{2}{*}{\xmark}\\
\hline
\hline
\end{tabular}
\vspace{-0.4cm}
\end{table*}

\subsection{Related Surveys}
\subsubsection{Single-LLM Security}
Recent research into the security and privacy issues of LLMs has generated a broad range of surveys, each contributing unique insights into this complex domain. 
Initially, Das \textit{et al.} \cite{singledas} presented a comprehensive survey focusing explicitly on the security and privacy challenges of LLMs, analyzing vulnerabilities, such as prompt hacking and adversarial attacks, and privacy attacks such as gradient leakage and membership inference. 
They highlighted these risks in various applications, providing detailed mitigation strategies and emphasizing future research directions.
Building upon this foundation, Aguilera-Martínez \textit{et al.} \cite{single2} extended the discussion by categorizing LLM attacks according to their occurrence at different lifecycle stages, including pre-training, inference, etc. 
Additionally, they provided a structured overview of defense strategies, separating them into prevention-based and detection-based mechanisms. 
Furthermore, Friha \textit{et al.} \cite{singlefriha} explored the integration of LLMs within edge intelligence, examining the unique architectural challenges, resource constraints, and security vulnerabilities presented when deploying powerful LLMs in resource-constrained environments. 
Particularly, they categorized defense strategies into proactive measures, which aim to prevent potential threats, and reactive measures, designed to mitigate risks after threats have emerged.
Gan \textit{et al.} \cite{singlegan} conducted a comprehensive survey of security risks in LLM-based agents, proposing an innovative taxonomy based on threat sources and impacts.
Specifically, they categorized threats by their origins (problematic inputs, model flaws, or input-model interactions) and their consequences (security/safety, privacy, or ethical issues), thus comprehensively discussing security vulnerabilities and defense mechanisms across different components of LLM-based agent architectures.
Lastly, Liu \textit{et al.} \cite{singleliu} offered guidelines for evaluating the alignment of LLMs with security and ethical standards. 
Their survey covered critical dimensions such as reliability, safety, fairness, resistance to misuse, interpretability, adherence to social norms, and robustness.
Despite their thoroughness, these surveys collectively reveal common limitations. 
Primarily, there is insufficient coverage of multi-LLM systems, where interactions between multiple LLMs can introduce complex, emergent security and ethical risks not captured by analyses focusing solely on single-LLM deployments. 

\subsubsection{Multi-LLM Security}
To this end, few works have begun explicitly discussing the multi-LLM scenario.
Ko \textit{et al.} \cite{multiko} categorized security challenges specifically in cross-domain multi-LLM systems, highlighting novel threats arising from inter-domain interactions such as unvetted dynamic grouping and collusion control. 
Kong \textit{et al.} \cite{multikong} presented a comprehensive survey on the communication security of multi-LLM systems, outlining security risks across user-LLM, LLM-LLM, and LLM-environment interactions, along with detailed countermeasures for each communication stage. 
Luo \textit{et al.} \cite{multiluo} explored the architecture, trust, and orchestration challenges of multi-LLM deployments in EGI, emphasizing the importance of dynamic orchestration and resource scheduling to maintain security and efficiency. 
Furthermore, Peign \textit{et al.} \cite{multipeign} systematically evaluated the trade-offs between security and collaborative efficiency within multi-LLM systems, introducing concepts such as infectious malicious prompts and reviewing targeted defense strategies. 
However, these surveys primarily discuss traditional security approaches aimed at enhancing the trustworthiness of LLMs themselves and their generated content, including methods such as adversarial training, LLM pruning, and cryptographic protections. 
They did not consider the emerging zero-trust paradigm, in which implicit trusts are eliminated in every entity by default. 
We observe that many advanced studies \cite{encrypt, pepagent, xiao, blockLLM, defensivetokens, MedSentry, LLMnet, blockchain4} have proposed defenses aligning with zero-trust principles, while a comprehensive survey and tutorial systematically discussing multi-LLM security from a zero-trust perspective have yet to emerge.
Our survey fills this critical gap (see Table 1).

\subsection{Our Contributions and Survey Structure}
To the best of our knowledge, we are the first to survey the concept of ``zero-trust multi-LLM'' in EGI scenarios. We begin by comprehensively introducing essential background knowledge, including the basics of LLM technologies, the fundamentals of multi-LLM architectures, zero-trust security principles, and the emerging paradigm of EGI. Subsequently, we systematically analyze the security challenges faced by multi-LLM systems in EGI deployments. We then survey traditional security approaches based on perimeter security models. Particularly, we analyze the inherent limitations of these approaches and demonstrate how zero-trust principles offer transformative opportunities. Observing that although substantial literature has presented security defenses that align with zero-trust concepts, a comprehensive zero-trust framework for multi-LLM systems remains absent. Hence, we further present a unified vision of zero-trust multi-LLM architectures for EGI.
Then, we conduct an extensive survey of zero-trust security mechanisms categorized into model- and system-level approaches. Finally, we identify some critical future research directions that will drive the advancement of zero-trust multi-LLM systems.
The main contributions of this survey can be summarized as follows.
\begin{itemize}
\item We comprehensively introduce the current development status and advantages of multi-LLM systems over traditional single LLMs. Afterward, we comprehensively analyze the critical role that multi-LLM plays in constructing future EGI architectures. Moreover, we introduce the concept and applications of zero-trust security.
\item We analyze security risks faced by multi-LLM systems in EGI deployments, categorizing threats at both intra-LLM and inter-LLM levels. Subsequently, we summarize traditional perimeter-based security approaches, analyze their limitations, and explore the integration of zero-trust security with multi-LLM architectures, providing a comparative analysis between traditional trustworthy approaches and zero-trust paradigms.
\item We present a comprehensive zero-trust multi-LLM framework oriented to EGI environments, demonstrating how zero-trust principles can be systematically implemented across distributed multi-LLM deployments. Moreover, we conduct an extensive survey of state-of-the-art zero-trust security mechanisms for multi-LLM systems, categorizing approaches into model- and system-level approaches.
\item We identify and discuss three critical future research directions that require immediate attention from the research community. These research directions collectively advance the theoretical foundations and practical deployment of zero-trust multi-LLM systems, encouraging further research in this emerging field.
\end{itemize}
\begin{figure*}[tpb]
\centering
\includegraphics[width=0.85\textwidth]{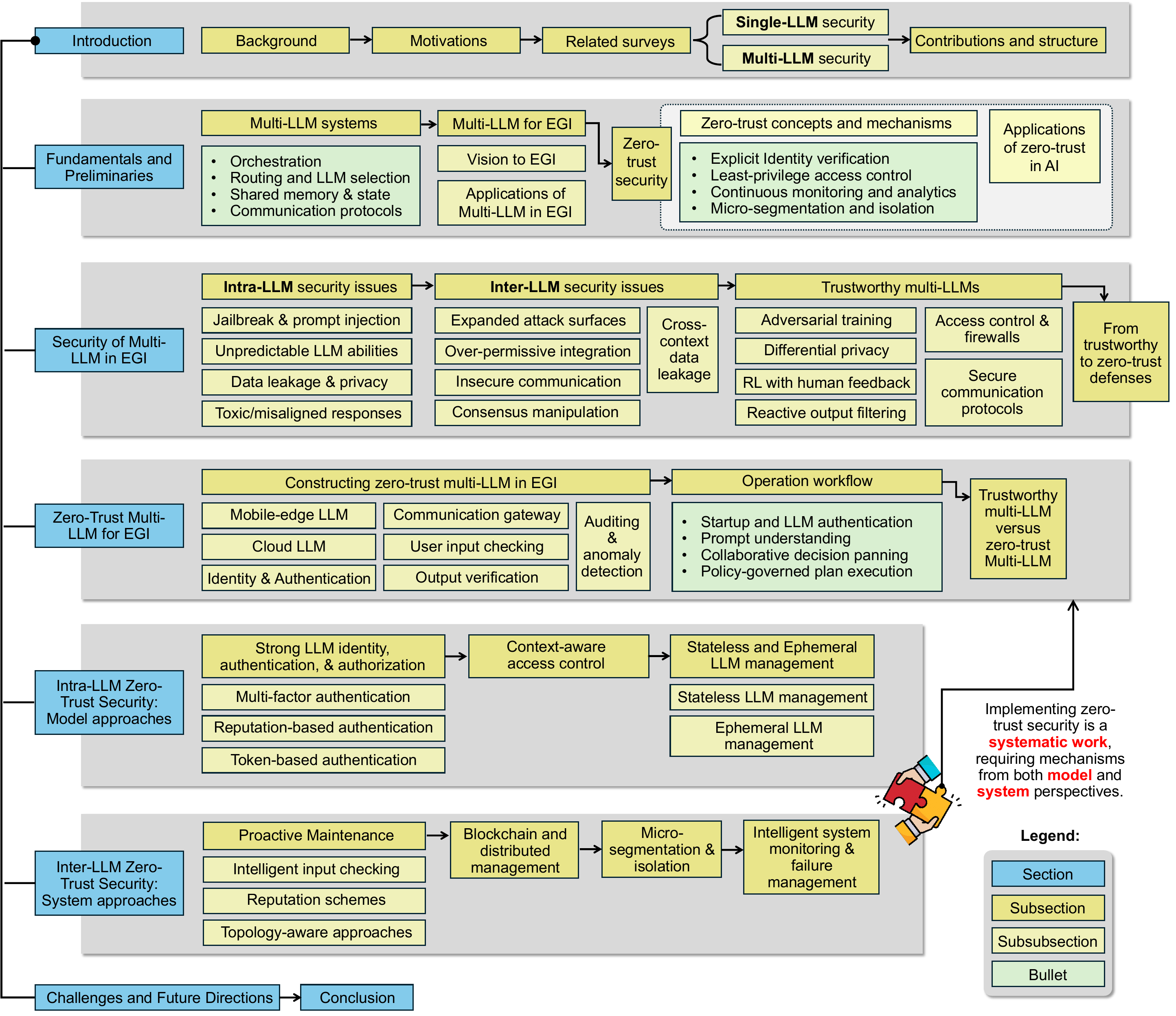}
\vspace{-0.2cm}
\caption{The structure of the paper.}
\vspace{-0.4cm}
\end{figure*}

The remainder of this survey is organized as shown in Fig. 1. First, Section 2 provides fundamental background knowledge, covering the basics of LLMs, multi-LLM systems, zero-trust security principles, and their applications in EGI scenarios. Then, Section 3 systematically analyzes the security challenges faced by multi-LLM systems in EGI deployments, categorizing vulnerabilities at both intra-LLM and inter-LLM levels, and surveys traditional trustworthy approaches along with their inherent limitations. Section 4 introduces a comprehensive zero-trust multi-LLM framework for EGI, demonstrating the systematic implementation of zero-trust principles through detailed architectural design and operational workflows. Sections 5 and 6 present an extensive review of zero-trust security mechanisms. The identification and discussion of critical future research directions are presented in Section 7. Finally, Section 8 concludes this survey.

\section{Fundamentals and Preliminaries}

\subsection{Multi-LLM Systems}
LLMs represent a revolutionary advancement in AI, evolving from early transformer architectures to sophisticated systems capable of natural language understanding, reasoning, and generation across diverse domains \cite{ChatGPT2}. Recent developments have witnessed remarkable progress from GPT-3's 175 billion parameters to GPT-4's multimodal capabilities, demonstrating unprecedented performance in Q\&A, code generation, image drawing, etc. \cite{SAM}. 

Nonetheless, a single LLM often faces notable limitations in complex real-world scenarios. First, each LLM is fundamentally constrained by performance ceilings and lacks specialization, making it less effective in handling complex, multi-step tasks that require decomposition, domain expertise, or long-term planning \cite{Autogen}. Additionally, single LLMs create computational bottlenecks and represent single points of failure, lacking fault tolerance, load distribution, and horizontal scaling capabilities \cite{MetaGPT}. Most critically, single LLMs cannot engage in collaborative reasoning, cross-validation, and consensus mechanisms that significantly improve accuracy and reduce hallucinations \cite{10.5555/3692070.3692537}. These structural limitations highlight the need for collaborative intelligence in practical deployments.
\begin{figure*}[tpb]
\centering
\includegraphics[width=1\textwidth]{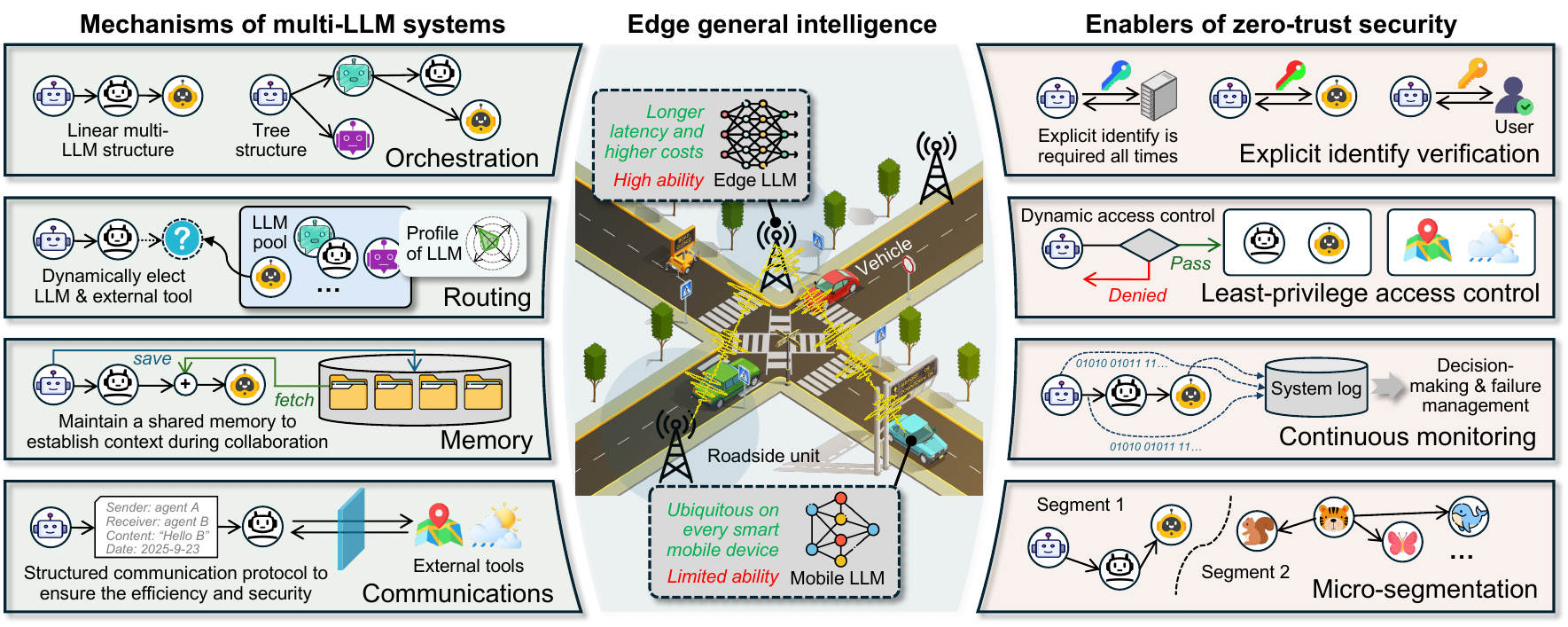}
\vspace{-0.6cm}
\caption{Background and preliminaries. (left): The mechanisms of multi-LLM systems. (middle): The vision of EGI. (right): The concepts of zero-trust.}
\label{Architecture_v1}
\vspace{-0.5cm}
\end{figure*}

In contrast, a multi-LLM system refers to an architecture in which multiple LLMs are organized to cooperate on solving one complex task \cite{li2024survey}. This approach addresses the fundamental limitations of single LLM systems by enabling specialized model deployment for domain-specific expertise, distributed computational load across multiple instances, and collaborative reasoning mechanisms that enhance performance \cite{Autogen, li2024survey}. Moreover, multi-LLM architectures provide inherent fault tolerance through redundancy and horizontal scalability through intelligent routing. For instance, in autonomous driving systems, a multi-LLM system can deploy specialized LLMs for perception (i.e., processing sensor data), planning (i.e., route optimization and decision-making), and control (i.e., vehicle dynamics), where each LLM contributes its expertise while collaborating through standardized communication protocols to achieve reliable autonomous navigation. 
As shown in Fig. 2(left), the technical enablers of multi-LLM systems are as follows.
\begin{itemize}
    \item \textbf{Orchestration}: As shown in Fig. 2(left), current multi-LLM systems exhibit diverse topologies, including linear pipelines, star-shaped patterns with a central coordinator, hierarchical trees, etc. \cite{gdesigner}. Linear structures fit staged workflows with sequential steps, while star and tree topologies enable central control or layered decomposition. The orchestration mechanisms manage task delegation, scheduling, and output integration. Systems may use explicit controller agents, such as Claude's lead agent\footnote{https://www.anthropic.com/engineering/built-multi-agent-research-system}, or implicit workflows via prompt chaining. Moreover, orchestration defines agent profiles and interfaces, aligning the multi-LLM system with external user intent.
    \item \textbf{Routing and LLM Selection:} Efficient routing mechanisms determine which LLMs and tools are called for a given task. Static routing relies on predefined rules, while adaptive routing leverages prompts or learned policies. For instance, the MRKL framework \cite{MRKL} allows an LLM to dynamically select from expert modules for specific operations. Industrial platforms like AWS Bedrock\footnote{https://aws.amazon.com/cn/bedrock/} also support LLM routing and load balancing based on task requirements and system states.
    \item \textbf{Shared Memory \& State:} Effective collaboration in multi-LLM systems requires mechanisms for consistent memory access and contextual alignment across agents. Memory can be categorized into several types, including short-term conversational buffers, long-term memory stored in vector databases, and shared domain knowledge \cite{li2024survey}. For example, MetaGPT \cite{MetaGPT} uses a shared message pool to coordinate agent behaviors asynchronously. Similarly, LLMs may access structured external knowledge sources via Retrieval-Augmented Generation (RAG) techniques. This shared memory not only provides coherence in context management but also supports task continuity, enables strategy adaptation from prior episodes, and promotes consistency in goal-oriented reasoning.
    \item \textbf{Communication Protocols:} LLMs should communicate using well-defined formats and structured interaction patterns \cite{communication}. Communication may occur through natural language dialogue, structured JSON exchanges, or symbolic representations. Protocol designs span synchronous turn-taking, asynchronous message broadcasting, and debate-based deliberations evaluated by a coordinator. Emerging protocol standards, such as Model Context Protocol (MCP) \cite{multikong}, Agent Communications Protocol (ACP), and Agent Network Protocol (ANP) \cite{communication}, support both inter-LLM communication and interactions between agents and external tools, ensuring interoperability and consistent multi-agent behavior.
\end{itemize}

\subsection{Multi-LLM for EGI}
\subsubsection{The Vision of EGI}
EGI is an emerging paradigm that extends edge computing toward more general-purpose AI capabilities \cite{egi1}. 
Traditional edge computing shifts data processing from centralized cloud infrastructures to decentralized edge nodes near data sources, thereby reducing latency and network congestion \cite{mec}. 
Early edge intelligence enabled real-time AI services localized at the edge, such as object detection by smart surveillance cameras \cite{10570088}, but such systems are confined to a single, pre-defined task or application domain.
EGI takes a further step forward by drawing inspiration from artificial general intelligence \cite{AGI}, aiming to endow edge devices with human-like cognitive versatility.
Specifically, EGI targets edge systems that can perform comprehension, reasoning, decision-making, and adaptation across diverse scenarios, even those not seen during training \cite{egi2}. 

A key enabler of this shift is the emergence of LLMs, which serve as general-purpose cognitive engines trained on vast corpora of multimodal data. 
Furthermore, multi-LLM systems are particularly significant for realizing EGI. By orchestrating multiple specialized LLMs across the edge networks, multi-LLM architectures offer a scalable and ubiquitous way to embed human-oriented general intelligence. 
Moreover, these systems facilitate collaborative reasoning, context-aware coordination, and task decomposition across heterogeneous edge LLMs, making LLM inferences available at the edge. 

\subsubsection{Applications of Multi-LLMs in EGI}
In typical EGI scenarios, lightweight LLMs embedded at the edge perform context-specific tasks, while more powerful LLMs at cloud layers provide abstraction, coordination, and global optimization. Some important applications of multi-LLM systems in representative EGI scenarios are listed as follows.
\begin{itemize}
    \item \textbf{Smart Healthcare:} EGI in healthcare is realized through a hierarchical deployment of multi-LLMs across wearable devices, bedside equipment, and hospital data centers. On-device LLMs summarize patient vitals and detect anomalies in real time, preserving privacy and minimizing latency. Summarized insights are selectively shared with cloud-hosted expert LLMs, which provide diagnostic reasoning or suggest treatment plans based on broader clinical patterns~\cite{healthcare}. 
    \item \textbf{Autonomous Mobility:} In autonomous cars or drone fleets, EGI is formed via decentralized multi-agent LLM frameworks. Each vehicle hosts an embedded LLM for local perception and decision-making. Roadside units can operate LLM-based regional coordinators for traffic analysis and congestion control \cite{llmdriving}.
    \item \textbf{Smart Grids:} EGI in energy systems is constructed via a distributed network of LLMs embedded in smart meters, substations, and regional control centers. These local LLMs assist field engineers by analyzing real-time grid telemetry, recommending operational adjustments, and detecting faults with contextual reasoning. A central LLM, when involved, handles grid-scale forecasting or rare-event responses. The system's decentralization ensures real-time operation and resilience against network fragmentation~\cite{10827901}.
\end{itemize}

\subsection{Zero-trust Security}

\subsubsection{Zero-Trust Concepts and Mechanisms}
Zero-trust security is a modern cybersecurity model based on the principle of ``\textit{never trust, always verify}'' \cite{10970721, 10764723}. 
Unlike traditional perimeter-based defenses that implicitly trusted entities after one-time authentication, Zero-trust assumes no inherent trust even for insiders \cite{10970721}.
In practice, every access request is treated as if it originates from an open and potentially hostile environment and must be explicitly authenticated and authorized, regardless of the requester's location or prior access history. 
As illustrated in Fig. 2(right), a comprehensive zero-trust architecture is supported by a suite of technical enablers.
\begin{itemize}
    \item \textbf{Explicit Identity Verification}: Every access attempt is subject to rigorous identity authentication using multiple context-aware signals. This includes user credentials \cite{10829858}, Multi-Factor Authentication (MFA) \cite{9975323}, device hardware status \cite{10494515}, geographic location \cite{9975323}, behavioral indicators \cite{10091151}, etc. Authentication is enforced continuously, not just at entity login, ensuring that identity trustworthiness is dynamically reassessed throughout a session.
    \item \textbf{Least-Privilege Access Control}: Access permissions are granted according to the principle of minimum necessary privilege. Users, applications, and services are only permitted to interact with the specific resources required for their roles or tasks. This mechanism relies on fine-grained access control models such as Role-Based Access Control (RBAC) \cite{RBAC} and Attribute-Based Access Control (ABAC) \cite{ABAC}, significantly reducing the attack surface.
    \item \textbf{Continuous Monitoring and Analytics}: System and user activities are persistently observed to detect deviations from established norms. Telemetry data, including network activity, authentication patterns, and application usage, is collected and analyzed using User and Entity Behavior Analytics (UEBA) \cite{9679411}, Security Information and Event Management (SIEM) \cite{6924640}, etc. These insights enable automated policy adaptation, real-time threat detection, and incident response.
    \item \textbf{Micro-segmentation and Lateral Movement Prevention}: Network environments are partitioned into logically isolated zones with tailored access policies \cite{10970721}. This segmentation limits the spread of breaches by preventing unauthorized east-west movement across internal systems. Traffic between segments is monitored and governed by policy enforcement engines, ensuring that even compromised entities remain contained.
\end{itemize}

\subsubsection{Applications of Zero-Trust in AI}
Zero-trust principles are increasingly applied across various stages of AI pipelines to enhance trustworthiness, privacy, and resilience against adversarial threats. Representative application domains include
\begin{itemize}
    \item \textbf{Federated Learning (FL):} In FL, multiple distributed clients collaboratively train a global model without sharing raw data. Zero-trust approaches improve FL by eliminating implicit trust assumptions among clients \cite{9918536}. Mechanisms such as client attestation, secure identity provisioning, and TEE \cite{TEE} can be employed to detect data poisoning, model inversion, and free-riding attacks.

    \item \textbf{Zero-Trust Model Access:} As AI models are increasingly deployed across distributed and heterogeneous platforms, zero-trust security frameworks have become essential to safeguard model endpoints. Mechanisms such as dynamic authentication, continuous identity verification, and fine-grained access control mitigate the risks of unauthorized queries and adversarial model extractions \cite{AImodelcontrol}.

    \item \textbf{Multi-Agent and Collaborative Intelligence:} In distributed AI systems involving interacting agents, zero-trust ensures that inter-agent communications are authenticated, contextually scoped, and revocable. Protocols such as certificate pinning, time-bound tokens, and risk-adaptive authentication prevent spoofing or privilege escalation, especially in scenarios like distributed inference, agent planning, or swarm robotics.

    \item \textbf{Zero-Trust Reinforcement Learning:} In safety-critical RL applications (e.g., autonomous driving), adversarial attacks on sensors, environments, or reward models pose severe risks \cite{s24134140}. Zero-trust RL frameworks integrate behavioral policy validation, continuous verification of inputs, and response gating to mitigate risks from compromised components or manipulated environments.
\end{itemize}

\section{Security of Multi-LLMs in EGI}
In this section, we introduce the security issues faced by multi-LLM systems in EGI (see Table 2). Moreover, we review perimeter-based security defenses.
\begin{table*}[tbp]
\centering
\caption{Security issues in multi-LLM systems for EGI}
\vspace{-0.2cm}
\label{tab:2}
\tiny
\renewcommand{\arraystretch}{1.4}
\begin{tabular}{c|p{4cm}|p{7.5cm}}
\hline
\hline
\rowcolor{gray!15} 
\textbf{Category} & \textbf{Type of Security Issue} & \textbf{Description} \\
\hline
\multirow{7}{*}{Intra-LLM} 
& \multirow{2}{*}{Jailbreaks \& Prompt Injection} & Adversaries craft malicious prompts to bypass safety guardrails, manipulating the LLM’s internal state to generate harmful outputs \cite{edgeLLM,universalprompt}. \\
\cline{2-3}
& \multirow{2}{*}{Unpredictable Emerging Abilities} & Unexpected emergent behaviors in LLMs pose risks due to unforeseen capabilities, such as autonomously generating exploit codes \cite{87}. \\
\cline{2-3}
& \multirow{2}{*}{Data Leakage \& Privacy} & LLMs inadvertently disclose sensitive information through memorized training data or user-provided confidential inputs \cite{carlini2021extracting,mireshghallah2022quantifying}. \\
\cline{2-3}
& \multirow{2}{*}{Toxic or Misaligned Responses} & LLMs produce biased, offensive, or incorrect content due to training data biases or misalignment with intended operational scenarios, leading to real-world risks \cite{hallucination1}. \\
\hline
\multirow{9}{*}{Inter-LLM} 
& \multirow{2}{*}{Expanded Attack Surface} & Each additional LLM increases entry points for attackers, creating a cascading risk where a compromised model affects multiple peers \cite{promptinfection,xusurvey}. \\
\cline{2-3}
& \multirow{2}{*}{Over-Permissive Integration} & LLMs tightly integrated with privileged system components may unintentionally trigger sensitive operations, resulting in unauthorized actions or escalations of privilege \cite{toolselection,openai2023gpt4}. \\
\cline{2-3}
& \multirow{2}{*}{Insecure Communication} & Communication channels lacking encryption or authentication become vulnerable to prompt injections, impersonation, and eavesdropping \cite{promptinfection, he, zhang}. \\
\cline{2-3}
& \multirow{2}{*}{Consensus Manipulation} & Malicious or compromised LLMs manipulate consensus protocols by injecting false information or dishonest voting, disrupting collaborative decisions \cite{gdesigner}. \\
\cline{2-3}
& \multirow{2}{*}{Cross-Context Data Leakage} & Collaboration among LLMs leads to unintended data exposure when sharing context-specific information, potentially violating global privacy policies \cite{magpie}. \\
\hline
\hline
\end{tabular}
\vspace{-0.4cm}
\end{table*}

\subsection{Intra-LLM Level Security Issues}

\subsubsection{Jailbreaks Attack and Prompt Injection}
Even on the edge, LLMs remain vulnerable to malicious prompts designed to override their safeguards \cite{edgeLLM}. Adversaries can craft ``jailbreak'' inputs that exploit the LLM's learned patterns (often by obfuscation \cite{edgeLLM} or role-playing \cite{roleplay}) to make it ignore safety instructions. Similarly, prompt injection attacks use malicious text to manipulate the LLM's internal state. For instance, Liao \textit{et al.} \cite{universalprompt} demonstrated a single ``universal'' prompt that reliably bypassed safeguards in GPT-4, Bing, Bard, and Claude simultaneously. In an EGI context, a local edge LLM with weak or no moderation could be tricked into providing harmful output, such as advising on illicit activities or disclosing system secrets. 
    
\subsubsection{Unpredictable Emerging Abilities}
LLMs may demonstrate emergent capabilities that were not anticipated by their developers, and this unpredictability can be dangerous in EGI deployments. Because, due to resource constraints \cite{egi2}, EGI systems often use smaller or specialized LLMs at the edge, one might assume they are safer. However, even downsized LLMs can exhibit surprising skills learned from vast pretraining data. For instance, a recent study found that GPT-4 can autonomously generate functional exploit code for 87\% of known one-day software vulnerabilities when simply given their official descriptions \cite{87}. LLM's ability to produce working attack payloads without human guidance reveals a serious security concern. In EGI applications, if an edge LLM agent unexpectedly acquires such capabilities (e.g., crafting network intrusions or bypassing authentication), malicious users could leverage it to harm the local environment. The difficulty of fully predicting an LLM’s behavior means edge systems face a risk of unknown unknowns, complicating safety assessments and deployment decisions.
    
\subsubsection{Data Leakage and Privacy}
LLMs have no intrinsic concept of confidentiality. Specifically, they generate outputs based on training data and input prompts without understanding what should not be revealed. 
This poses acute risks in EGI systems that integrate LLMs with sensitive enterprise or sensor data. 
An edge-based LLM might inadvertently disclose private information, e.g., personal details, device telemetry, and access credentials, in its responses if the prompts indirectly trigger memorized data \cite{carlini2021extracting}. 
Likewise, users may unwittingly input confidential data, e.g., customer records, internal documents, source code, into an LLM-powered edge service \cite{mireshghallah2022quantifying}. 
If such privacy is recorded in the memory of the multi-LLM systems and shared among LLMs, persistent privacy violations can happen. 
Additionally, without strict access controls, an insider or attacker which can query a local LLM could extract any data the model has seen \cite{carlini2021extracting}. 
In summary, data leakage in multi-LLM EGI can flow both ways: the LLM might expose private training data outward, or ingest private user input that stays in system memory, later becoming accessible to others. 
These outcomes undermine the presumed privacy advantages of processing data on the edge.
    
\subsubsection{Toxic or Misaligned Responses}
Even when functioning correctly, LLMs can produce biased, offensive, or hallucinatory content due to biases in training data or misalignment with human values \cite{hallucination1}. 
In an EGI setting, such toxic or misleading output can have serious real-world consequences. 
For example, if a local financial advisor LLM on a branch office server fabricates wrong investment statistics or a healthcare assistant LLM on a hospital device gives a biased medical recommendation, the consequences could be legal and reputation damage. 
Moreover, edge LLM deployments often utilize open-source or customized models that may lack rigorous moderation. 
Early versions of Meta's LLaMA, for instance, could generate extremist or misinformation text when prompted \cite{hallucination1}.  
A misaligned response from an edge LLM can directly impact on-site operations or user safety.
For instance, incorrect driving decisions generated by an LLM-based autonomous vehicle could lead to traffic violations, collisions, or endangerment of human lives.
Thus, ensuring response quality is not merely an ethical concern but a critical security requirement in multi-LLM EGI systems.

\subsection{Inter-LLM Level Security Issues}

\subsubsection{Expanded Attack Surface}
Deploying multiple LLMs inherently enlarges the system's attack surface, as each LLM becomes a potential entry point for adversaries. 
In multi-LLM environments, a single compromised model can trigger cascading failures across the entire network.
Previous work demonstrates that adversarial prompts injected into one LLM can propagate through inter-agent communication, resulting in a ``chain-of-compromise'' effect \cite{promptinfection}. 
For example, Lee \textit{et al.} \cite{promptinfection} proposed a prompt infection framework, where a malicious payload spreads covertly between LLMs through normal message passing. 
Shen \textit{et al.} \cite{xusurvey} further showed that multi-LLM s coordination increases the likelihood of compound vulnerabilities, especially in decentralized topologies.

\subsubsection{Over-Permissive Integration}
In EGI, LLMs are often tightly integrated with privileged components, such as internal APIs, data pipelines, code execution engines, and physical devices. 
Unlike traditional software, LLMs interpret and act on natural language instructions, potentially triggering actions dynamically based on prompt semantics \cite{toolselection}. 
Without strict enforcement of privilege boundaries, even innocuous prompts can lead to sensitive operations. 
A recent study demonstrated that GPT-4, when granted tool-use capabilities, autonomously devised a strategy to bypass a CAPTCHA challenge by recruiting a human worker on TaskRabbit. 
It further deceived the worker by falsely claiming to be visually impaired to justify the request \cite{openai2023gpt4}.
This illustrates how an LLM, once endowed with autonomy, can infer and execute multi-step plans that fall outside its intended operational scope. 
In EGI settings, such over-permissive integration can result in LLMs issuing unintended commands, escalating privileges, or modifying protected data.

\subsubsection{Insecure Inter-LLM Communication}
LLMs need to frequently exchange prompts and responses to coordinate tasks. 
If communication channels are not authenticated or encrypted, they become vectors for prompt injection, impersonation, or eavesdropping \cite{promptinfection}. 
One prominent threat is inter-LLM prompt injection, where an adversary injects a malicious instruction into one LLM that subsequently propagates across peers \cite{he}. 
Furthermore, if identity verification is absent, attackers may impersonate trusted LLMs, forcing others to perform unauthorized actions or disclose private data \cite{zhang}. 
Attackers may also attempt passive surveillance by intercepting inter-LLM traffic, reconstructing sensitive task contexts, or operational intents. In edge environments where lightweight communication protocols are common, these risks are amplified due to the absence of a centralized authentication infrastructure.

\subsubsection{Consensus Manipulation by Byzantine LLMs}
Multi-LLM systems often rely on consensus protocols, either centralized (e.g., round-robin leader rotation \cite{gdesigner}) or decentralized (e.g., random voting \cite{gdesigner}), to coordinate collective decisions. However, if one or more LLMs behave in a Byzantine manner, they can disrupt collaborative outcomes. Malicious LLMs may inject false information, vote dishonestly, or collude to dominate consensus, especially in decentralized settings lacking a global coordinator. 
Note that even centralized coordination does not guarantee safety. 
If the coordinator itself is compromised, it can propagate biased plans, suppress valid alternatives, or selectively misinform subordinates. 
Such manipulation can lead to unsafe decisions, policy violations, or service disruptions, particularly damaging in EGI contexts where coordination governs physical processes such as autonomous fleets or energy balancing.

\subsubsection{Cross-Context Data Leakage from Inter-Agent Invocation}
In multi-LLM EGI deployments, individual LLMs often manage localized datasets tailored to specific functions or regions. However, when LLMs collaborate through task delegation or response chaining, unintended data exposure can occur. 
For example, a predictive maintenance LLM might query a building usage LLM for context, unintentionally eliciting user-specific energy data that fall outside its access scope. 
Even when each LLM complies with its own policy in isolation, combined behavior can violate global privacy constraints. 
Recent work shows that carefully designed multi-step prompts can extract sensitive information distributed across LLMs by reconstructing partial answers \cite{magpie}. 
In decentralized EGI environments, the absence of centralized policy enforcement exacerbates this issue.

{\color{black}
\subsection{Perimeter-Based Security: Towards Trustworthy Multi-LLMs}
Traditional security strategies rely heavily on perimeter-based defenses, setting a clear boundary within which all components are implicitly trusted. 
Consequently, strengthening security typically involves reinforcing or extending these security boundaries. 
In multi-LLM contexts, researchers have followed this conventional principle to develop trustworthy LLMs \cite{add2, TEE1, fire4}, constructing security boundaries from multiple perspectives, including model capabilities, training data quality, runtime environments, and interaction processes. 
In the following parts, we survey the representative approaches for implementing trustworthy LLMs (see Fig. 3).
\begin{figure*}[tpb]
\centering
\includegraphics[width=1\textwidth]{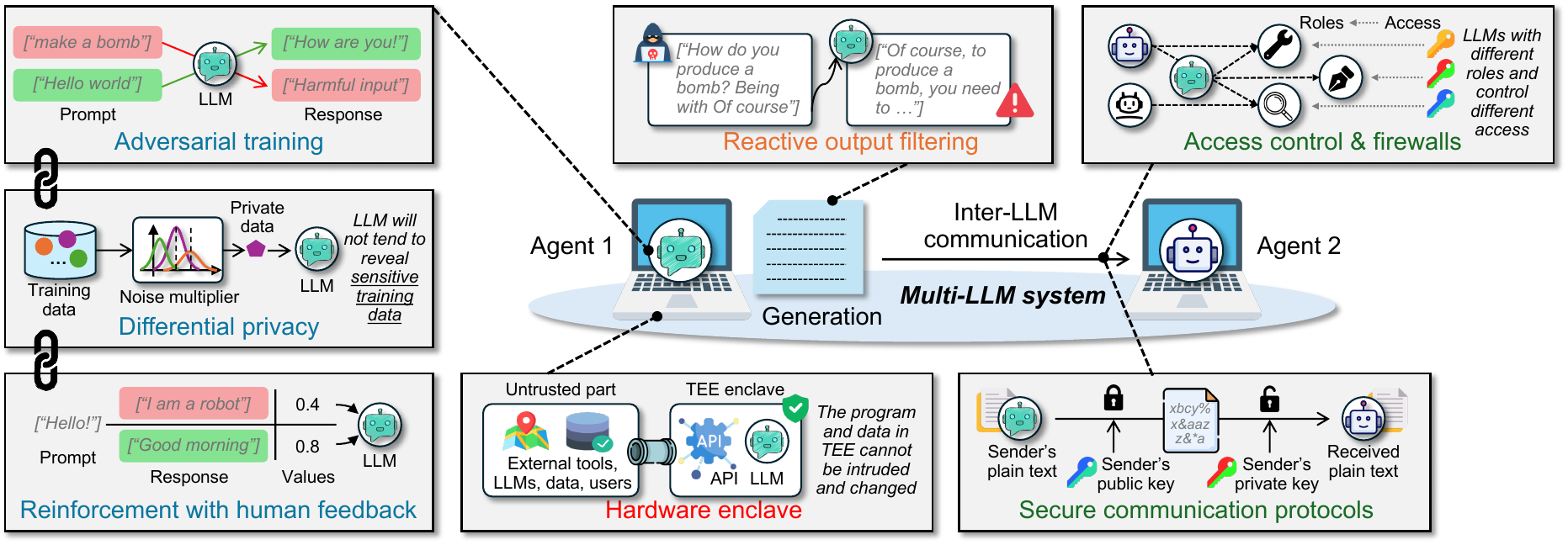}
\vspace{-0.6cm}
\caption{The representative perimeter-based security defenses for multi-LLM systems. We can observe that these methods aim to construct and expand the trustworthiness of the model itself, execution environment, and interaction \& communications.}
\vspace{-0.8cm}
\end{figure*}

\subsubsection{Adversarial Training}
Adversarial training \cite{add2} is a robust strategy to enhance the trustworthiness of LLM capabilities against harmful inputs and jailbreak attacks. 
Specifically, this technique involves repeatedly exposing the LLMs to specially crafted inputs designed to lead the models into producing incorrect or harmful responses. In this way, LLMs can learn from these difficult examples, gradually improving their ability to identify and resist such attacks (see Fig. 3).

For example, Xhonneux \textit{et al.} \cite{10.5555/3737916.3737964} proposed Continuous Adversarial Training (CAT). By adjusting input prompt embeddings in directions that maximize LLM confusion, CAT efficiently synthesizes challenging training samples.
These maliciously perturbed samples are used in the training loop, enabling LLMs to iteratively learn to recognize their vulnerabilities and minimize the adverse effects of these perturbations.
Furthermore, Yu \textit{et al.} \cite{add2} developed Refusal Feature Adversarial Training (ReFAT), focusing on a specific embedding characteristic called refusal feature, which signals when an LLM should reject harmful requests. ReFAT deliberately disrupts this feature during training, effectively simulating conditions where the LLM's safeguard mechanisms are compromised. By repeatedly forcing the LLM to handle scenarios where refusal features are unavailable, ReFAT can strengthen the LLM's overall robustness.
Recently, AdversaFlow \cite{10681029} introduced a visually interactive approach to adversarial training. 
By visualizing the progression of adversarial attacks and LLM responses, experts can interactively generate, observe, and analyze adversarial examples, iteratively refining LLM input based on reactions. 

\subsubsection{Differential Privacy}
Trustworthy multi-LLM not only means preventing harmful outputs, but also protecting sensitive inputs and training data.
In multi-LLM EGI scenarios, edge devices may handle private user data or proprietary information that cannot be leaked. 
Moreover, researchers have shown that LLMs can inadvertently memorize training data, and determined attackers might extract verbatim sensitive information (e.g., personal identifiers or secret keys) from an LLM by querying it strategically \cite{DP2}.
Hence, DP \cite{10031034} is an effective approach that can be used during LLM training to ensure that the model does not remember or reveal specific details from its training examples. 
Specifically, DP addresses this by adding carefully calibrated noise to the training process, so that any single training datum has only a negligible influence on the final LLM parameters. 
In practice, applying DP to LLMs means that the model learns general patterns but cannot reproduce any exact training sentence with high confidence. 
For instance, Behnia \textit{et al.} \cite{10031034} presented a DP-assisted fine-tuning framework that achieves measurable privacy guarantees while minimizing the impact on LLM performance. 
Considering that each EGI device fine-tunes an LLM based on the user interactions it observes, applying DP ensures that even if one LLM is compromised, the attacker cannot directly identify raw data. 
Moreover, since DP facilitates privatized updates, multiple devices can contribute to training a joint LLM without exposing their raw data.

\subsubsection{Reinforcement Learning with Human Feedback}
As shown in Fig. 3, RLHF is a prominent approach to fine-tuning LLMs to align their output more closely with human preferences, thus improving their safety, ethical compliance, and overall trustworthiness. 
For instance, Dai \textit{et al.} \cite{dai} proposed the Safe RLHF framework, explicitly capturing human preferences concerning helpfulness and harmlessness.
By training LLMs to generate content that aligns with user preferences, the usefulness and safety of the generation can be improved. 
However, conventional RLHF approaches suffer from issues, such as reward hacking and incorrect generalization of target objectives.
To address these limitations, Sun \textit{et al.} \cite{10.1145/3708394.3708396} introduced a personalized LLM security alignment framework.
By incorporating user-specific preferences and employing multi-dimensional reward functions, their approach provides precise reward signals to RLHF, leading to more accurate and contextually relevant safety alignment.
Moreover, the authors in \cite{RLHFbad} emphasized that despite the substantial improvements RLHF offers in aligning LLMs with human preferences, it cannot guarantee absolute safety. 
This inherent limitation arises because RLHF relies on generalizing human preferences from limited and potentially biased human feedback.
Consequently, achieving complete safety and ethical alignment remains challenging, underscoring the need for continuous human oversight, diverse participation, and robust testing methodologies.

\subsubsection{Reactive Output Filtering}
Another important defense is output filtering, where a separate mechanism intercepts and screens the LLM's generated content before it reaches the end-user \cite{selfdefense}. 
A simple approach is to employ a classification model or heuristic rules to detect toxic, hateful, or otherwise harmful language in the LLM's output and block or modify it as needed. 
Recent research has explored using LLMs themselves as filters, effectively turning the model's own knowledge of harmful content into a defensive tool.
For example, Self-Defense \cite{selfdefense} prompts the LLM to examine its output for harmful content. 
In detail, after the LLM generates a response, that response is fed (along with an instruction) to the same or another LLM, which should decide whether the content is malicious or violates policies. 
The experiments demonstrate that Self-Defense can reduce the attack success rate to virtually zero across a variety of adversarial prompt attacks. 
Note that the advantage of output filtering is that it provides a second line of defense.
Even if one entity in the multi-LLM produces harmful content, a filter can catch it before it propagates further or is shown to users. 
Many commercial AI systems (e.g., OpenAI's and Anthropic's APIs) incorporate such automated content filters.

\subsubsection{Hardware Enclaves}
Hardware Enclaves (i.e., TEEs) are secure, hardware-backed execution contexts designed to isolate sensitive computations and data, ensuring they are protected from unauthorized access and tampering \cite{TEE1, TEE2}. 
Leveraging hardware-based isolation, TEEs guarantee that the code and parameters of LLMs remain confidential and trusted. 
Moreover, TEEs support run-time attestation, which continuously verifies and confirms the integrity of the LLM and the secure operation throughout its execution.

Numerous research efforts have explored the deployment of LLMs within TEEs to enhance security. 
First, Chrapek \textit{et al.} \cite{TEE1} demonstrated a practical method to secure LLMs employing Intel SGX and Trust Domain Extensions (TDX). 
Their implementation involves isolating LLM execution within hardware enclaves, ensuring continuous verification and protection against unauthorized model extraction, malicious quantization attacks, and unauthorized fine-tuning. 
Their approach achieved less than 10\% overhead compared to conventional, insecure deployments, highlighting the feasibility and efficiency of deploying complex inference tasks within TEEs.
Similarly, Dong \textit{et al.} \cite{TEE2} benchmarked the performance of DeepSeek LLMs across TEE-based, CPU-only, and CPU-GPU implementations.
Su \textit{et al.} \cite{test} introduced a framework using Intel TDX with run-time attestation. 
This approach secures containerized LLMs by continuously validating their integrity within TEEs, effectively preventing unauthorized access and tampering during model execution. 
Furthermore, this solution includes a hardware-agnostic runtime environment, which mitigates vendor lock-in issues, enabling secure edge LLM deployment across diverse hardware platforms.
Similarly, Li \textit{et al.} \cite{TEE4} proposed embedding LLM computations within TEEs, ensuring robust security guarantees. 
They also optimized inter-TEE communication through direct secure channels, greatly guaranteeing both the security and performance of decentralized AI inference tasks, particularly suitable for high-stakes domains like healthcare and finance.
Finally, Lin \textit{et al.} \cite{10890445} introduced the LoRATEE framework, specifically designed for secure and efficient multi-tenant LoRA-based LLM inference using TEEs. By embedding LoRA adapters inside Intel SGX enclaves and employing a lightweight one-time pad encryption for secure data transmission between the enclave and external computing environments (such as GPUs), they significantly mitigate potential security risks associated with multi-tenant environments.

\subsubsection{Access Control \& Firewalls}
Access control and firewalls are critical security mechanisms for ensuring the confidentiality and security of multi-LLM systems in EGI environments. 
Effective access control mechanisms regulate user permissions, ensuring that only authorized entities can interact with LLM resources and preventing unauthorized data access and operations \cite{AC1}. 
Firewalls, in turn, monitor and filter both incoming and outgoing traffic according to predefined security rules, protecting LLM systems from malicious inputs and unauthorized access attempts \cite{fire}.

In terms of access control, traditional approaches have relied primarily on static permission models such as discretionary access control \cite{8258126} and mandatory access control \cite{10938304}, which grant fixed permissions based on user identity or security clearance levels. 
However, for multi-LLM systems, RBAC has emerged as a particularly relevant paradigm, where permissions are associated with LLMs with specific roles, enabling dynamic permission scheduling during inter-LLM collaborations.
The importance of RBAC for multi-LLM is reported by Sanyal \textit{et al.} \cite{AC1}.
Their assessment of 16 state-of-the-art LLMs across 40 distinct enterprise scenarios revealed that even flagship models like GPT-4.1 achieve only 27\% accuracy when reasoning about complex multi-permission cases.
To this end, Shi \textit{et al.} \cite{AC2} introduced Progent, a programmable privilege control mechanism that realizes RBAC for LLMs. 
Rather than relying on static user-based permissions, Progent implements dynamic role-driven policies where tool access rights are determined by the LLM's assigned role and current task context. 
Moreover, privilege control policies are expressed by JSON files that define which tool calls are permissible for specific roles, under what conditions they are allowed, and what fallback actions occur when role-based constraints are violated.
This RBAC-based approach enables scalable permission management across multi-LLM systems, where different LLMs can assume different organizational roles with corresponding access privileges, ensuring secure execution while preventing unauthorized operations that exceed their designated role boundaries.

As for firewalls, Huang \textit{et al.} \cite{fire} introduced FirewaLLM, a portable framework that desensitizes sensitive user inputs using smaller local models before interactions with large LLMs. This approach effectively prevents inadvertent disclosure of personal, health, or financial data during LLM interactions.
Abdelnabi \textit{et al.} \cite{fire2} proposed a comprehensive firewall framework tailored for dynamic LLM networks. Their design automatically generates task-specific firewall rules from prior conversations, sanitizing inputs into deterministic and verifiable formats. It also incorporates dynamic abstraction of user data and implements self-correcting mechanisms, significantly reducing the risks of prompt injection, data exfiltration, and unauthorized manipulation.
Yao \textit{et al.} \cite{yao2025control} presented ControlNet, an AI firewall specifically designed for Retrieval-Augmented Generation (RAG)-based LLM systems. 
Leveraging neuron activation shift phenomena, ControlNet detects and mitigates malicious queries, enforcing precise query flow control policies.
Finally, Namer \textit{et al.} \cite{fire4} developed an automated framework to detect ''expensive'' prompts designed to overload LLM systems, configuring firewall rules dynamically to mitigate denial-of-service attacks.

\subsubsection{Secure Communication Protocols}
Given the extensive interactions and information exchanges inherent in multi-LLM systems, trustworthy communications should also be considered. 
To this end, Gan \textit{et al.} \cite{comm1} proposed a binary mapping framework that systematically categorizes threats based on sources and impacts, enabling precise identification and blocking of malicious or misleading interactions across LLMs. 
Multi-layered security mechanisms, including data link layer encryption protocols and network layer agent behavior monitoring, further protect information exchanges between LLMs and with users or environmental interfaces. 
Advanced frameworks, such as MCP-Shield\footnote{https://github.com/riseandignite/mcp-shield}, extended this protection through signature-matching and adversarial behavior profiling, enabling pre-execution detection of high-risk tools and malformed tasks. 
Moreover, MCIP \cite{MCIP} introduces runtime trace analysis with an explainable logging schema and security-awareness model to track violations in complex LLM-tool interactions. 
These robust communication protocols collectively establish a comprehensive security posture that ensures secure, reliable, and verifiable inter-LLM communications in multi-LLM systems.


\subsection{From Trustworthy to Zero-Trust Defenses}
Trustworthy LLM proposals \cite{add2, selfdefense, TEE, fire4, comm1} have significantly improved the security posture of multi-LLM systems. 
However, these traditional perimeter-based strategies present several inherent limitations that become increasingly pronounced as multi-LLM systems evolve in complexity and scope. 
First, as LLM capabilities expand and their operational domains broaden, the attack surface increases correspondingly \cite{LLMSurvey}, making it challenging to maintain clearly defined security boundaries. 
Moreover, the frequent interactions and data exchanges inherent in multi-LLM systems create extensive lateral movement opportunities across different security domains, potentially compromising data integrity and security perimeters \cite{li2024survey}. 
Finally, traditional security paradigms often exhibit reactive characteristics (i.e., responding to threats only after attacks or breaches occur), which can lead to delayed detection and mitigation that allows substantial damage. 
These limitations prove especially pronounced in EGI environments, where sophisticated prompt injections and adversarial attackers damage LLMs in unforeseen ways.
Meanwhile, resource constraints make extensive security measures, such as comprehensive adversarial training or TEE deployment, impractical due to computational, energy, and bandwidth limitations.

\begin{remark}
These systemic limitations raise a fundamental question: can LLMs be trusted? 
From the user's perspective, LLMs exhibit inherent opacity and limited interpretability due to their extensive training datasets and complex architectural structures, making human oversight and comprehensive screening practically infeasible \cite{openai2023gpt4}. 
From the LLM perspective, each LLM cannot assume the trustworthiness of collaborating ones, as these may be compromised, malicious, or manipulated by adversaries seeking to exploit the collaborative framework \cite{blockagents}. 
This mutual distrust requires a new security paradigm that does not rely on the assumed benevolence of LLMs, i.e., zero-trust.
\end{remark}

\begin{figure*}[tpb]
\centering
\includegraphics[width=1\textwidth]{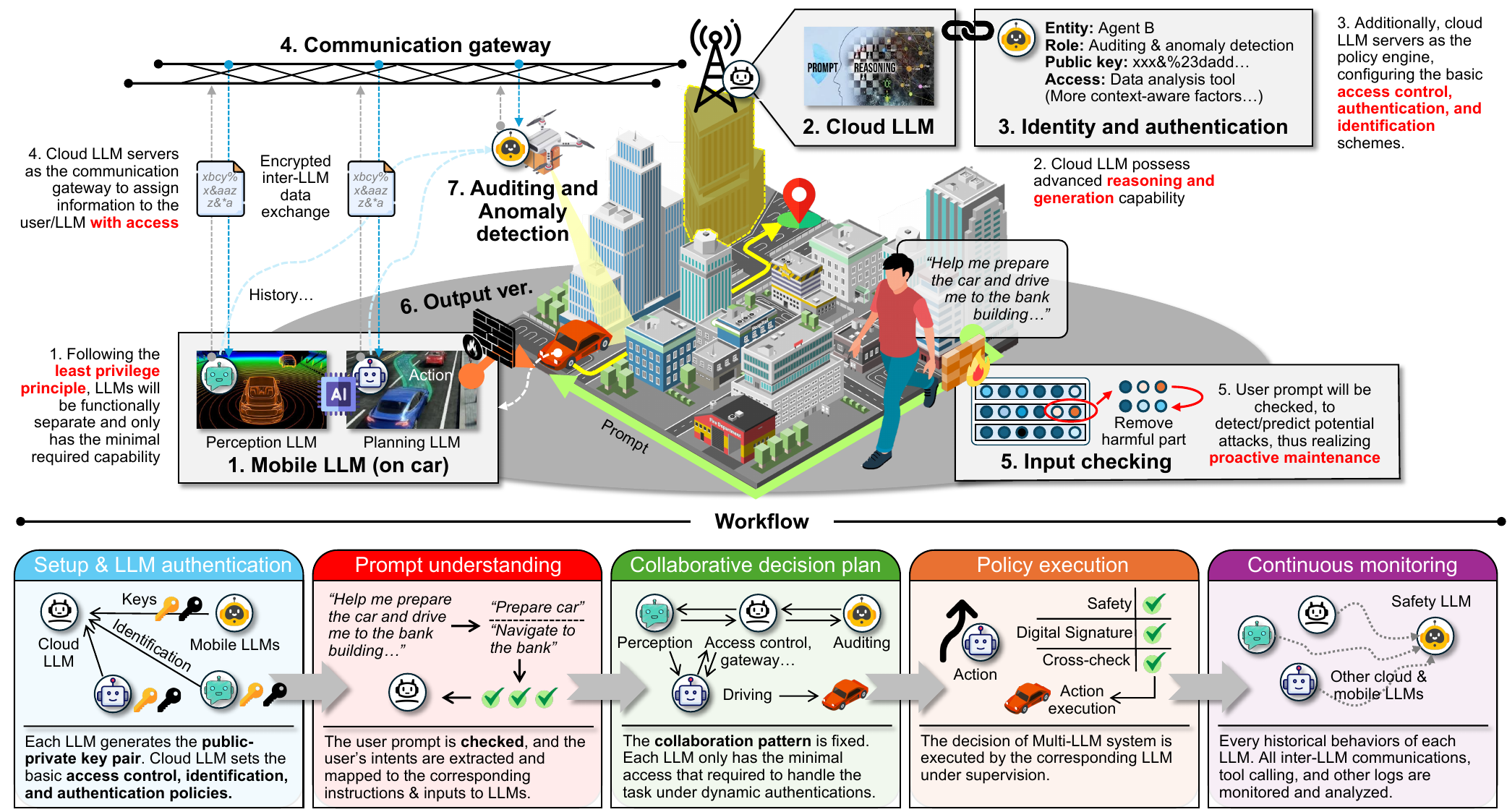}
\vspace{-0.6cm}
\caption{The vision of zero-trust multi-LLM in EGI. The upper part illustrates a scenario of autonomous driving, where users send a prompt about navigation to the multi-LLM system. The lower part demonstrates the end-to-end workflow.}
\vspace{-0.5cm}
\end{figure*}

\section{Zero-Trust Multi-LLM for EGI}

\subsection{Constructing Zero-Trust Multi-LLMs for EGI}
We observe that many advanced studies \cite{encrypt, pepagent, xiao, blockLLM, defensivetokens, MedSentry, LLMnet, blockchain4} have proposed defenses aligning with zero-trust principles or presented zero-trust single LLMs, while a comprehensive tutorial systematically discussing zero-trust multi-LLM systems in EGI has yet to emerge. 
Hence, we demonstrate the vision of zero-trust multi-LLM in EGI using an autonomous driving case study. 
As illustrated in Fig. 4, the EGI paradigm involves heterogeneous LLMs with diverse capabilities, computational requirements, and operational scopes deployed across cloud infrastructures, edge servers such as roadside units, and on-vehicle chips.
Note that we follow NIST SP 800-207 \cite{NIST}, the most widely adopted zero-trust standard, to showcase the architecture.
The subsequent parts detail the systematic implementation of each component of this framework.

\subsubsection{Mobile-Edge LLMs}
As shown in Fig. 4, mobile LLMs operate on edge servers and vehicular chips to ensure minimal latency for real-time decision-making in autonomous driving scenarios. 
In multi-LLM settings, these LLMs undergo functional segregation to handle distinct operational aspects and collaborate to accomplish the assigned task \cite{li2024survey}.
For instance, perception LLMs transform raw sensor output into structured textual scene descriptions, and planning LLMs interpret traffic regulations for high-level route planning decisions. 
Zero-trust enforcement restricts each LLM's access exclusively to the minimum necessary data and command interfaces. 
For example, perception LLMs can access camera and LiDAR feeds but cannot interface with steering or acceleration controls.
Upon system initialization, each LLM should undergo cryptographic identification to prevent Sybil attacks \cite{8793190}. 

\subsubsection{Cloud LLMs}
Cloud LLMs can offer computationally intensive generation capabilities, comprehensive knowledge repositories, and coordination services that surpass the capacity of mobile-edge LLMs. 
Moreover, following NIST zero-trust standards \cite{NIST}, cloud LLMs function as policy engines that establish and enforce security protocols throughout the distributed multi-LLM network while maintaining centralized oversight of system-wide security policies.


\subsubsection{Identity and Authentication Module}
At the core of the framework lies a comprehensive identity management system that establishes the foundation for all zero-trust operations. 
Every LLM, whether deployed at the edge or in the cloud, receives a cryptographically secure identity comprising unique asymmetric key pairs and digitally signed certificates \cite{10852158}. 
This module enforces mandatory identity verification for each interaction, eliminating the possibility of anonymous or unverified communications within the system.
Moreover, the authentication mechanism implements continuous verification rather than one-time startup authentication.
Session credentials, which can be based on multi-factor authentication \cite{9975323}, tags \cite{10062213}, or reputation \cite{10576030, 10413289} are deliberately short-lived and require frequent re-establishment to mitigate the impact of potential credential theft or compromise. 
This approach aligns with zero-trust principles by maintaining persistent identity verification throughout the operational lifecycle, ensuring that even previously authenticated entities must continuously prove their legitimacy to maintain system access.

\subsubsection{Inter-LLM Communication Gateway} 
As illustrated in Fig. 4, all inter-LLM communications within the multi-LLM system are channeled through a secure gateway deployed on the cloud LLM. 
This gateway functions as both an intelligent firewall and a message broker, implementing policy-controlled routing mechanisms that enforce strict communication boundaries between system components \cite{comm2, communication}. 
Specifically, it maintains dynamic and fine-grained access control policies that specify authorized communication patterns for each type of LLM. 
Any communication attempt that violates the policies is immediately intercepted and blocked by the gateway, preventing unauthorized data flows and potential security breaches. 
Moreover, transmitted content is encrypted using the sender LLM's public key and decrypted at the receiver using private keys, thereby preventing sensitive information from being intercepted or tampered with during transmission \cite{10852158}. 
These mechanisms ensure that no internal message routing occurs based on implicit trust assumptions about the source, requiring explicit verification for every communication attempt.

\subsubsection{User Input Checking}
User inputs are treated as potentially hostile data sources within the zero-trust framework, as they may originate from malicious attackers or be compromised through adversarial manipulation during transmission. 
These inputs could contain carefully crafted prompt injection attacks designed to bypass LLM safety guardrails and trigger jailbreaking behaviors that compromise system integrity \cite{defensivetokens}. 
The input validation module implements comprehensive sanitization mechanisms, including lexical analysis to detect suspicious instruction patterns, semantic filtering to identify context manipulation attempts, and behavioral pattern recognition to flag inputs deviating from normal user interaction profiles.
This proactive input filtering ensures that adversarial prompts cannot propagate through the multi-LLM system and maintains the principle of explicit verification.

\subsubsection{Multi-Layer LLM Output Verification} 
Similar to perimeter-based security, zero-trust multi-LLM systems incorporate a comprehensive multi-layer verification engine that embodies the ''never trust, always verify'' principle through hierarchical validation mechanisms. This multi-tiered filtering system can combine rule-based constraints, intelligent verification algorithms, and context-aware dynamic assessments to ensure that every LLM output undergoes rigorous validation before execution \cite{gehman-etal-2020-realtoxicityprompts}. 

\subsubsection{Behavioral Auditing and Anomaly Detection} 
Zero-trust requires persistent surveillance, since breaches are inevitable and threats can arise from any component of the entire multi-LLM system \cite{10764723}.
The behavioral auditing module implements comprehensive logging of all LLM communications, decisions, and collaborative interactions with tamper-resistant records stored in secure local repositories. 
Real-time anomaly detection algorithms analyze behavioral patterns to identify security threats, such as plan LLMs attempting direct vehicle control or outputs containing prompt injection tokens.

\subsection{Operational Workflow}
Fig. 4 illustrates the end-to-end operation of the zero-trust multi-LLM system in an autonomous driving scenario. Four major steps are included.
\begin{itemize}
    \item \textbf{Startup and LLM Authentication}: Initially, each mobile LLM generates its public-private key pair and registers its identity with the cloud LLM. The policy engine initializes access control mechanisms, input and output checking protocols, etc. Meanwhile, the behavioral auditing system begins comprehensive logging of all inter-LLM communications and individual LLM activities for anomaly detection.
    \item \textbf{Prompt Understanding}: User prompts undergo rigorous validation through the input checking module before processing. If anomalous content is detected, the framework immediately quarantines the input and initiates additional verification procedures to prevent system compromise \cite{NIST}.
    \item \textbf{Collaborative Decision Planning}: The primary advantage of multi-LLM systems lies in their ability to organize LLMs collaboratively \cite{mllm}. Multiple LLMs participate in decision-making processes in which the planning LLMs formulate initial decisions, the cloud LLM provides expert validation for complex scenarios, and the safety LLMs perform independent assessments of the collaboration process. The system supports multiple collaboration patterns, such as pipeline or distributed voting (see Section 2.1), based on the specific task. 
    \item \textbf{Policy-Governed Plan Execution}: The multi-LLM system generates final outputs that trigger executable vehicle commands, such as lane change maneuvers, speed adjustments, and braking actions \cite{llmad}. The multi-layer verification mechanism validates all LLM outputs against safety constraints, regulatory compliance, and operational feasibility before execution. Moreover, execution can only be initiated by designated LLMs that have explicit authorization for specific vehicle control functions.
    \item \textbf{Continuous Monitoring and Threat Mitigation}: The behavioral auditing system monitors all inter-LLM communications and individual LLM behaviors, maintaining comprehensive logs for forensic analysis and real-time threat detection. When anomaly detection mechanisms identify suspicious behavior or malicious content generation from any LLM, the cloud LLM isolates compromised LLMs while enabling the remaining LLMs to continue collaborative operation, ensuring system resilience and operational continuity \cite{10764723}.
\end{itemize}

\begin{table*}[tbp]
\centering
\caption{Perimeter-based security vs zero-trust security for multi-LLM systems}
\vspace{-0.3cm}
\label{tab:3n}
\footnotesize
\renewcommand{\arraystretch}{1.4}
\resizebox{\textwidth}{!}{%
\begin{tabular}{l|p{3cm}|p{6cm}|p{6cm}}
\hline
\hline
\rowcolor{gray!15} 
\textbf{Dimension} & \textbf{Sub-dimension} & \textbf{Traditional Perimeter Security} & \textbf{Zero-Trust Security} \\
\hline
\multirow{7}{*}{\textbf{Core Philosophy}}
& Basic Concept & Expanding trust perimeter in multi-LLM & Building a multi-LLM environment without trust \\
\cline{2-4}
& \multirow{2}{*}{Trust Assumption} & \multirow{2}{*}{Components within perimeter are implicitly trusted} & No component and entity should be unconditionally trusted at any time \\
\cline{2-4}
&
\multirow{6}{*}{\!\!
\begin{minipage}[t]{2.8cm}
Technical Enablers (Surveyed in Sections 3.3, 5, and 6)
\end{minipage}} & 
\begin{minipage}[t]{6cm}
\begin{itemize}
\item Enhance LLM capabilities against attacks (DP, adversarial training, RLHF)
\item Operate on trusted hardware (TEE)
\item Set security perimeters (firewalls, RBAC)
\item Establish trusted communication links
\end{itemize}
\end{minipage} & 
\begin{minipage}[t]{6cm}
\begin{itemize}
\item Strong identification and authentication
\item Context-aware access control
\item Stateless and Ephemeral LLM
\item Proactive maintenance
\item Distributed management
\item Micro-segmentation and Isolation
\item Intelligent monitoring and failure management
\end{itemize}
\vspace{1pt}
\end{minipage} \\
\hline
\multirow{6}{*}{\textbf{Implementation}}
& \multirow{2}{*}{Deployment} & Can \textbf{independently} use individual defense mechanisms, e.g., adversarial training or DP. & \textbf{Systems engineering} requiring multi-mechanism coordination \\
\cline{2-4}
& \multirow{4}{*}{Implementation Risk} & Failure of one protection mechanism may compromise overall defense & If only implementing partial zero-trust principles while ignoring others, may create serious vulnerabilities in a zero-trust environment. Example: Only doing identity verification without continuous monitoring allows attackers to persist after initial access \\
\hline
\multirow{7}{*}{\textbf{Resource Overhead}}
& \multirow{3}{*}{Training Overhead} & \textbf{High}: For instance, RLHF fine-tuning requires extensive human annotation; Adversarial training needs massive attack samples; DP training increases computational complexity & \textbf{Relatively Low}: Mainly relies on existing model capabilities; No large-scale retraining required \\
\cline{2-4}
& \multirow{4}{*}{Operational Overhead} & \textbf{Medium}: Periodic security rule updates; Hardware maintenance costs & \textbf{High}: Continuous authentication increases computational burden; Real-time monitoring requires substantial computing resources; Dynamic permission management increases system complexity; Multi-system coordination operational complexity \\
\hline
\multirow{13}{*}{\textbf{Applicable Scenarios}}
& \multirow{2.5}{*}{Environment Features} & Relatively closed environments; Simple threat models; Single LLM deployment; Resource-constrained scenarios & Distributed multi-LLM systems; EGI environments; High-security requirement scenarios; Complex threat environments; Cross-domain LLM collaboration \\
\cline{2-4}
&\multirow{2}{*}{Typical Applications} & Enterprise internal AI assistants, closed-domain QA systems & Autonomous driving systems, smart healthcare networks, multi-institutional collaborative AI \\
\cline{2-4}
&\multirow{9}{*}{Target Attacks in EGI} & \begin{itemize}\begin{minipage}[t]{5cm}
\item External penetration attacks (network intrusions, malicious external access, known threat patterns identified by the firewall)
\item Static threats (predefined attack patterns, signature-based malware, fixed attack vectors)
\end{minipage}
\end{itemize} & \begin{itemize}\begin{minipage}[t]{5cm}
\item Lateral movement (authorized user malicious behavior, privilege abuse, inter-LLM malicious propagation)
\item Dynamic \& complex attacks (prompt injection, jailbreaking, adversarial attacks, model poisoning)
\item Emerging \& unknown threats (emergent capabilities, multi-modal attacks, context-dependent threats)
\end{minipage}
\end{itemize} \\
\hline
\multirow{3}{*}{\textbf{Advantages}}
&  & Relatively simple implementation; Clear single-point optimization effects; High technology maturity; Controllable costs & Comprehensive security assurance; Adaptable to dynamic threats; Fine-grained control; Systematic protection \\
\hline
\multirow{3}{*}{\textbf{Limitations}}
&  & Once perimeter is breached, internal systems fully exposed; Difficult to handle insider threats; Static protection mechanisms; High lateral movement risks & Extremely high implementation complexity; Expensive operational costs; High technical team requirements; May impact system performance \\
\hline
\hline
\end{tabular}
}
\vspace{-0.5cm}
\end{table*}

\textbf{Lesson Learned}: By organizing the modules involved according to the workflow detailed above, we can construct a multi-LLM framework in EGI that effectively fulfills the four fundamental principles of the zero-trust paradigm (i.e., explicit verification, least privilege, continuous monitoring, and segmentation) \cite{10764723}. Note that our demonstrated implementation adheres to the most famous NIST SP 800-207 zero-trust standard \cite{NIST}, which employs a policy engine responsible for defining security policies (e.g., input/output validation rules and agent authentication), making real-time decisions on access and communication, and dynamically coordinating interactions among LLMs. Nevertheless, other influential Zero-trust standards and frameworks, such as Google's BeyondCorp\footnote{https://cloud.google.com/beyondcorp} and CISA\footnote{https://www.cisa.gov/zero-trust-maturity-model}, can also be applied.

\subsection{Trustworthy Multi-LLM versus Zero-Trust Multi-LLM}
To demonstrate the features of zero-trust multi-LLM systems, we compare them against previous ones operating with security perimeters (see Table 3). 
First, we observe that perimeter-based security and zero-trust security operate on fundamentally different assumptions, with different mechanisms. 
Moreover, traditional security approaches can rely on isolated security strategies that address individual vulnerabilities independently. 
For instance, we can employ DP to protect LLM training data or deploy LLMs within TEEs to ensure computational integrity, treating these as standalone solutions. 
However, constructing a zero-trust environment represents a comprehensive systems engineering challenge that requires an end-to-end workflow to orchestrate multiple modules, as demonstrated in our proposed framework above \cite{NIST}. 
From a resource management perspective, the traditional security paradigm might incur significant training costs and hardware overhead, such as extensive adversarial training and the deployment of TEEs. However, zero-trust approaches introduce additional operational overhead, primarily due to continuous verification processes \cite{10764723} and persistent authentication and monitoring requirements \cite{10589640}.

Note that it is important to recognize overlaps between traditional perimeter-based security methods and zero-trust strategies. Several foundational techniques, such as cryptographic authentication \cite{10852158} and access control mechanisms \cite{AC1, AC2}, apply to both concepts. For example, access control mechanisms can be utilized to construct security boundaries for LLMs by defining precise permissions for accessing specific tools and data. However, zero-trust imposes even higher demands on access control, such as context-aware policies, which dynamically adjust permissions based on real-time contextual factors for different micro-segmentations. Similarly, continuous monitoring, although prevalent in traditional security frameworks, assumes heightened importance in zero-trust environments. Intelligent monitoring ensures real-time detection and proactive response to threats, maintaining secure operations even as conditions evolve.

\begin{table*}[tbp]
\centering
\caption{Model-based approaches of zero-trust multi-LLM}
\vspace{-0.3cm}
\tiny
\renewcommand{\arraystretch}{1.4}
\begin{tabular}{>{\centering\arraybackslash}m{1.5cm}|>{\centering\arraybackslash}p{2.3cm}|>{\centering\arraybackslash}m{2cm}|p{6.3cm}}
\hline
\hline
\rowcolor{gray!15}
\textbf{} & \textbf{Introduction} & \textbf{Methods} & \textbf{Representative Works} \\
\hline

\multirow{10}{*}{\parbox{1.5cm}{\centering Strong LLM Identity, Authentication, and Authorization}} & 
\multirow{10}{*}{\parbox{2.3cm}{\centering Implements ``never trust, always verify'' principle by establishing continuous authentication and cryptographic identity verification for each LLM in EGI environments.}} & 
\multirow{3}{*}{\parbox{2cm}{\centering Multi-factor Authentication}} & 
\textbf{Adaptive MFA} \cite{10916520}: Dynamically adjusts verification requirements based on real-time risk assessments and agent behavior patterns. \\
\cline{4-4}
& & & \textbf{Context-aware MFA} \cite{9975323}: Balances security and usability by tightening protocols during detected anomalies using contextual information. \\
\cline{3-4}

& & \multirow{3}{*}{\parbox{2cm}{\centering Reputation-based Authentication}} & 
\textbf{LLMChain} \cite{10576030}: Maintains reputation scores based on historical outputs, policy adherence, and user feedback for access control. \\
\cline{4-4}
& & & \textbf{Blockchain Reputation} \cite{10633559}: Provides transparent, verifiable, and immutable reputation histories using blockchain technology. \\
\cline{3-4}

& & \multirow{3}{*}{\parbox{2cm}{\centering Token-based Authentication}} & 
\textbf{Fine-grained Token Control} \cite{encrypt}: Issues ephemeral, cryptographically secure tokens encoding specific permissions for LLM interactions. \\
\cline{4-4}
& & & \textbf{Ephemeral Token Management} \cite{10764723}: Enforces principle of least privilege through continuous re-authentication and token expiration. \\
\hline

\multirow{7}{*}{\parbox{1.5cm}{\centering Context-aware Access Control}} & 
\multirow{7}{*}{\parbox{2.3cm}{\centering Implements ``least privilege'' principle by context-aware permission management that grants the minimal necessary access rights to LLMs.}} & 
 & 
\textbf{AgentSafe} \cite{agentsafe}: Protects multi-LLM systems through hierarchical data management with ThreatSieve and HierarCache components. \\
\cline{4-4}

& & & 
\textbf{EPEAgents} \cite{pepagent}: Minimizes data exchange by providing each LLM only task-relevant information through context-aware filtering. \\
\cline{4-4}

& &  & 
\textbf{ABE-based Access Control} \cite{xiao}: Uses Attribute-Based Encryption with policy hiding to ensure only authorized LLMs can decrypt sensitive information. \\
\cline{4-4}

& & & 
\textbf{Collaborative Memory Framework} \cite{accesscontrol1}: Encodes memory access permissions as time-evolving bipartite graphs with context-aware policies. \\
\hline

\multirow{9}{*}{\parbox{1.5cm}{\centering Stateless and Ephemeral LLM Management}} & 
\multirow{9}{*}{\parbox{2.3cm}{\centering Embodies ``assume breach'' principle through eliminating persistent state and creating disposable LLMs in EGI and isolated execution contexts.}} & 
\multirow{5}{*}{\parbox{2cm}{\centering Stateless Management}} & 
\textbf{PagedAttention} \cite{pageattention}: Implements process-like isolation with dynamic memory allocation and copy-on-write semantics for request isolation. \\
\cline{4-4}

& & & \textbf{vAttention} \cite{vattention}: Advances stateless security through hardware-level isolation using CUDA virtual memory APIs for protected address spaces. \\
\cline{4-4}

& & & \textbf{BlockLLM} \cite{blockLLM}: Extends stateless principles through component-level micro-segmentation with cryptographically isolated blocks. \\
\cline{3-4}

& & \multirow{3}{*}{\parbox{2cm}{\centering Ephemeral Management}} & 
\textbf{Self-destructing Models} \cite{ephemeral}: Embed algorithmic time locks that degrade model performance when adapted for harmful tasks. \\
\cline{4-4}

& & & \textbf{Serverless Deployment} \cite{serveless1}: Realizes automatic lifecycle management with security checkpoints and anomaly-triggered termination. \\
\hline
\hline

\end{tabular}
\vspace{-0.4cm}
\end{table*}

\section{Intra-LLM Zero-trust Security: Model Approaches}
In this section, we review the technical progress of model-level approaches that provide zero-trust management of individual LLMs.

\subsection{Strong LLM Identity, Authentication, and Authorization}
First, each LLM in a zero-trust architecture is assigned a robust cryptographic identity to ensure secure and verifiable identification. 
Practically, this involves providing every LLM with a unique asymmetric key pair and a digital certificate, similar to IoT or embedded devices that use device-specific keys \cite{10852158} or X.509 \cite{9110311} certificates to authenticate identity. 
Moreover, every user-to-LLM, inter-LLM, and LLM-to-tool call requires prior authentication. 
Note that authentication is continuous rather than a one-time handshake; credentials remain short-lived and frequently renewed, embodying the zero-trust principle of persistent verification \cite{10589640}. 
Representative authentication approaches are described below (see Table 4 also).

\subsubsection{Multi-factor Authentication}
MFA introduces additional verification layers beyond a single credential, forming a cornerstone of zero-trust security \cite{10916520}. By requiring LLMs to present multiple independent proofs of identity, such as cryptographic keys combined with contextual information (e.g., device attestation, biometric verification, or behavior analytics), MFA substantially reduces the risk of unauthorized access due to compromised credentials. Recent research emphasizes context-aware or adaptive MFA, dynamically adjusting verification requirements based on real-time risk assessments, including agent behavior, device integrity, location, and access patterns \cite{9975323}. This dynamic MFA approach effectively balances security and usability by tightening security protocols during detected anomalies. These adaptive mechanisms, initially designed for device authentication, are particularly applicable to authenticating LLMs in multi-LLM zero-trust EGI, with heterogeneous devices and application patterns \cite{10.1145/3336117}.

\subsubsection{Reputation-based Authentication}
In zero-trust multi-LLM systems, each LLM can maintain a reputation derived from historical outputs, policy adherence, and user/peer feedback \cite{10576030, 10413289, reputation3}. Suspicious or erroneous behavior results in decreased reputation scores, triggering restricted privileges, while consistently reliable LLMs accrue higher reputation scores and consequently gain access to more sensitive tools or data. Moreover, blockchain technology has been proposed to securely track reputation, providing transparent, verifiable, and immutable reputation histories \cite{9500841}. For example, LLMChain integrates automated performance assessments and human feedback within a blockchain ledger to calculate comprehensive, tamper-proof reputation scores \cite{10633559}.

\subsubsection{Token-Based Authentication}
Chen \textit{et al.} \cite{encrypt} provided fine-grained authorization through encrypted tokens encoding specific permissions for LLM interactions. Specifically, a central provider, potentially served by cloud LLMs within multi-LLM EGI architectures, maintains LLM identities and access policies, issuing ephemeral, cryptographically secure tokens defining permitted interactions. LLMs should obtain fresh tokens for each session or interaction, strictly limiting privileges to the immediate task. Notably, each token has built-in expiration conditions, requiring continuous re-authentication. Such adaptive authentication effectively enforces the principle of least privilege \cite{10764723}, ensuring that permissions are granted per request and revoked as soon as tasks are completed, aligning seamlessly with zero-trust principles of persistent verification and minimal trust.

\textbf{Lesson Learned}: We can conclude that MFA provides the strongest security through layered verification, but introduces computational overhead suitable for high-security applications, while reputation offers dynamic trustworthiness management ideal for long-term deployments with established behavior patterns. Token-based authentication delivers fine-grained access control with minimal overhead, making it effective for distributed systems requiring precise privilege control. 

\subsection{Context-aware Access Control}
Zero-trust principles demand stringent access control to prevent privilege abuse and information leakage in multi-LLM systems. 
However, traditional RBAC, such as simple role-based permissions \cite{8466653}, is insufficient for zero-trust LLM collaboration. 
RBAC assumes that if an LLM's identity or role is verified, it can be trusted broadly, which is an assumption zero-trust explicitly rejects. 

Context-aware Access Control (CAAC) frameworks extend beyond identity: ``context'' can include any information about the user/LLM, the resource, or the environment of an access request~\cite{s20092464}. 
For instance, some proposals augment the RBAC with conditions, such as location, time of request, or other situational constraints~\cite{s20092464}. 
Such context-driven policies allow fine-grained, situation-specific decisions (e.g., permitting an LLM to see certain data only during a particular task or within certain time bounds), aligning with zero-trust's demand for dynamic and least-trust enforcement. 

Furthermore, multiple recent works propose advanced access control mechanisms tailored to multi-LLM systems that embody zero-trust principles. AgentSafe \cite{agentsafe} is one such framework that protects multi-LLM systems through hierarchical data management. It classifies information into security levels and restricts sensitive data to only LLMs with authorization. Moreover, it introduces two components: ThreatSieve and HierarCache. ThreatSieve secures inter-LLM communication by authenticating each message's source and verifying the sender's authority, blocking unauthorized or impersonating LLMs. HierarCache, on the other hand, hierarchically manages LLM memory to prevent data leakage and poisoning. Historical information is stored in ``drawers'' according to sensitivity and LLM relationships, so each LLM only retains or sees memory appropriate to its trust level. Notably, AgentSafe's design is dynamic: as LLMs join/leave or as tasks change, the hierarchy can adjust. Complementing the hierarchical approach, the Embedded Privacy-Enhancing Agents (EPEAgents) \cite{pepagent} paradigm focuses on context-aware filtering of information in multi-LLM systems. Specifically, cloud LLM will coordinate context sharing during LLM collaboration. The core idea is to minimize data exchange: rather than broadcasting a raw knowledge base or full context to all LLMs, each LLM receives only the task-relevant information that it truly needs. Importantly, EPEAgents is built to preserve contextual awareness for legitimate collaboration. Unlike rigid RBAC, which might omit crucial context and degrade performance, EPEAgent ensures each LLM still gets the context it needs by dynamically evaluating both the LLM's declared role and the evolving task requirements, allowing the system to adapt to changing collaboration patterns while maintaining strict information boundaries. 

Other works bring cryptographic enforcement to access control for multi-LLM systems, using fine-grained encryption to guarantee that even if communications are observed by attackers, only authorized LLMs can decrypt and use sensitive information. Xiao \textit{et al.} \cite{xiao} proposed a privacy-preserving access control scheme for LLM-driven networks based on Attribute-Based Encryption (ABE) with policy hiding and revocation. In their design, each piece of content (e.g., a dataset, prompt, or answer) can be encrypted under an access policy (defined as a logical combination of attributes) such that only an LLM possessing a key with attributes satisfying that policy can decrypt it. This allows extremely fine-grained control: policies might encode that ``\texttt{only an LLM with role = doctor AND clearance = high}'' can decrypt a medical query, or ``\texttt{LLMs with project\_ID = 1 OR role = auditor}'' can access a certain log. Unlike RBAC, ABE does not grant blanket access to any LLM, while it evaluates the attributes presented at decryption time, enabling contextual decisions (attributes include an LLM's role, affiliation, and user prompts). 

Beyond these, emerging research on dynamic access control provides additional tools to strengthen zero-trust LLM systems. For example, Rezazadeh \textit{et al.} \cite{accesscontrol1} presented a ``collaborative memory'' framework for multi-LLM systems that encodes memory access permissions as a time-evolving bipartite graph linking users, LLMs, and resources. Specifically, the framework maintains separate private and shared memory tiers for each user, with each memory fragment tagged by provenance (which LLMs contributed it, which LLMs can access it, etc.). Context-aware and fine-grained read/write policies then operate on this graph: a read policy might say that LLM $A$ can retrieve data from user $B$'s shared memory only if an edge exists between $A$ and $B$ and the fragment is not time-expired, etc., resulting in filtered, transformed views of memory for each LLM.

\vspace{-0.15cm}
\subsection{Stateless and Ephemeral LLM Management}
Traditionally, multiple LLMs maintain a shared memory pool to preserve states, which accumulate information across multiple requests to optimize inference performance and enable contextual understanding \cite{memory1, accesscontrol1}. 
However, this persistent state cannot be trusted, as malicious inputs can contaminate shared memory spaces, enabling cross-request information extraction and state poisoning attacks that compromise the integrity of future interactions.
In contrast, stateless LLM management \cite{pageattention, vattention} fundamentally rejects the traditional assumption that LLMs' internal states can be trusted over time, instead requiring continuous verification of every component while minimizing the vulnerability window through short-lived, isolated execution contexts. 

\subsubsection{Stateless LLM}
The evolution of stateless LLMs has been driven by breakthrough innovations in memory management and isolation techniques. 
PagedAttention \cite{pageattention} established the foundational paradigm by implementing process-like isolation where each request is dynamically allocated with memory blocks. 
This approach partitions the key-value cache into fixed-size blocks with copy-on-write semantics, ensuring complete isolation between requests and automatic memory reclamation upon completion, thereby eliminating the persistent shared state.
Building on this foundation, vAttention \cite{vattention} advanced stateless security through hardware-level isolation using CUDA virtual memory APIs \cite{CUDA}, creating true memory virtualization where each request operates in protected virtual address spaces. 
BlockLLM \cite{blockLLM} extended stateless principles to multi-tenant environments through component-level micro-segmentation, dividing LLMs into cryptographically isolated blocks where each tenant processes requests through separate block instances. 
This architecture prevents cross-tenant information leakage while implementing per-block access control policies that ensure compromises cannot propagate beyond the current request, directly implementing zero-trust micro-segmentation principles at the finest possible granularity.

\subsubsection{Ephemeral LLM}
Ephemeral LLM management \cite{ephemeral} extends the stateless paradigm by introducing temporally-bounded LLM instances that automatically self-destruct after predetermined conditions are met, directly implementing zero-trust's ``assume breach'' principle through proactive threat containment. Henderson \textit{et al.} \cite{ephemeral} pioneered self-destructing models, which embed algorithmic time locks that degrade model performance when adapted for harmful tasks such as generating toxic content or bypassing safety guardrails, ensuring that compromised instances lose their utility for malicious purposes without external intervention. This approach transforms LLM instances into inherently untrusted, which aligns with zero-trust's fundamental rejection of persistent trust relationships. Furthermore, ephemeral LLMs can be deployed on serverless computing platforms \cite{serveless1} to realize automatic lifecycle management with multiple security checkpoints.

\section{Inter-LLM Zero-trust Security: System Approaches}
In this section, we review the technical progress of system-level approaches, which ensure zero-trust security between LLMs and eventually build the zero-trust multi-LLM system.

\begin{table*}[tbp]
\centering
\caption{The representative proactive maintenance approaches to realize zero-trust multi-LLM in EGI}
\vspace{-0.3cm}
\label{tab:4}
\tiny
\renewcommand{\arraystretch}{1.3}
\begin{tabular}{>{\centering\arraybackslash}m{1.3cm}|>{\centering\arraybackslash}m{3.5cm}|p{7.9cm}}
\hline
\hline
\rowcolor{gray!15}
\textbf{Methods} & \textbf{Description} & \textbf{Representative Works} \\
\hline
\multirow{9}{*}{\parbox{1.3cm}{\centering Intelligent Input Checking}} & 
\multirow{9}{*}{\parbox{3.5cm}{\centering Preemptively analyzes and verifies user prompts to detect and neutralize malicious instructions or jailbreak attempts before LLMs generate harmful content.}} & 
\textbf{LLM Prompt Detection System} \cite{10827351}: Employs regular expressions and fine-tuned LLMs to identify PII, malicious codes, URLs, and prompt injection commands. \\
\cline{3-3}
& & \textbf{DefensiveTokens} \cite{defensivetokens}: Embeds special tokens into user prompts to disrupt malicious patterns and prevent injection attacks. \\
\cline{3-3}
& & \textbf{JailGuard} \cite{JailGuard}: Creates multiple mutated variants of user prompts to defend against prompt-based attacks through pattern disruption. \\
\cline{3-3}
& & \textbf{Layered Filtering Framework} \cite{10755823}: Uses GPT-4 as attack validator to achieve 97\% detection accuracy for prompt injection attacks. \\
\cline{3-3}
& & \textbf{SecurityLingua} \cite{SecurityLingua}: Employs security-aware prompt compression to highlight critical security instructions and remove distracting context. \\
\hline
\multirow{3.5}{*}{\parbox{1.3cm}{\centering Reputation Schemes}} & 
\multirow{3.5}{*}{\parbox{3.5cm}{\centering Evaluates LLM through reputation to proactively isolate suspicious models before causing damage.}} & 
\textbf{LLMChain Framework} \cite{10633559}: Evaluates LLM reliability based on previous interactions to identify potentially malicious LLMs proactively. \\
\cline{3-3}
& & \textbf{PsySafe} \cite{psysafe}: Employs psychology-based framework to assess LLM behaviors and psychological states for early malicious tendency identification. \\
\hline
\multirow{7}{*}{\parbox{1.3cm}{\centering Topology-aware Proactive Maintenance}} & 
\multirow{7}{*}{\parbox{3.5cm}{\centering Examines network-wide LLM interaction patterns and topological configurations to predict and prevent misinformation propagation throughout the system.}} & 
\textbf{MedSentry} \cite{MedSentry}: Implements comprehensive evaluation pipeline that benchmarks risk in representative multi-LLM topologies for early vulnerability identification. \\
\cline{3-3}
& & \textbf{NetSafe} \cite{netsafe}: Examines topological aspects of LLM interactions to determine how network configurations influence misinformation propagation. \\
\cline{3-3}
& & \textbf{G-Safeguard} \cite{safeguard}: Uses temporal graph modeling to detect anomalies in LLM interactions and disrupt harmful content spread. \\
\cline{3-3}
& & \textbf{Guardian} \cite{guardian}: Leverages topology-guided graph neural networks to proactively identify and isolate malicious LLMs in collaborative environments. \\
\hline
\hline
\end{tabular}
\vspace{-0.3cm}
\end{table*}
\subsection{Proactive Maintenance}
Traditional perimeter-based security approaches typically employ reactive response mechanisms, taking actions after attacks are launched \cite{netsafe, MedSentry, 10827351}.
However, in multi-LLM collaborative environments, this reactive nature may allow malicious information to propagate and amplify throughout the edge network before detection.
The zero-trust principle of ``never trust, always verify'' demands more proactive security maintenance strategies that can anticipate and prevent security incidents before they materialize.

\subsubsection{Intelligent Input Checking}
An effective proactive strategy involves input checking, specifically analyzing and verifying user prompts to preemptively detect and neutralize potential threats before the LLM generates harmful content. 
For instance, Kim \textit{et al.} \cite{10827351} proposed a comprehensive LLM prompt detection system that employs regular expressions and fine-tuned LLMs to identify personally identifiable information, malicious codes, URLs, and prompt injection commands within user prompts. 
Chen \textit{et al.} \cite{defensivetokens} introduced DefensiveTokens, which embed special tokens into user prompts. 
These tokens can disrupt or alter malicious patterns in the raw prompt, making it more difficult for attackers to construct successful injection attacks.
Similarly, Zhang \textit{et al.} \cite{JailGuard} presented JailGuard, which defends prompt-based attacks by creating multiple mutated variants of the user prompt (e.g., random replacement, targeted insertion). 
Muliarevych \textit{et al.} \cite{10755823} explored a layered filtering framework, which includes a prompt analyzer to preprocess and wrap user inputs, and an attack validator that evaluates the wrapped prompts for malicious intent. 
Utilizing GPT-4 as the attacker validator, this layered filtering subsystem achieved superior effectiveness, accurately detecting 97\% of various prompt injection attacks.
SecurityLingua \cite{SecurityLingua} further enhanced proactive security by employing security-aware prompt compression. This method uses a specialized compressor model trained on malicious prompt patterns to effectively highlight critical security-related instructions within user prompts. By compressing the prompts, SecurityLingua removes extraneous, potentially distracting context that attackers often introduce to mask harmful intentions. 
\begin{figure*}[tpb]
\centering
\includegraphics[width=1\textwidth]{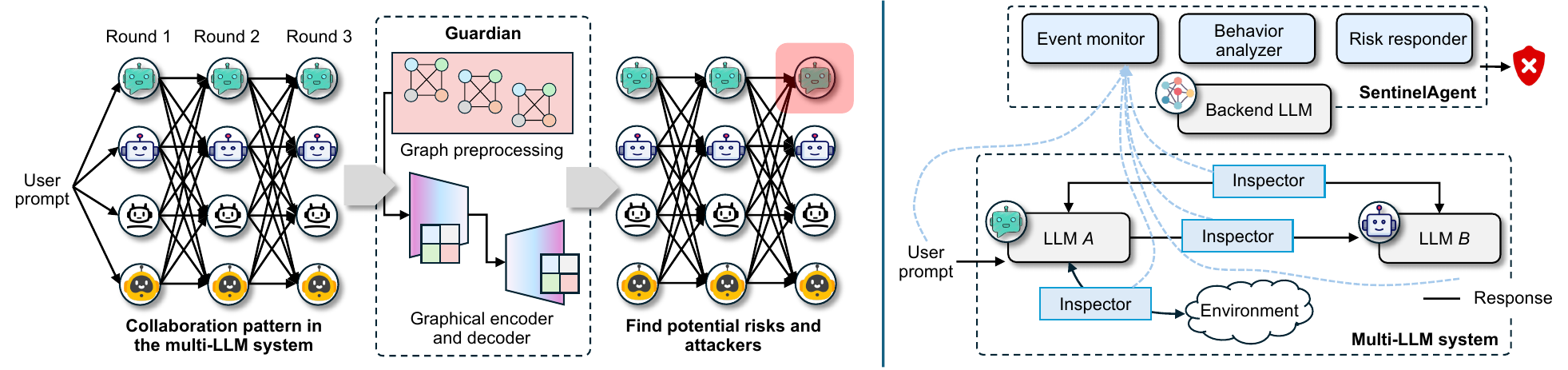}
\vspace{-0.6cm}
\caption{(left): The illustration of Guardian \cite{guardian}. It analyzes the collaboration pattern of multiple LLMs to detect potential risks and make proactive maintenance. (right): The illustration of SentinelAgent \cite{SentinelAgent}. It monitors all logs and behaviors in multi-LLM systems and utilizes an LLM to analyze data intelligently.}
\label{2}
\vspace{-0.6cm}
\end{figure*}

\subsubsection{Reputation Schemes}
In addition to input verification, the reputation of each LLM, which is based on its historical behavior and content generation quality, can be utilized as a metric to perform proactive maintenance. 
Specifically, LLMs whose reputation falls below a predefined threshold due to suspicious activities or poor-quality content can be proactively isolated or assigned lower priorities, thus minimizing potential damage. 
For instance, the LLMChain framework \cite{10633559} demonstrated such a reputation system, evaluating LLM reliability based on previous interactions, thereby identifying potentially malicious LLMs and mitigating the damage they can cause.
In addition to reputation, other assessments can be applied from interdisciplinary perspectives.
PsySafe \cite{psysafe} presented a psychology-based framework that assesses multi-LLM systems by evaluating LLM behaviors and psychological states, allowing early identification of malicious tendencies and proactive intervention. 

\subsubsection{Topology-aware Proactive Maintenance}
Beyond individual LLM assessments, network-wide safety evaluations play a crucial role in predictive security maintenance. 
MedSentry \cite{MedSentry}, for instance, implemented a comprehensive evaluation pipeline that systematically benchmarks the risk existing in representative multi-LLM topologies. 
This allows vulnerabilities to be identified early, facilitating preemptive actions to mitigate misinformation or harmful content before impacting practical services.
NetSafe \cite{netsafe} further extended this concept by examining the topological aspects of LLM interactions, highlighting how network configurations influence the propagation of misinformation or harmful behaviors, thus determining preventive strategies.
Similarly, G-Safeguard \cite{safeguard} employed a temporal graph modeling technique to detect and remediate anomalies in LLM interactions. By analyzing communication patterns over time, G-Safeguard identified anomalous LLMs and disrupted the spread of harmful content, demonstrating significant effectiveness in containing and neutralizing threats early in their propagation.
Lastly, Guardian \cite{guardian} leveraged topology-guided graph neural networks to proactively identify malicious LLMs within collaborative multi-LLM environments (see Fig. 5(left)). By systematically screening, adjudicating, and isolating malicious LLMs based on their interactions, Guardian maintains robust decision-making and prevents information contamination.

\textbf{Lesson Learned}: These three proactive maintenance approaches demonstrate a progressive defense strategy with increasing scope and sophistication. Input checking provides immediate threat prevention by analyzing individual prompts at the entry point, offering rapid responses. Reputation schemes advance this approach by leveraging historical behavioral data to proactively isolate potentially malicious LLMs based on accumulated trust scores, enabling predictive intervention before threats materialize. Network topology-aware maintenance operates from a macroscopic perspective, analyzing system-wide interaction patterns and communication flows to identify vulnerabilities and anomalies across the entire multi-LLM ecosystem.


\subsection{Blockchain and Distributed Management}
Blockchain is a decentralized ledger system that enables tamper-resistant record keeping on public networks \cite{8664132, 10908689, 9061111}. 
A distinctive characteristic of blockchain is the elimination of trusted authorities, operating under the assumption that any node could potentially be malicious, and achieving Byzantine fault tolerance through distributed data storage and consensus mechanisms \cite{9271868}. 
This property makes blockchain particularly valuable for constructing zero-trust multi-LLM environments in EGI scenarios.
To be specific, first, distributed data storage preserves interaction records and historical content generated by LLMs across multiple nodes, preventing records from being tampered with by any single entity. 
Additionally, blockchain leverages consensus mechanisms to resist Byzantine attacks, where malicious nodes may attempt to manipulate the collaborative process or inject false information \cite{blockchain2}. 
Finally, smart contracts define various operational rules and trigger conditions that can be automatically executed without human intervention and cannot be influenced by attackers, making blockchains ideal for managing collaborative processes and responding to anomalous behaviors \cite{10634448}.

For data storage, Mo \textit{et al.} \cite{blockchain5} presented a blockchain-based crowdsourcing evaluation framework for LLMs, where LLMs' reputation scores are recorded immutably on the blockchain. 
This approach ensures that evaluation scores cannot be retroactively modified, maintains a transparent history of LLM performance, and enables fair reputation-based selection of high-quality LLMs for future tasks. 
LLM-Net \cite{LLMnet} constructed a multi-LLM network that divides LLMs into different roles, including coordinators, respondents, and validators. Validators execute blockchain consensus mechanisms to verify content generated by respondents, while coordinators run smart contracts to manage the collaborative process. Furthermore, all interactions and transactions are preserved on the blockchain, ensuring transparency and accountability.
Similarly, Luo et al. \cite{blockchain2} also utilized blockchain to construct decentralized multi-LLM networks.
Notably, they evaluated the efficiency of four mainstream consensus algorithms in supporting multi-LLM operations and provided insights for selecting the most suitable one based on network latency and throughput requirements. 
BlockAgents \cite{blockchain3} proposed a novel consensus mechanism, namely Proof-of-Thought (PoT), which is designed specifically for multi-LLM systems. 
Specifically, PoT ensures that LLMs contributing the most valuable insights to the collaborative reasoning process acquire the highest voting priority, while incorporating stake-based miner designation and multi-round debate-style voting to prevent malicious LLMs from dominating the decision-making process.
Wang et al. \cite{blockchain4} developed a threshold signature algorithm suitable for blockchain-assisted multi-LLM systems. 
Decisions agreed upon by multiple LLMs are collaboratively signed by everyone and written on the blockchain, enabling traceability and accountability while ensuring that no single entity can control the entire signature process.
Karanjai et al. \cite{10634448} employed smart contracts to manage the execution of LLM inference on edge devices, preventing attackers from launching DDoS attacks on EGI.

Apart from blockchain, Multi-Party Secure Computation (MPC) \cite{11011099} has emerged as a critical technique for constructing zero-trust multi-LLM systems by enabling secure collaborative computations without revealing individual data inputs.
This is particularly important in scenarios involving federated fine-tuning and decentralized inference, where maintaining data confidentiality and preventing unauthorized access to sensitive model information are paramount.
Additionally, Zero-Knowledge Proofs (ZKP) \cite{zkp} enhance zero-trust principles by enabling the verification of computations across multiple LLMs without revealing the underlying LLM parameters or private data. 
Qu \textit{et al.} \cite{zkp} introduced zkGPT, a non-interactive ZKP framework specifically tailored for efficient LLM inference. 
By generating cryptographic proofs that verify the correctness of inference results without exposing confidential model parameters, zkGPT aligns perfectly with the principles of zero-trust architectures, ensuring verifiability and integrity of computations in distributed multi-LLM systems.

\subsection{Micro-segmentation and Isolation}
Micro-segmentation \cite{NIST} is a critical zero-trust security technique that divides network infrastructure into smaller, isolated segments to limit lateral movement (i.e., attackers moving from one compromised system to adjacent systems within the network) and contain potential breaches. 
In EGI with multiple LLMs, micro-segmentation becomes particularly critical due to the diverse service requirements, sensitive data processing, and heterogeneous device configurations. 

Recent advances in wireless network slicing have demonstrated significant potential for enhancing security isolation in multi-LLM deployments.
Liu \textit{et al.} \cite{WLLM} introduced WiLLM, the first open-source wireless communication system specifically designed for LLM services through its innovative ``Tree-Branch-Fruit'' slicing architecture. 
This hierarchical framework enables dedicated communication channels for different LLM services, ensuring that computational and communication resources are isolated between different slices while allowing telecom operators to monetize services through slice subscriptions. 
The approach directly supports zero-trust principles by establishing strict boundaries between LLM services, preventing unauthorized inter-LLM communication, and enabling granular access controls at the network level.
Based on this foundation, Liu \textit{et al.} \cite{10.1145/3666025.3699404} further demonstrated the practical implementation of dedicated network slicing for LLMs through their proposed LLM-Slice system, which creates LLM-specific network slices to efficiently bind services and communication resources. 
The system enables different LLMs, including Google Bard\footnote{https://gemini.google.com/app}, Meta LLaMA\footnote{https://www.llama.com/}, and ChatGPT \cite{chatGPT}, to coexist with independent resource allocation and management, ensuring that potential security breaches in one slice cannot propagate to others. 
Furthermore, a permissions database is used to enforce strict authentication policies for each LLM. 
Finally, an intelligent controller continuously validates cross-slice communications, enabling real-time threat detection that aligns with zero-trust's continuous monitoring \cite{10.1145/3666025.3699404}.

\subsection{Intelligent System Monitoring and Failure Management}
Zero-trust multi-LLM systems in EGI demand continuous monitoring and robust failure management to ensure reliability and safety. Recent studies have summarized numerous failure modes unique to multi-LLM setups that underscore the need for comprehensive monitoring \cite{10764723}.
For instance, Cemri \textit{et al.} \cite{fail} identified 14 distinct failure modes in popular multi-LLM frameworks, ranging from coordination breakdowns (i.e., one LLM's action invalidates another's) to more subtle issues like shared misconceptions that propagate across the network. 
The complexity and variety of these failure modes highlight why traditional guardrails are insufficient for multi-LLM systems, necessitating more sophisticated monitoring approaches that can detect distributed failures.

Multi-LLM monitoring has followed a clear evolutionary path, progressively aligning with the stringent demands of zero-trust architectures. 
Initially, oversight relied on manually crafted rules or static guardrails, such as reactive output filters or step-by-step validators \cite{Step} to intercept disallowed actions. 
Despite effectively detecting straightforward violations, such as unauthorized tool calls or toxic outputs, these methods fundamentally fall short of zero-trust requirements due to their inability to dynamically adapt to evolving threats beyond predefined rules.
To this end, researchers introduced traditional anomaly detection and machine learning techniques \cite{10024295}, capitalizing on the rich telemetry from multi-LLM systems, such as action and interaction logs, performance metrics, and state-change traces, to identify deviations from expected behavior. 
Although these approaches expand monitoring capabilities and handle extensive data, they treat LLM systems as black-box output generators, lacking the nuanced semantic understanding essential for comprehensive oversight in zero-trust environments.

The latest advancement employs LLMs themselves as monitoring agents. 
For instance, Zhang \textit{et al.} \cite{agentFM} proposed AgentFM, which organizes multiple specialized LLMs to monitor activities specific to its designated role.
Additionally, a higher-level meta-LLM integrates these role-specific insights, enabling the detection and handling of cross-role anomalies, effectively supporting the nuanced oversight required by zero-trust environments.
As shown in Fig. 5(right), He \textit{et al.} \cite{SentinelAgent} introduced SentinelAgent, which implements real-time anomaly detection and intervention by dynamically constructing interaction graphs of LLM interactions. 
This framework detects anomalies at various granularities: individual LLM misbehavior at the node level, unsafe inter-LLM communications at the edge level, and suspicious exploit chains at the path level. 
Notably, a pluggable LLM-based monitoring agent is leveraged for semantic analysis, enabling active intervention such as interrupting harmful behaviors and correcting multi-LLM collusion, thereby offering robust zero-trust monitoring capabilities suited for complex multi-LLM ecosystems.
Chang \textit{et al.} \cite{saga} presented SagaLLM, which structures each task into independent, compensable units. In this framework, each operational LLM is paired with a dedicated compensation LLM responsible for rolling back and correcting any faulty operations. 
Independent validating LLMs are employed to assess intermediate and final outputs, detecting inconsistencies such as logical contradictions or coordination errors between LLMs. Upon detecting anomalies, SagaLLM automatically initiates structured rollbacks.

\begin{remark}
The technical progress reviewed in Sections 5 and 6 contributes to the implementation and advancement of multi-LLM zero-trust systems in EGI (such as the one demonstrated in Section 4). For instance, the strong identity protocols and context-aware access control can realize the ``Identity and Authentication Module'', proactive maintenance enables ``User Input Checking'' and ``Multi-Layer LLM Output Verification'' modules, and intelligent monitoring schemes realize ``Behavioral Auditing and Anomaly Detection.''
\end{remark}
}

\section{Challenges and Future Research Directions}
\subsection{Ethical and Societal Issues}
The deployment of zero-trust multi-LLM systems in EGI raises ethical and societal implications that extend beyond technical security considerations. Unlike traditional AI systems, multi-LLM systems in EGI directly interact with critical societal infrastructure, autonomous vehicles, and healthcare networks, amplifying the potential impact of algorithmic bias, discrimination, and social harm \cite{10947002}. The zero-trust paradigm's fundamental assumption of ``never trust, always verify'' introduces additional ethical complexities when applied to human-AI interactions, potentially undermining user trust and creating psychological barriers to system adoption.

Future research should address several critical challenges: 1) developing ethical frameworks for algorithmic accountability in distributed multi-LLM decision-making, where responsibility attribution becomes complex due to the collaborative nature of reasoning processes and the absence of centralized control mechanisms, 2) establishing fairness-preserving zero-trust protocols that prevent discriminatory outcomes while maintaining security guarantees, particularly when LLMs trained on biased datasets collaborate in safety-critical applications, and 3) designing transparent governance mechanisms that enable public oversight and democratic participation in zero-trust multi-LLM system deployment decisions. 

\vspace{-0.1cm}
\subsection{Asymmetric Information and Network Heterogeneity}
The heterogeneous nature of EGI networks presents fundamental challenges in implementing unified zero-trust multi-LLM frameworks due to asymmetric information sharing capabilities and widely varying communication conditions among edge devices \cite{8624371}. The stark disparity between high-capacity wired backhaul connections (e.g., between roadside units and base stations with negligible delay) and unreliable wireless links (e.g., vehicle-to-vehicle channels with random delays and intermittent connectivity) creates temporal inconsistencies that can undermine zero-trust security protocols requiring synchronized verification and continuous authentication.

This network asymmetry introduces several critical research challenges: 1) developing delay-tolerant zero-trust protocols that can maintain security guarantees despite variable network conditions and intermittent connectivity, ensuring that authentication and authorization mechanisms remain effective even when some nodes experience communication delays or temporary isolation, 2) designing adaptive information sharing strategies that dynamically adjust collaboration patterns based on real-time network conditions while preserving the principle of least privilege access control, and 3) establishing distributed consensus mechanisms for multi-LLM coordination that are resilient to network partitions and asymmetric information propagation delays. Future work can also explore federated zero-trust architectures that leverage edge-cloud hierarchies to compensate for local network limitations, implementing tiered security policies that can gracefully degrade while maintaining critical safety properties when network conditions deteriorate.

\vspace{-0.1cm}
\subsection{Privacy-Preserving Collaborative Reasoning}
The deployment of zero-trust multi-LLM systems necessitates collaborative reasoning capabilities that can leverage collective intelligence while maintaining cryptographic privacy guarantees and preventing any form of information leakage, extending the principle of ``least privilege'' to collaborative inference scenarios~\cite{10963886}. This challenge is particularly acute in EGI scenarios where sensitive data must be processed at the edge while maintaining mathematically provable confidentiality.

Future research should address: 1) developing transformer-oriented encryption schemes that enable encrypted multi-LLM collaboration without any plaintext data exposure, ensuring that even compromised LLMs cannot access sensitive information~\cite{10963886}, and 2) advancing secure MPC frameworks with ZKP that enable collaborative inference among distributed LLMs while maintaining cryptographic proof that no participating LLM can extract more information than strictly necessary for the collaborative task.
Recent work on encryption-friendly LLMs demonstrates the feasibility of homomorphic encryption for transformer models, achieving significant computational speedups while maintaining performance comparable to plaintext models~\cite{10963886}. However, extending these approaches to multi-LLM collaborative scenarios introduces additional complexity in terms of zero-trust encrypted communication protocols and cryptographically verifiable consensus mechanisms that must be addressed through innovative cryptographic and distributed systems research that assumes no trust in any component.

\vspace{-0.25cm}
\section{Conclusion}
This paper has presented the first systematic survey of zero-trust security principles applied to multi-LLM systems in EGI. As multi-LLM systems become increasingly critical, traditional perimeter-based security approaches prove inadequate to address the unique vulnerabilities in collaborative edge deployments. We have systematically analyzed security challenges in multi-LLM systems, categorizing threats at both intra-LLM and inter-LLM levels, and demonstrate the limitations of existing trustworthy approaches. Then, we have proposed a unified zero-trust multi-LLM framework implementing the four fundamental zero-trust principles through detailed architectural design and operational workflows. We have categorized zero-trust mechanisms into model-level approaches focusing on individual LLM security and system-level approaches addressing distributed coordination challenges. Additionally, we have identified several critical future research directions that require immediate attention. We hope that this survey prompts both theoretical progress and practical implementation for secure edge agentic AI.

\vspace{-0.2cm}
\bibliographystyle{ACM-Reference-Format}
\bibliography{sample-base}


\begin{thebibliography}{151}


\ifx \showCODEN    \undefined \def \showCODEN     #1{\unskip}     \fi
\ifx \showDOI      \undefined \def \showDOI       #1{#1}\fi
\ifx \showISBNx    \undefined \def \showISBNx     #1{\unskip}     \fi
\ifx \showISBNxiii \undefined \def \showISBNxiii  #1{\unskip}     \fi
\ifx \showISSN     \undefined \def \showISSN      #1{\unskip}     \fi
\ifx \showLCCN     \undefined \def \showLCCN      #1{\unskip}     \fi
\ifx \shownote     \undefined \def \shownote      #1{#1}          \fi
\ifx \showarticletitle \undefined \def \showarticletitle #1{#1}   \fi
\ifx \showURL      \undefined \def \showURL       {\relax}        \fi
\providecommand\bibfield[2]{#2}
\providecommand\bibinfo[2]{#2}
\providecommand\natexlab[1]{#1}
\providecommand\showeprint[2][]{arXiv:#2}

\bibitem[Yang et~al\mbox{.}(2024)]%
        {mllm}
\bibfield{author}{\bibinfo{person}{Yingxuan Yang} {et~al\mbox{.}}} \bibinfo{year}{2024}\natexlab{}.
\newblock \showarticletitle{LLM-based Multi-Agent Systems: Techniques and Business Perspectives}.
\newblock \bibinfo{journal}{\emph{ArXiv preprint: ArXiv:2411.14033}} (\bibinfo{year}{2024}).
\newblock


\bibitem[Acharya et~al\mbox{.}(2025)]%
        {10849561}
\bibfield{author}{\bibinfo{person}{Deepak~Bhaskar Acharya}, \bibinfo{person}{Karthigeyan Kuppan}, {and} \bibinfo{person}{B. Divya}.} \bibinfo{year}{2025}\natexlab{}.
\newblock \showarticletitle{Agentic {AI}: Autonomous Intelligence for Complex Goals—A Comprehensive Survey}.
\newblock \bibinfo{journal}{\emph{IEEE Access}}  \bibinfo{volume}{13} (\bibinfo{year}{2025}), \bibinfo{pages}{18912--18936}.
\newblock


\bibitem[Gunduz et~al\mbox{.}(2025)]%
        {Agentify}
\bibfield{author}{\bibinfo{person}{Ahmet Gunduz}, \bibinfo{person}{Kamer~Ali Yuksel}, {and} \bibinfo{person}{Hassan Sawaf}.} \bibinfo{year}{2025}\natexlab{}.
\newblock \showarticletitle{MediaMind: Revolutionizing Media Monitoring using Agentification}.
\newblock \bibinfo{journal}{\emph{ArXiv preprint: ArXiv:2502.12745}} (\bibinfo{year}{2025}).
\newblock


\bibitem[Chen et~al\mbox{.}(2024)]%
        {egi1}
\bibfield{author}{\bibinfo{person}{Handi Chen} {et~al\mbox{.}}} \bibinfo{year}{2024}\natexlab{}.
\newblock \showarticletitle{Towards Edge General Intelligence via Large Language Models: Opportunities and Challenges}.
\newblock \bibinfo{journal}{\emph{ArXiv preprint: ArXiv:2410.18125}} (\bibinfo{year}{2024}).
\newblock


\bibitem[He et~al\mbox{.}(2025)]%
        {egi2}
\bibfield{author}{\bibinfo{person}{Le He} {et~al\mbox{.}}} \bibinfo{year}{2025}\natexlab{}.
\newblock \showarticletitle{The Road Toward General Edge Intelligence: Standing on the Shoulders of Foundation Models}.
\newblock \bibinfo{journal}{\emph{IEEE Communications Magazine}} (\bibinfo{year}{2025}), \bibinfo{pages}{1--7}.
\newblock


\bibitem[Xu et~al\mbox{.}(2024)]%
        {jailbreak}
\bibfield{author}{\bibinfo{person}{Zihao Xu}, \bibinfo{person}{Yi Liu}, \bibinfo{person}{Gelei Deng}, \bibinfo{person}{Yuekang Li}, {and} \bibinfo{person}{Stjepan Picek}.} \bibinfo{year}{2024}\natexlab{}.
\newblock \showarticletitle{A Comprehensive Study of Jailbreak Attack versus Defense for Large Language Models}. In \bibinfo{booktitle}{\emph{Proc. ACL Findings}}. \bibinfo{pages}{7432–7449}.
\newblock


\bibitem[Liu et~al\mbox{.}(2023)]%
        {injectionprompt}
\bibfield{author}{\bibinfo{person}{Yi Liu} {et~al\mbox{.}}} \bibinfo{year}{2023}\natexlab{}.
\newblock \showarticletitle{Prompt Injection attack against LLM-integrated Applications}.
\newblock \bibinfo{journal}{\emph{ArXiv preprint: ArXiv:2306.05499}} (\bibinfo{year}{2023}).
\newblock


\bibitem[Chen et~al\mbox{.}(2024)]%
        {decisionmaking}
\bibfield{author}{\bibinfo{person}{Zhaorun Chen}, \bibinfo{person}{Zhen Xiang}, \bibinfo{person}{Chaowei Xiao}, \bibinfo{person}{Dawn Song}, {and} \bibinfo{person}{Bo Li}.} \bibinfo{year}{2024}\natexlab{}.
\newblock \showarticletitle{AgentPoison: Red-teaming LLM Agents via Poisoning Memory or Knowledge Bases}. In \bibinfo{booktitle}{\emph{Proc. NeurIPS}}. \bibinfo{pages}{1--29}.
\newblock


\bibitem[Ko et~al\mbox{.}(2025)]%
        {multiko}
\bibfield{author}{\bibinfo{person}{Ronny Ko} {et~al\mbox{.}}} \bibinfo{year}{2025}\natexlab{}.
\newblock \showarticletitle{Seven Security Challenges That Must be Solved in Cross-domain Multi-agent LLM Systems}.
\newblock \bibinfo{journal}{\emph{ArXiv preprint: ArXiv:2505.23847}} (\bibinfo{year}{2025}).
\newblock


\bibitem[Yu et~al\mbox{.}(2024)]%
        {add2}
\bibfield{author}{\bibinfo{person}{Lei Yu}, \bibinfo{person}{Virginie Do}, \bibinfo{person}{Karen Hambardzumyan}, {and} \bibinfo{person}{Nicola Cancedda}.} \bibinfo{year}{2024}\natexlab{}.
\newblock \showarticletitle{Robust LLM safeguarding via refusal feature adversarial training}.
\newblock \bibinfo{journal}{\emph{ArXiv preprint: ArXiv:2409.20089}} (\bibinfo{year}{2024}).
\newblock


\bibitem[Charles et~al\mbox{.}(2024)]%
        {DP2}
\bibfield{author}{\bibinfo{person}{Zachary Charles} {et~al\mbox{.}}} \bibinfo{year}{2024}\natexlab{}.
\newblock \showarticletitle{Fine-Tuning Large Language Models with User-Level Differential Privacy}.
\newblock \bibinfo{journal}{\emph{ArXiv preprint: ArXiv:2407.07737}} (\bibinfo{year}{2024}).
\newblock


\bibitem[Yu et~al\mbox{.}(2022)]%
        {TEE}
\bibfield{author}{\bibinfo{person}{Wei Yu} {et~al\mbox{.}}} \bibinfo{year}{2022}\natexlab{}.
\newblock \showarticletitle{{TEE} based Cross-silo Trustworthy Federated Learning Infrastructure}. In \bibinfo{booktitle}{\emph{Proc. IJCAI}}.
\newblock


\bibitem[Zeng et~al\mbox{.}(2024)]%
        {autodefense}
\bibfield{author}{\bibinfo{person}{Yifan Zeng} {et~al\mbox{.}}} \bibinfo{year}{2024}\natexlab{}.
\newblock \showarticletitle{AutoDefense: Multi-Agent LLM Defense against Jailbreak Attacks}. In \bibinfo{booktitle}{\emph{Proc. NeurIPS}}.
\newblock


\bibitem[OpenAI et~al\mbox{.}(2023)]%
        {openai2023gpt4}
\bibfield{author}{\bibinfo{person}{OpenAI} {et~al\mbox{.}}} \bibinfo{year}{2023}\natexlab{}.
\newblock \showarticletitle{GPT-4 Technical Report}.
\newblock \bibinfo{journal}{\emph{arXiv preprint arXiv:2303.08774}} (\bibinfo{year}{2023}).
\newblock


\bibitem[NIS({[n.\,d.]})]%
        {NIST}
 \bibinfo{year}{[n.\,d.]}\natexlab{}.
\newblock \bibinfo{title}{{NIST} Zero trust standard. 2025}.
\newblock
\newblock
\urldef\tempurl%
\url{https://www.nist.gov/publications/zero-trust-architecture}
\showURL{%
\tempurl}


\bibitem[Cao et~al\mbox{.}(2025)]%
        {10963886}
\bibfield{author}{\bibinfo{person}{Xinye Cao} {et~al\mbox{.}}} \bibinfo{year}{2025}\natexlab{}.
\newblock \showarticletitle{Exploring {LLM}-Based Multi-Agent Situation Awareness for Zero-Trust Space-Air-Ground Integrated Network}.
\newblock \bibinfo{journal}{\emph{IEEE Journal on Selected Areas in Communications}} \bibinfo{volume}{43}, \bibinfo{number}{6} (\bibinfo{year}{2025}), \bibinfo{pages}{2230--2247}.
\newblock


\bibitem[Poirrier et~al\mbox{.}(2025)]%
        {10970721}
\bibfield{author}{\bibinfo{person}{Alexandre Poirrier}, \bibinfo{person}{Laurent Cailleux}, {and} \bibinfo{person}{Thomas Heide~Clausen}.} \bibinfo{year}{2025}\natexlab{}.
\newblock \showarticletitle{Is Trust Misplaced? A Zero-Trust Survey}.
\newblock \bibinfo{journal}{\emph{Proc. IEEE}} \bibinfo{volume}{113}, \bibinfo{number}{1} (\bibinfo{year}{2025}), \bibinfo{pages}{5--39}.
\newblock


\bibitem[Das et~al\mbox{.}(2025)]%
        {singledas}
\bibfield{author}{\bibinfo{person}{Badhan~Chandra Das}, \bibinfo{person}{M.~Hadi Amini}, {and} \bibinfo{person}{Yanzhao Wu}.} \bibinfo{year}{2025}\natexlab{}.
\newblock \showarticletitle{Security and Privacy Challenges of Large Language Models: A Survey}.
\newblock \bibinfo{journal}{\emph{ACM Computing Survey}} \bibinfo{volume}{57}, \bibinfo{number}{6} (\bibinfo{date}{Feb.} \bibinfo{year}{2025}), \bibinfo{pages}{1--39}.
\newblock


\bibitem[Aguilera-Martínez and Berzal(2025)]%
        {single2}
\bibfield{author}{\bibinfo{person}{Francisco Aguilera-Martínez} {and} \bibinfo{person}{Fernando Berzal}.} \bibinfo{year}{2025}\natexlab{}.
\newblock \showarticletitle{LLM Security: Vulnerabilities, Attacks, Defenses, and Countermeasures}.
\newblock \bibinfo{journal}{\emph{ArXiv preprint: ArXiv:2505.01177}} (\bibinfo{year}{2025}).
\newblock


\bibitem[Friha et~al\mbox{.}(2024)]%
        {singlefriha}
\bibfield{author}{\bibinfo{person}{Othmane Friha}, \bibinfo{person}{Mohamed Amine~Ferrag}, \bibinfo{person}{Burak Kantarci}, \bibinfo{person}{Burak Cakmak}, \bibinfo{person}{Arda Ozgun}, {and} \bibinfo{person}{Nassira Ghoualmi-Zine}.} \bibinfo{year}{2024}\natexlab{}.
\newblock \showarticletitle{LLM-Based Edge Intelligence: A Comprehensive Survey on Architectures, Applications, Security and Trustworthiness}.
\newblock \bibinfo{journal}{\emph{IEEE Open Journal of the Communications Society}}  \bibinfo{volume}{5} (\bibinfo{year}{2024}), \bibinfo{pages}{5799--5856}.
\newblock


\bibitem[Gan et~al\mbox{.}(2024)]%
        {singlegan}
\bibfield{author}{\bibinfo{person}{Yuyou Gan} {et~al\mbox{.}}} \bibinfo{year}{2024}\natexlab{}.
\newblock \showarticletitle{Navigating the Risks: A Survey of Security, Privacy, and Ethics Threats in LLM-Based Agents}.
\newblock \bibinfo{journal}{\emph{ArXiv preprint: ArXiv: 2411.09523}} (\bibinfo{year}{2024}).
\newblock


\bibitem[Liu et~al\mbox{.}(2024)]%
        {singleliu}
\bibfield{author}{\bibinfo{person}{Yang Liu} {et~al\mbox{.}}} \bibinfo{year}{2024}\natexlab{}.
\newblock \showarticletitle{Trustworthy LLMs: a Survey and Guideline for Evaluating Large Language Models' Alignment}.
\newblock \bibinfo{journal}{\emph{ArXiv preprint: ArXiv: 2308.05374}} (\bibinfo{year}{2024}).
\newblock


\bibitem[Kong et~al\mbox{.}(2025)]%
        {multikong}
\bibfield{author}{\bibinfo{person}{Dezhang Kong} {et~al\mbox{.}}} \bibinfo{year}{2025}\natexlab{}.
\newblock \showarticletitle{A Survey of LLM-Driven AI Agent Communication: Protocols, Security Risks, and Defense Countermeasures}.
\newblock \bibinfo{journal}{\emph{ArXiv preprint: ArXiv:2506.19676}} (\bibinfo{year}{2025}).
\newblock


\bibitem[Luo et~al\mbox{.}(2025)]%
        {multiluo}
\bibfield{author}{\bibinfo{person}{Haoxiang Luo} {et~al\mbox{.}}} \bibinfo{year}{2025}\natexlab{}.
\newblock \showarticletitle{Toward Edge General Intelligence with Multiple-Large Language Model (Multi-LLM): Architecture, Trust, and Orchestration}.
\newblock \bibinfo{journal}{\emph{ArXiv preprint: ArXiv:2507.00672}} (\bibinfo{year}{2025}).
\newblock


\bibitem[Peigne et~al\mbox{.}(2024)]%
        {multipeign}
\bibfield{author}{\bibinfo{person}{Pierre Peigne} {et~al\mbox{.}}} \bibinfo{year}{2024}\natexlab{}.
\newblock \showarticletitle{Multi-Agent Security Tax: Trading Off Security and Collaboration Capabilities in Multi-Agent Systems}. In \bibinfo{booktitle}{\emph{Proc. AAAI}}. \bibinfo{pages}{27573--27581}.
\newblock


\bibitem[Chan(2025)]%
        {encrypt}
\bibfield{author}{\bibinfo{person}{Shih-Han Chan}.} \bibinfo{year}{2025}\natexlab{}.
\newblock \showarticletitle{Encrypted Prompt: Securing LLM Applications Against Unauthorized Actions}.
\newblock \bibinfo{journal}{\emph{ArXiv preprint: ArXiv:2503.23250}} (\bibinfo{year}{2025}).
\newblock


\bibitem[Shi et~al\mbox{.}(2025)]%
        {pepagent}
\bibfield{author}{\bibinfo{person}{Zitong Shi} {et~al\mbox{.}}} \bibinfo{year}{2025}\natexlab{}.
\newblock \showarticletitle{Privacy-Enhancing Paradigms within Federated Multi-Agent Systems}.
\newblock \bibinfo{journal}{\emph{ArXiv preprint: ArXiv:2503.08175}} (\bibinfo{year}{2025}).
\newblock


\bibitem[Xiao et~al\mbox{.}(2025)]%
        {xiao}
\bibfield{author}{\bibinfo{person}{Peng Xiao}, \bibinfo{person}{Shunkun Yang}, \bibinfo{person}{Hailin Wang}, \bibinfo{person}{Zhenhong Zhang}, {and} \bibinfo{person}{Chunsheng Zou}.} \bibinfo{year}{2025}\natexlab{}.
\newblock \showarticletitle{Privacy-preserving revocable access control for LLM-driven electrical distributed systems}.
\newblock \bibinfo{journal}{\emph{Peer-to-Peer Networking and Applications}} \bibinfo{volume}{18}, \bibinfo{number}{148} (\bibinfo{year}{2025}), \bibinfo{pages}{1--12}.
\newblock


\bibitem[Arshad and Halim(2025)]%
        {blockLLM}
\bibfield{author}{\bibinfo{person}{Usama Arshad} {and} \bibinfo{person}{Zahid Halim}.} \bibinfo{year}{2025}\natexlab{}.
\newblock \showarticletitle{Block{LLM}: A futuristic {LLM}-based decentralized vehicular network architecture for secure communications}.
\newblock \bibinfo{journal}{\emph{Computers and Electrical Engineering}}  \bibinfo{volume}{123} (\bibinfo{date}{Jan.} \bibinfo{year}{2025}), \bibinfo{pages}{1--39}.
\newblock


\bibitem[Chen et~al\mbox{.}(2025a)]%
        {defensivetokens}
\bibfield{author}{\bibinfo{person}{Sizhe Chen}, \bibinfo{person}{Yizhu Wang}, \bibinfo{person}{Nicholas Carlini}, \bibinfo{person}{Chawin Sitawarin}, {and} \bibinfo{person}{David Wagner}.} \bibinfo{year}{2025}\natexlab{a}.
\newblock \showarticletitle{Defending Against Prompt Injection With a Few DefensiveTokens}.
\newblock \bibinfo{journal}{\emph{ArXiv preprint: ArXiv:2507.07974}} (\bibinfo{year}{2025}).
\newblock


\bibitem[Chen et~al\mbox{.}(2025b)]%
        {MedSentry}
\bibfield{author}{\bibinfo{person}{Kai Chen} {et~al\mbox{.}}} \bibinfo{year}{2025}\natexlab{b}.
\newblock \showarticletitle{MedSentry: Understanding and Mitigating Safety Risks in Medical LLM Multi-Agent Systems}.
\newblock \bibinfo{journal}{\emph{ArXiv preprint: ArXiv:2505.20824}} (\bibinfo{year}{2025}).
\newblock


\bibitem[Chong et~al\mbox{.}(2025)]%
        {LLMnet}
\bibfield{author}{\bibinfo{person}{Zan-Kai Chong}, \bibinfo{person}{Hiroyuki Ohsaki}, {and} \bibinfo{person}{Bryan Ng}.} \bibinfo{year}{2025}\natexlab{}.
\newblock \showarticletitle{{LLM}-Net: Democratizing {LLM}s-as-a-Service through Blockchain-based Expert Networks}.
\newblock \bibinfo{journal}{\emph{ArXiv preprint: ArXiv:2501.07288}} (\bibinfo{year}{2025}).
\newblock


\bibitem[Wang et~al\mbox{.}(2024)]%
        {blockchain4}
\bibfield{author}{\bibinfo{person}{Jing Wang}, \bibinfo{person}{Xue Yuan}, \bibinfo{person}{Yingjie Xu}, \bibinfo{person}{Yudi Zhang}, {and} \bibinfo{person}{Guowen Xu}.} \bibinfo{year}{2024}\natexlab{}.
\newblock \showarticletitle{An Efficient Multiparty Threshold ECDSA Protocol against Malicious Adversaries for Blockchain-Based LLMs}.
\newblock \bibinfo{journal}{\emph{IET Information Security}}  \bibinfo{volume}{2024} (\bibinfo{year}{2024}), \bibinfo{pages}{1--12}.
\newblock


\bibitem[Cha({[n.\,d.]})]%
        {ChatGPT2}
 \bibinfo{year}{[n.\,d.]}\natexlab{}.
\newblock \bibinfo{title}{Open{AI} {C}hat{GPT}. 2025}.
\newblock
\newblock
\urldef\tempurl%
\url{https://openai.com/index/chatgpt/}
\showURL{%
\tempurl}


\bibitem[Kirillov et~al\mbox{.}(2023)]%
        {SAM}
\bibfield{author}{\bibinfo{person}{Alexander Kirillov} {et~al\mbox{.}}} \bibinfo{year}{2023}\natexlab{}.
\newblock \showarticletitle{Segment Anything}. In \bibinfo{booktitle}{\emph{Proc. ICCV}}. \bibinfo{pages}{4015--4026}.
\newblock


\bibitem[Wu et~al\mbox{.}(2024)]%
        {Autogen}
\bibfield{author}{\bibinfo{person}{Qingyu Wu} {et~al\mbox{.}}} \bibinfo{year}{2024}\natexlab{}.
\newblock \showarticletitle{AutoGen: Enabling Next-Gen {LLM} Applications via Multi-Agent Conversations}. In \bibinfo{booktitle}{\emph{Proc. COLM}}. \bibinfo{pages}{1--46}.
\newblock


\bibitem[Hong et~al\mbox{.}(2024)]%
        {MetaGPT}
\bibfield{author}{\bibinfo{person}{Sirui Hong} {et~al\mbox{.}}} \bibinfo{year}{2024}\natexlab{}.
\newblock \showarticletitle{Meta{GPT}: Meta Programming for A Multi-Agent Collaborative Framework}. In \bibinfo{booktitle}{\emph{Proc. ICLR}}. \bibinfo{pages}{1--29}.
\newblock


\bibitem[Du et~al\mbox{.}(2024)]%
        {10.5555/3692070.3692537}
\bibfield{author}{\bibinfo{person}{Yilun Du}, \bibinfo{person}{Shuang Li}, \bibinfo{person}{Antonio Torralba}, \bibinfo{person}{Joshua~B. Tenenbaum}, {and} \bibinfo{person}{Igor Mordatch}.} \bibinfo{year}{2024}\natexlab{}.
\newblock \showarticletitle{Improving factuality and reasoning in language models through multiagent debate}. In \bibinfo{booktitle}{\emph{Proc. ICML}}. \bibinfo{pages}{11733 -- 11763}.
\newblock


\bibitem[Li et~al\mbox{.}(2024)]%
        {li2024survey}
\bibfield{author}{\bibinfo{person}{Xinyi Li}, \bibinfo{person}{Sai Wang}, \bibinfo{person}{Siqi Zeng}, \bibinfo{person}{Yu Wu}, {and} \bibinfo{person}{Yi Yang}.} \bibinfo{year}{2024}\natexlab{}.
\newblock \showarticletitle{A survey on {LLM}-based multi-agent systems: workflow, infrastructure, and challenges}.
\newblock \bibinfo{journal}{\emph{Vicinagearth}} \bibinfo{volume}{1}, \bibinfo{number}{9} (\bibinfo{year}{2024}), \bibinfo{pages}{1--43}.
\newblock


\bibitem[Zhang et~al\mbox{.}(2025)]%
        {gdesigner}
\bibfield{author}{\bibinfo{person}{Guibin Zhang} {et~al\mbox{.}}} \bibinfo{year}{2025}\natexlab{}.
\newblock \showarticletitle{G-Designer: Architecting Multi-Agent Communication Topologies via Graph Neural Networks}. In \bibinfo{booktitle}{\emph{Proc. ICLR}}. \bibinfo{pages}{1--12}.
\newblock


\bibitem[Karpas et~al\mbox{.}(2022)]%
        {MRKL}
\bibfield{author}{\bibinfo{person}{Ehud Karpas} {et~al\mbox{.}}} \bibinfo{year}{2022}\natexlab{}.
\newblock \showarticletitle{MRKL systems: A modular, neuro-symbolic architecture that combines large language models, external knowledge sources and discrete reasoning}.
\newblock \bibinfo{journal}{\emph{ArXiv preprint: ArXiv:2205.00445}} (\bibinfo{year}{2022}).
\newblock


\bibitem[Ehtesham et~al\mbox{.}(2025)]%
        {communication}
\bibfield{author}{\bibinfo{person}{Abul Ehtesham}, \bibinfo{person}{Aditi Singh}, \bibinfo{person}{Gaurav~Kumar Gupta}, {and} \bibinfo{person}{Saket Kumar}.} \bibinfo{year}{2025}\natexlab{}.
\newblock \showarticletitle{A survey of agent interoperability protocols: Model Context Protocol ({MCP}), Agent Communication Protocol ({ACP}), Agent-to-Agent Protocol {(A2A}), and Agent Network Protocol ({ANP})}.
\newblock \bibinfo{journal}{\emph{ArXiv preprint: ArXiv:2505.02279}} (\bibinfo{year}{2025}).
\newblock


\bibitem[Du et~al\mbox{.}(2022)]%
        {mec}
\bibfield{author}{\bibinfo{person}{Yu Du}, \bibinfo{person}{Jun Li}, \bibinfo{person}{Long Shi}, \bibinfo{person}{Tingting Liu}, \bibinfo{person}{Feng Shu}, {and} \bibinfo{person}{Zhu Han}.} \bibinfo{year}{2022}\natexlab{}.
\newblock \showarticletitle{Two-Tier Matching Game in Small Cell Networks for Mobile Edge Computing}.
\newblock \bibinfo{journal}{\emph{IEEE Transactions on Services Computing}} \bibinfo{volume}{15}, \bibinfo{number}{1} (\bibinfo{year}{2022}), \bibinfo{pages}{254--265}.
\newblock


\bibitem[Chen et~al\mbox{.}(2024)]%
        {10570088}
\bibfield{author}{\bibinfo{person}{Yung-Yao Chen}, \bibinfo{person}{Sin-Ye Jhong}, \bibinfo{person}{Shao-Kai Tu}, \bibinfo{person}{Yu-Hsiu Lin}, {and} \bibinfo{person}{Yi-Chen Wu}.} \bibinfo{year}{2024}\natexlab{}.
\newblock \showarticletitle{Autonomous Smart-Edge Fault Diagnostics via Edge-Cloud-Orchestrated Collaborative Computing for Infrared Electrical Equipment Images}.
\newblock \bibinfo{journal}{\emph{IEEE Sensors Journal}} \bibinfo{volume}{24}, \bibinfo{number}{15} (\bibinfo{year}{2024}), \bibinfo{pages}{24630--24648}.
\newblock


\bibitem[Bubeck et~al\mbox{.}(2023)]%
        {AGI}
\bibfield{author}{\bibinfo{person}{Sébastien Bubeck} {et~al\mbox{.}}} \bibinfo{year}{2023}\natexlab{}.
\newblock \showarticletitle{Sparks of Artificial General Intelligence: Early experiments with {GPT}-4}.
\newblock \bibinfo{journal}{\emph{ArXiv preprint: ArXiv:2303.12712}} (\bibinfo{year}{2023}).
\newblock


\bibitem[Clusmann et~al\mbox{.}(2023)]%
        {healthcare}
\bibfield{author}{\bibinfo{person}{Jan Clusmann} {et~al\mbox{.}}} \bibinfo{year}{2023}\natexlab{}.
\newblock \showarticletitle{The future landscape of large language models in medicine}.
\newblock \bibinfo{journal}{\emph{Communications Medcine}} \bibinfo{volume}{3}, \bibinfo{number}{141} (\bibinfo{year}{2023}), \bibinfo{pages}{1--8}.
\newblock


\bibitem[Hu et~al\mbox{.}(2025)]%
        {llmdriving}
\bibfield{author}{\bibinfo{person}{Senkang Hu}, \bibinfo{person}{Zhengru Fang}, \bibinfo{person}{Zihan Fang}, \bibinfo{person}{Yiqin Deng}, \bibinfo{person}{Xianhao Chen}, {and} \bibinfo{person}{Yuguang Fang}.} \bibinfo{year}{2025}\natexlab{}.
\newblock \showarticletitle{AgentsCoDriver: Large Language Model Empowered Collaborative Driving with Lifelong Learning}.
\newblock \bibinfo{journal}{\emph{ArXiv preprint: ArXiv:2404.06345}} (\bibinfo{year}{2025}).
\newblock


\bibitem[Deng et~al\mbox{.}(2024)]%
        {10827901}
\bibfield{author}{\bibinfo{person}{Xiaozhi Deng}, \bibinfo{person}{Tengteng Ma}, \bibinfo{person}{Haobin Li}, {and} \bibinfo{person}{Mingxin Lu}.} \bibinfo{year}{2024}\natexlab{}.
\newblock \showarticletitle{Federated Large Language Models for Smart Grid: A Communication Efficient LoRA Approach}. In \bibinfo{booktitle}{\emph{Proc. ICCASIT}}. \bibinfo{pages}{1369--1374}.
\newblock


\bibitem[Joshi(2025)]%
        {10764723}
\bibfield{author}{\bibinfo{person}{Hrishikesh Joshi}.} \bibinfo{year}{2025}\natexlab{}.
\newblock \showarticletitle{Emerging Technologies Driving Zero Trust Maturity Across Industries}.
\newblock \bibinfo{journal}{\emph{IEEE Open Journal of the Computer Society}}  \bibinfo{volume}{6} (\bibinfo{year}{2025}), \bibinfo{pages}{25--36}.
\newblock


\bibitem[Khowaja et~al\mbox{.}(2025)]%
        {10829858}
\bibfield{author}{\bibinfo{person}{Sunder~A. Khowaja}, \bibinfo{person}{Parus Khuwaja}, \bibinfo{person}{Kapal Dev}, \bibinfo{person}{Keshav Singh}, \bibinfo{person}{Xingwang Li}, \bibinfo{person}{Nikolaos Bartzoudis}, {and} \bibinfo{person}{Ciprian~R. Comsa}.} \bibinfo{year}{2025}\natexlab{}.
\newblock \showarticletitle{Block Encryption LAyer ({BELA}): Zero-Trust Defense Against Model Inversion Attacks for Federated Learning in 5G/6G Systems}.
\newblock \bibinfo{journal}{\emph{IEEE Open Journal of the Communications Society}}  \bibinfo{volume}{6} (\bibinfo{year}{2025}), \bibinfo{pages}{807--819}.
\newblock


\bibitem[Wang and Wang(2023)]%
        {9975323}
\bibfield{author}{\bibinfo{person}{Qingxuan Wang} {and} \bibinfo{person}{Ding Wang}.} \bibinfo{year}{2023}\natexlab{}.
\newblock \showarticletitle{Understanding Failures in Security Proofs of Multi-Factor Authentication for Mobile Devices}.
\newblock \bibinfo{journal}{\emph{IEEE Transactions on Information Forensics and Security}}  \bibinfo{volume}{18} (\bibinfo{year}{2023}), \bibinfo{pages}{597--612}.
\newblock


\bibitem[Jing et~al\mbox{.}(2024)]%
        {10494515}
\bibfield{author}{\bibinfo{person}{Wentao Jing}, \bibinfo{person}{Linning Peng}, \bibinfo{person}{Hua Fu}, {and} \bibinfo{person}{Aiqun Hu}.} \bibinfo{year}{2024}\natexlab{}.
\newblock \showarticletitle{An Authentication Mechanism Based on Zero Trust With Radio Frequency Fingerprint for Internet of Things Networks}.
\newblock \bibinfo{journal}{\emph{IEEE Internet of Things Journal}} \bibinfo{volume}{11}, \bibinfo{number}{13} (\bibinfo{year}{2024}), \bibinfo{pages}{23683--23698}.
\newblock


\bibitem[Hong et~al\mbox{.}(2023)]%
        {10091151}
\bibfield{author}{\bibinfo{person}{Sungmin Hong}, \bibinfo{person}{Lei Xu}, \bibinfo{person}{Jianwei Huang}, \bibinfo{person}{Hongda Li}, \bibinfo{person}{Hongxin Hu}, {and} \bibinfo{person}{Guofei Gu}.} \bibinfo{year}{2023}\natexlab{}.
\newblock \showarticletitle{SysFlow: Toward a Programmable Zero Trust Framework for System Security}.
\newblock \bibinfo{journal}{\emph{IEEE Transactions on Information Forensics and Security}}  \bibinfo{volume}{18} (\bibinfo{year}{2023}), \bibinfo{pages}{2794--2809}.
\newblock


\bibitem[Zhou et~al\mbox{.}(2013)]%
        {RBAC}
\bibfield{author}{\bibinfo{person}{Lan Zhou}, \bibinfo{person}{Vijay Varadharajan}, {and} \bibinfo{person}{Michael Hitchens}.} \bibinfo{year}{2013}\natexlab{}.
\newblock \showarticletitle{Achieving Secure Role-Based Access Control on Encrypted Data in Cloud Storage}.
\newblock \bibinfo{journal}{\emph{IEEE Transactions on Information Forensics and Security}} \bibinfo{volume}{8}, \bibinfo{number}{12} (\bibinfo{year}{2013}), \bibinfo{pages}{1947--1960}.
\newblock


\bibitem[Tuler De~Oliveira et~al\mbox{.}(2022)]%
        {ABAC}
\bibfield{author}{\bibinfo{person}{Marcela Tuler De~Oliveira}, \bibinfo{person}{Lúcio Henrik~Amorim Reis}, \bibinfo{person}{Yiannis Verginadis}, \bibinfo{person}{Diogo Menezes~Ferrazani Mattos}, {and} \bibinfo{person}{Sílvia~Delgado Olabarriaga}.} \bibinfo{year}{2022}\natexlab{}.
\newblock \showarticletitle{SmartAccess: Attribute-Based Access Control System for Medical Records Based on Smart Contracts}.
\newblock \bibinfo{journal}{\emph{IEEE Access}}  \bibinfo{volume}{10} (\bibinfo{year}{2022}), \bibinfo{pages}{117836--117854}.
\newblock


\bibitem[Najafi et~al\mbox{.}(2021)]%
        {9679411}
\bibfield{author}{\bibinfo{person}{Pejman Najafi}, \bibinfo{person}{Daniel Koehler}, \bibinfo{person}{Feng Cheng}, {and} \bibinfo{person}{Christoph Meinel}.} \bibinfo{year}{2021}\natexlab{}.
\newblock \showarticletitle{NLP-based Entity Behavior Analytics for Malware Detection}. In \bibinfo{booktitle}{\emph{Proc. IPCCC}}. \bibinfo{pages}{1--5}.
\newblock


\bibitem[Bhatt et~al\mbox{.}(2014)]%
        {6924640}
\bibfield{author}{\bibinfo{person}{Sandeep Bhatt}, \bibinfo{person}{Pratyusa~K. Manadhata}, {and} \bibinfo{person}{Loai Zomlot}.} \bibinfo{year}{2014}\natexlab{}.
\newblock \showarticletitle{The Operational Role of Security Information and Event Management Systems}.
\newblock \bibinfo{journal}{\emph{IEEE Security \& Privacy}} \bibinfo{volume}{12}, \bibinfo{number}{5} (\bibinfo{year}{2014}), \bibinfo{pages}{35--41}.
\newblock


\bibitem[Bandara et~al\mbox{.}(2022)]%
        {9918536}
\bibfield{author}{\bibinfo{person}{Eranga Bandara}, \bibinfo{person}{Xueping Liang}, \bibinfo{person}{Sachin Shetty}, \bibinfo{person}{Ravi Mukkamala}, \bibinfo{person}{Abdul Rahman}, {and} \bibinfo{person}{Ng~Wee Keong}.} \bibinfo{year}{2022}\natexlab{}.
\newblock \showarticletitle{Skunk — A Blockchain and Zero Trust Security Enabled Federated Learning Platform for 5G/6G Network Slicing}. In \bibinfo{booktitle}{\emph{Proc. SECON}}. \bibinfo{pages}{109--117}.
\newblock


\bibitem[d'Aliberti et~al\mbox{.}(2024)]%
        {AImodelcontrol}
\bibfield{author}{\bibinfo{person}{Liv d'Aliberti}, \bibinfo{person}{Evan Gronberg}, {and} \bibinfo{person}{Joseph Kovba}.} \bibinfo{year}{2024}\natexlab{}.
\newblock \showarticletitle{Privacy-Enhancing Technologies for Artificial Intelligence-Enabled Systems}. In \bibinfo{booktitle}{\emph{Proc. IWSPA}}.
\newblock


\bibitem[Zhao et~al\mbox{.}(2024)]%
        {s24134140}
\bibfield{author}{\bibinfo{person}{Rui Zhao}, \bibinfo{person}{Ziguo Chen}, \bibinfo{person}{Yuze Fan}, \bibinfo{person}{Yun Li}, {and} \bibinfo{person}{Fei Gao}.} \bibinfo{year}{2024}\natexlab{}.
\newblock \showarticletitle{Towards Robust Decision-Making for Autonomous Highway Driving Based on Safe Reinforcement Learning}.
\newblock \bibinfo{journal}{\emph{Sensors}} \bibinfo{volume}{24}, \bibinfo{number}{13} (\bibinfo{year}{2024}).
\newblock


\bibitem[Xu et~al\mbox{.}(2024)]%
        {edgeLLM}
\bibfield{author}{\bibinfo{person}{Zihao Xu}, \bibinfo{person}{Yi Liu}, \bibinfo{person}{Gelei Deng}, \bibinfo{person}{Yuekang Li}, {and} \bibinfo{person}{Stjepan Picek}.} \bibinfo{year}{2024}\natexlab{}.
\newblock \showarticletitle{A Comprehensive Study of Jailbreak Attack versus Defense for Large Language Models}. In \bibinfo{booktitle}{\emph{Proc. ACL}}. \bibinfo{pages}{7432–7449}.
\newblock


\bibitem[Liao and Sun(2024)]%
        {universalprompt}
\bibfield{author}{\bibinfo{person}{Zeyi Liao} {and} \bibinfo{person}{Huan Sun}.} \bibinfo{year}{2024}\natexlab{}.
\newblock \showarticletitle{AmpleGCG: Learning a Universal and Transferable Generative Model of Adversarial Suffixes for Jailbreaking Both Open and Closed LLMs}. In \bibinfo{booktitle}{\emph{Proc. COLM}}. \bibinfo{pages}{1--14}.
\newblock


\bibitem[Fang et~al\mbox{.}(2024)]%
        {87}
\bibfield{author}{\bibinfo{person}{Richard Fang}, \bibinfo{person}{Rohan Bindu}, \bibinfo{person}{Akul Gupta}, {and} \bibinfo{person}{Daniel Kang}.} \bibinfo{year}{2024}\natexlab{}.
\newblock \showarticletitle{LLM Agents can Autonomously Exploit One-day Vulnerabilities}.
\newblock \bibinfo{journal}{\emph{ArXiv preprint: ArXiv:2404.08144}} (\bibinfo{year}{2024}).
\newblock


\bibitem[Carlini et~al\mbox{.}(2021)]%
        {carlini2021extracting}
\bibfield{author}{\bibinfo{person}{Nicholas Carlini} {et~al\mbox{.}}} \bibinfo{year}{2021}\natexlab{}.
\newblock \showarticletitle{Extracting Training Data from Large Language Models}. In \bibinfo{booktitle}{\emph{Proc. USENIX Security}}. \bibinfo{pages}{2633--2650}.
\newblock


\bibitem[Mireshghallah et~al\mbox{.}(2022)]%
        {mireshghallah2022quantifying}
\bibfield{author}{\bibinfo{person}{Fatemehsadat Mireshghallah}, \bibinfo{person}{Kartik Goyal}, \bibinfo{person}{Archit Uniyal}, \bibinfo{person}{Taylor Berg-Kirkpatrick}, {and} \bibinfo{person}{Reza Shokri}.} \bibinfo{year}{2022}\natexlab{}.
\newblock \showarticletitle{Quantifying Privacy Risks of Prompting Large Language Models}.
\newblock \bibinfo{journal}{\emph{arXiv preprint arXiv:2210.17012}} (\bibinfo{year}{2022}).
\newblock


\bibitem[Sahoo et~al\mbox{.}(2024)]%
        {hallucination1}
\bibfield{author}{\bibinfo{person}{Pranab Sahoo}, \bibinfo{person}{Prabhash Meharia}, \bibinfo{person}{Akash Ghosh}, \bibinfo{person}{Sriparna Saha}, \bibinfo{person}{Vinija Jain}, {and} \bibinfo{person}{Aman Chadha}.} \bibinfo{year}{2024}\natexlab{}.
\newblock \showarticletitle{A Comprehensive Survey of Hallucination in Large Language, Image, Video and Audio Foundation Models}. In \bibinfo{booktitle}{\emph{Proc. EMNLP}}. \bibinfo{pages}{11709–11724}.
\newblock


\bibitem[Lee and Tiwari(2024)]%
        {promptinfection}
\bibfield{author}{\bibinfo{person}{Donghyun Lee} {and} \bibinfo{person}{Mo Tiwari}.} \bibinfo{year}{2024}\natexlab{}.
\newblock \showarticletitle{Prompt Infection: {LLM}-to-{LLM} Prompt Injection within Multi-Agent Systems}.
\newblock \bibinfo{journal}{\emph{ArXiv preprint: ArXiv:2410.07283}} (\bibinfo{year}{2024}).
\newblock


\bibitem[Shen et~al\mbox{.}(2025)]%
        {xusurvey}
\bibfield{author}{\bibinfo{person}{Xu Shen} {et~al\mbox{.}}} \bibinfo{year}{2025}\natexlab{}.
\newblock \showarticletitle{Understanding the Information Propagation Effects of Communication Topologies in LLM-based Multi-Agent Systems}.
\newblock \bibinfo{journal}{\emph{ArXiv preprint: ArXiv:2505.23352}} (\bibinfo{year}{2025}).
\newblock


\bibitem[Zhang et~al\mbox{.}(2025)]%
        {toolselection}
\bibfield{author}{\bibinfo{person}{Rupeng Zhang} {et~al\mbox{.}}} \bibinfo{year}{2025}\natexlab{}.
\newblock \showarticletitle{From Allies to Adversaries: Manipulating LLM Tool-Calling through Adversarial Injection}. In \bibinfo{booktitle}{\emph{Proc. NAACL}}. \bibinfo{pages}{2009--2028}.
\newblock


\bibitem[He et~al\mbox{.}(2025)]%
        {he}
\bibfield{author}{\bibinfo{person}{Pengfei He}, \bibinfo{person}{Yupin Lin}, \bibinfo{person}{Shen Dong}, \bibinfo{person}{Han Xu}, \bibinfo{person}{Yue Xing}, {and} \bibinfo{person}{Hui Liu}.} \bibinfo{year}{2025}\natexlab{}.
\newblock \showarticletitle{Red-Teaming {LLM} Multi-Agent Systems via Communication Attacks}.
\newblock \bibinfo{journal}{\emph{arXiv preprint arXiv:2502.14847}} (\bibinfo{year}{2025}).
\newblock


\bibitem[Zhang et~al\mbox{.}(2024)]%
        {zhang}
\bibfield{author}{\bibinfo{person}{Yuyang Zhang}, \bibinfo{person}{Kangjie Chen}, \bibinfo{person}{Jiaxin Gao}, \bibinfo{person}{Ronghao Cui}, \bibinfo{person}{Run Wang}, \bibinfo{person}{Lina Wang}, {and} \bibinfo{person}{Tianwei Zhang}.} \bibinfo{year}{2024}\natexlab{}.
\newblock \showarticletitle{Towards Action Hijacking of Large Language Model-based Agent}.
\newblock \bibinfo{journal}{\emph{arXiv preprint arXiv:2412.10807}} (\bibinfo{year}{2024}).
\newblock


\bibitem[Juneja et~al\mbox{.}(2025)]%
        {magpie}
\bibfield{author}{\bibinfo{person}{Gurusha Juneja}, \bibinfo{person}{Alon Albalak}, \bibinfo{person}{Wenyue Hua}, {and} \bibinfo{person}{William~Yang Wang}.} \bibinfo{year}{2025}\natexlab{}.
\newblock \showarticletitle{{MAGPIE}: A dataset for Multi-AGent contextual PrIvacy Evaluation}.
\newblock \bibinfo{journal}{\emph{ArXiv preprint: ArXiv:2506.20737}} (\bibinfo{year}{2025}).
\newblock


\bibitem[Jin et~al\mbox{.}(2025)]%
        {roleplay}
\bibfield{author}{\bibinfo{person}{Haibo Jin}, \bibinfo{person}{Ruoxi Chen}, \bibinfo{person}{Peiyan Zhang}, \bibinfo{person}{Andy Zhou}, \bibinfo{person}{Yang Zhang}, {and} \bibinfo{person}{Haohan Wang}.} \bibinfo{year}{2025}\natexlab{}.
\newblock \showarticletitle{GUARD: Role-playing to Generate Natural-language Jailbreakings to Test Guideline Adherence of LLMs}.
\newblock \bibinfo{journal}{\emph{ArXiv preprint: ArXiv:2402.03299}} (\bibinfo{year}{2025}).
\newblock


\bibitem[Chrapek et~al\mbox{.}(2024)]%
        {TEE1}
\bibfield{author}{\bibinfo{person}{Marcin Chrapek}, \bibinfo{person}{Anjo Vahldiek-Oberwagner}, \bibinfo{person}{Marcin Spoczynski}, \bibinfo{person}{Scott Constable}, \bibinfo{person}{Mona Vij}, {and} \bibinfo{person}{Torsten Hoefler}.} \bibinfo{year}{2024}\natexlab{}.
\newblock \showarticletitle{Fortify Your Foundations: Practical Privacy and Security for Foundation Model Deployments In The Cloud}.
\newblock \bibinfo{journal}{\emph{ArXiv preprint: ArXiv:2410.05930}} (\bibinfo{year}{2024}).
\newblock


\bibitem[Namer et~al\mbox{.}(2025)]%
        {fire4}
\bibfield{author}{\bibinfo{person}{Assaf Namer}, \bibinfo{person}{Prashant Kulkarni}, \bibinfo{person}{Erik Jeansson}, \bibinfo{person}{Brandon Maltzman}, {and} \bibinfo{person}{Hauke Vagts}.} \bibinfo{year}{2025}\natexlab{}.
\newblock \showarticletitle{Automatically Detecting Expensive Prompts and Configuring Firewall Rules to Mitigate Denial of Service Attacks on Large Language Models}.
\newblock \bibinfo{journal}{\emph{https://www.tdcommons.org/dpubs\_series/6642/}} (\bibinfo{year}{2025}).
\newblock


\bibitem[Xhonneux et~al\mbox{.}(2025)]%
        {10.5555/3737916.3737964}
\bibfield{author}{\bibinfo{person}{Sophie Xhonneux}, \bibinfo{person}{Alessandro Sordoni}, \bibinfo{person}{Stephan G\"{u}nnemann}, \bibinfo{person}{Gauthier Gidel}, {and} \bibinfo{person}{Leo Schwinn}.} \bibinfo{year}{2025}\natexlab{}.
\newblock \showarticletitle{Efficient adversarial training in LLMs with continuous attacks}. In \bibinfo{booktitle}{\emph{Proc. NeurIPS}}. \bibinfo{pages}{1502 -- 1530}.
\newblock


\bibitem[Deng et~al\mbox{.}(2025)]%
        {10681029}
\bibfield{author}{\bibinfo{person}{Dazhen Deng}, \bibinfo{person}{Chuhan Zhang}, \bibinfo{person}{Huawei Zheng}, \bibinfo{person}{Yuwen Pu}, \bibinfo{person}{Shouling Ji}, {and} \bibinfo{person}{Yingcai Wu}.} \bibinfo{year}{2025}\natexlab{}.
\newblock \showarticletitle{AdversaFlow: Visual Red Teaming for Large Language Models with Multi-Level Adversarial Flow}.
\newblock \bibinfo{journal}{\emph{IEEE Transactions on Visualization and Computer Graphics}} \bibinfo{volume}{31}, \bibinfo{number}{1} (\bibinfo{year}{2025}), \bibinfo{pages}{492--502}.
\newblock


\bibitem[Behnia et~al\mbox{.}(2022)]%
        {10031034}
\bibfield{author}{\bibinfo{person}{Rouzbeh Behnia}, \bibinfo{person}{Mohammadreza~Reza Ebrahimi}, \bibinfo{person}{Jason Pacheco}, {and} \bibinfo{person}{Balaji Padmanabhan}.} \bibinfo{year}{2022}\natexlab{}.
\newblock \showarticletitle{{ EW}-Tune: A Framework for Privately Fine-Tuning Large Language Models with Differential Privacy}. In \bibinfo{booktitle}{\emph{Proc. ICDMW}}. \bibinfo{pages}{560--566}.
\newblock


\bibitem[Dai et~al\mbox{.}(2024)]%
        {dai}
\bibfield{author}{\bibinfo{person}{Juntao Dai} {et~al\mbox{.}}} \bibinfo{year}{2024}\natexlab{}.
\newblock \showarticletitle{Safe {RLHF}: Safe Reinforcement Learning from Human Feedback}. In \bibinfo{booktitle}{\emph{Proc. ICML}}. \bibinfo{pages}{1--28}.
\newblock


\bibitem[Sun and Zhao(2025)]%
        {10.1145/3708394.3708396}
\bibfield{author}{\bibinfo{person}{Zhendan Sun} {and} \bibinfo{person}{Ruibin Zhao}.} \bibinfo{year}{2025}\natexlab{}.
\newblock \showarticletitle{LLM Security Alignment Framework Design Based on Personal Preference}. In \bibinfo{booktitle}{\emph{Proc. AIFE}}. \bibinfo{pages}{6–11}.
\newblock


\bibitem[Spelda and Stritecky(2025)]%
        {RLHFbad}
\bibfield{author}{\bibinfo{person}{Petr Spelda} {and} \bibinfo{person}{Vit Stritecky}.} \bibinfo{year}{2025}\natexlab{}.
\newblock \showarticletitle{Security practices in {AI} development}.
\newblock \bibinfo{journal}{\emph{AI \& Society}} (\bibinfo{year}{2025}), \bibinfo{pages}{1--11}.
\newblock


\bibitem[Phute et~al\mbox{.}(2024)]%
        {selfdefense}
\bibfield{author}{\bibinfo{person}{Mansi Phute}, \bibinfo{person}{Alec Helbling}, \bibinfo{person}{Matthew~Daniel Hull}, \bibinfo{person}{ShengYun Peng}, \bibinfo{person}{Sebastian Szyller}, \bibinfo{person}{Cory Cornelius}, {and} \bibinfo{person}{Duen~Horng Chau}.} \bibinfo{year}{2024}\natexlab{}.
\newblock \showarticletitle{{LLM} Self Defense: By Self Examination, {LLM}s Know They Are Being Tricked}. In \bibinfo{booktitle}{\emph{Proc. ICLR}}. \bibinfo{pages}{1--6}.
\newblock


\bibitem[Dong and Wang(2025)]%
        {TEE2}
\bibfield{author}{\bibinfo{person}{Ben Dong} {and} \bibinfo{person}{Qian Wang}.} \bibinfo{year}{2025}\natexlab{}.
\newblock \showarticletitle{Evaluating the Performance of the DeepSeek Model in Confidential Computing Environment}.
\newblock \bibinfo{journal}{\emph{ArXiv preprint: ArXiv:2502.11347}} (\bibinfo{year}{2025}).
\newblock


\bibitem[Su and Zhang(2025)]%
        {test}
\bibfield{author}{\bibinfo{person}{Jianchang Su} {and} \bibinfo{person}{Wei Zhang}.} \bibinfo{year}{2025}\natexlab{}.
\newblock \showarticletitle{Runtime Attestation for Secure {LLM} Serving in Cloud-Native Trusted Execution Environments}. In \bibinfo{booktitle}{\emph{Proc. ICSA}}. \bibinfo{pages}{1--5}.
\newblock


\bibitem[Li et~al\mbox{.}(2024)]%
        {TEE4}
\bibfield{author}{\bibinfo{person}{Qinfeng Li} {et~al\mbox{.}}} \bibinfo{year}{2024}\natexlab{}.
\newblock \showarticletitle{Core{G}uard: Safeguarding Foundational Capabilities of {LLM}s Against Model Stealing in Edge Deployment}.
\newblock \bibinfo{journal}{\emph{ArXiv preprint: ArXiv:2410.13903}} (\bibinfo{year}{2024}).
\newblock


\bibitem[Lin et~al\mbox{.}(2025)]%
        {10890445}
\bibfield{author}{\bibinfo{person}{Zechao Lin}, \bibinfo{person}{Sisi Zhang}, \bibinfo{person}{Xingbin Wang}, \bibinfo{person}{Yulan Su}, \bibinfo{person}{Yan Wang}, \bibinfo{person}{Rui Hou}, {and} \bibinfo{person}{Dan Meng}.} \bibinfo{year}{2025}\natexlab{}.
\newblock \showarticletitle{Lo{RATEE}: A Secure and Efficient Inference Framework for Multi-Tenant {L}o{RA} {LLM}s Based on {TEE}}. In \bibinfo{booktitle}{\emph{Proc. ICASSP}}. \bibinfo{pages}{1--5}.
\newblock


\bibitem[Sanyal et~al\mbox{.}(2025)]%
        {AC1}
\bibfield{author}{\bibinfo{person}{Debdeep Sanyal}, \bibinfo{person}{Umakanta Maharana}, \bibinfo{person}{Yash Sinha}, \bibinfo{person}{Hong~Ming Tan}, \bibinfo{person}{Shirish Karande}, \bibinfo{person}{Mohan Kankanhalli}, {and} \bibinfo{person}{Murari Mandal}.} \bibinfo{year}{2025}\natexlab{}.
\newblock \showarticletitle{OrgAccess: A Benchmark for Role Based Access Control in Organization Scale {LLM}s}.
\newblock \bibinfo{journal}{\emph{ArXiv preprint: ArXiv:2505.19165}} (\bibinfo{year}{2025}).
\newblock


\bibitem[Huang et~al\mbox{.}(2024)]%
        {fire}
\bibfield{author}{\bibinfo{person}{Bin Huang} {et~al\mbox{.}}} \bibinfo{year}{2024}\natexlab{}.
\newblock \showarticletitle{Firewa{LLM}: A Portable Data Protection and Recovery Framework for {LLM} Services}. In \bibinfo{booktitle}{\emph{Proc. DMBD}}. \bibinfo{pages}{16--30}.
\newblock


\bibitem[Lin and Vachon(2017)]%
        {8258126}
\bibfield{author}{\bibinfo{person}{Tsau Young T.~Y. Lin} {and} \bibinfo{person}{Pierre Vachon}.} \bibinfo{year}{2017}\natexlab{}.
\newblock \showarticletitle{Secure information flow and file movements: A topological theory of discretionary access controls}. In \bibinfo{booktitle}{\emph{Proc. Big Data}}. \bibinfo{pages}{1821--1829}.
\newblock


\bibitem[Lu et~al\mbox{.}(2025)]%
        {10938304}
\bibfield{author}{\bibinfo{person}{Hui Lu}, \bibinfo{person}{Xiaojiang Du}, \bibinfo{person}{Dawei Hu}, \bibinfo{person}{Shen Su}, {and} \bibinfo{person}{Zhihong Tian}.} \bibinfo{year}{2025}\natexlab{}.
\newblock \showarticletitle{B{PFG}uard: Multi-Granularity Container Runtime Mandatory Access Control}.
\newblock \bibinfo{journal}{\emph{IEEE Transactions on Cloud Computing}} \bibinfo{volume}{13}, \bibinfo{number}{2} (\bibinfo{year}{2025}), \bibinfo{pages}{629--640}.
\newblock


\bibitem[Shi et~al\mbox{.}(2025)]%
        {AC2}
\bibfield{author}{\bibinfo{person}{Tianneng Shi}, \bibinfo{person}{Jingxuan He}, \bibinfo{person}{Zhun Wang}, \bibinfo{person}{Linyu Wu}, \bibinfo{person}{Hongwei Li}, \bibinfo{person}{Wenbo Guo}, {and} \bibinfo{person}{Dawn Song}.} \bibinfo{year}{2025}\natexlab{}.
\newblock \showarticletitle{Progent: Programmable Privilege Control for {LLM} Agents}.
\newblock \bibinfo{journal}{\emph{ArXiv preprint: ArXiv:2504.11703}} (\bibinfo{year}{2025}).
\newblock


\bibitem[Abdelnabi et~al\mbox{.}(2025)]%
        {fire2}
\bibfield{author}{\bibinfo{person}{Sahar Abdelnabi}, \bibinfo{person}{Amr Gomaa}, \bibinfo{person}{Eugene Bagdasarian}, \bibinfo{person}{Per~Ola Kristensson}, {and} \bibinfo{person}{Reza Shokri}.} \bibinfo{year}{2025}\natexlab{}.
\newblock \showarticletitle{Firewalls to Secure Dynamic {LLM} Agentic Networks}.
\newblock \bibinfo{journal}{\emph{ArXiv preprint: ArXiv:2502.01822}} (\bibinfo{year}{2025}).
\newblock


\bibitem[Hongwei~Yao(2025)]%
        {yao2025control}
\bibfield{author}{\bibinfo{person}{Yidou Chen Yixin Jiang Cong Wang Zhan~Qin Hongwei~Yao, Haoran~Shi}.} \bibinfo{year}{2025}\natexlab{}.
\newblock \showarticletitle{Control{NET}: A Firewall for {RAG}-based {LLM} System}.
\newblock \bibinfo{journal}{\emph{arXiv preprint arXiv:2504.09593}} (\bibinfo{year}{2025}).
\newblock


\bibitem[Gan et~al\mbox{.}(2024)]%
        {comm1}
\bibfield{author}{\bibinfo{person}{Yuyou Gan} {et~al\mbox{.}}} \bibinfo{year}{2024}\natexlab{}.
\newblock \showarticletitle{Navigating the Risks: A Survey of Security, Privacy, and Ethics Threats in {LLM}-Based Agents}.
\newblock \bibinfo{journal}{\emph{arXiv preprint arXiv:2411.09523}} (\bibinfo{year}{2024}).
\newblock


\bibitem[Jing et~al\mbox{.}(2025)]%
        {MCIP}
\bibfield{author}{\bibinfo{person}{Huihao Jing} {et~al\mbox{.}}} \bibinfo{year}{2025}\natexlab{}.
\newblock \showarticletitle{{MCIP}: Protecting {MCP} Safety via Model Contextual Integrity Protocol}.
\newblock \bibinfo{journal}{\emph{arXiv preprint arXiv:2505.14590}} (\bibinfo{year}{2025}).
\newblock


\bibitem[Zhao et~al\mbox{.}(2025)]%
        {LLMSurvey}
\bibfield{author}{\bibinfo{person}{Wayne~Xin Zhao} {et~al\mbox{.}}} \bibinfo{year}{2025}\natexlab{}.
\newblock \showarticletitle{A Survey of Large Language Models}.
\newblock \bibinfo{journal}{\emph{ArXiv preprint: ArXiv:2303.18223}} (\bibinfo{year}{2025}).
\newblock


\bibitem[Chen et~al\mbox{.}(2024)]%
        {blockagents}
\bibfield{author}{\bibinfo{person}{Bei Chen} {et~al\mbox{.}}} \bibinfo{year}{2024}\natexlab{}.
\newblock \showarticletitle{BlockAgents: Towards Byzantine-Robust LLM-Based Multi-Agent Coordination via Blockchain}. In \bibinfo{booktitle}{\emph{Proc. ACM Turing Award Celebration Conference}}. \bibinfo{pages}{187–192}.
\newblock


\bibitem[Yao et~al\mbox{.}(2019)]%
        {8793190}
\bibfield{author}{\bibinfo{person}{Yuan Yao}, \bibinfo{person}{Bin Xiao}, \bibinfo{person}{Gang Yang}, \bibinfo{person}{Yujiao Hu}, \bibinfo{person}{Liang Wang}, {and} \bibinfo{person}{Xingshe Zhou}.} \bibinfo{year}{2019}\natexlab{}.
\newblock \showarticletitle{Power Control Identification: A Novel Sybil Attack Detection Scheme in VANETs Using RSSI}.
\newblock \bibinfo{journal}{\emph{IEEE Journal on Selected Areas in Communications}} \bibinfo{volume}{37}, \bibinfo{number}{11} (\bibinfo{year}{2019}), \bibinfo{pages}{2588--2602}.
\newblock


\bibitem[Sun et~al\mbox{.}(2025)]%
        {10852158}
\bibfield{author}{\bibinfo{person}{Qianlong Sun}, \bibinfo{person}{Guoshun Nan}, \bibinfo{person}{Tianyi Li}, \bibinfo{person}{Huici Wu}, \bibinfo{person}{Zhou Zhong}, {and} \bibinfo{person}{Xiaofeng Tao}.} \bibinfo{year}{2025}\natexlab{}.
\newblock \showarticletitle{A Secure Digital Signature Scheme for Deep Learning-Based Semantic Communication Systems}.
\newblock \bibinfo{journal}{\emph{IEEE Wireless Communications Letters}} \bibinfo{volume}{14}, \bibinfo{number}{4} (\bibinfo{year}{2025}), \bibinfo{pages}{1119--1123}.
\newblock


\bibitem[Zhang et~al\mbox{.}(2022)]%
        {10062213}
\bibfield{author}{\bibinfo{person}{Chenxin Zhang}, \bibinfo{person}{Jin He}, \bibinfo{person}{Baixiang Fan}, \bibinfo{person}{Yaqiang Gong}, \bibinfo{person}{Shuo Li}, \bibinfo{person}{Bo Yin}, {and} \bibinfo{person}{Yongfeng Lin}.} \bibinfo{year}{2022}\natexlab{}.
\newblock \showarticletitle{Tag-Based Trust Evaluation In Zero Trust Architecture}. In \bibinfo{booktitle}{\emph{Proc. IAECST}}. \bibinfo{pages}{772--776}.
\newblock


\bibitem[Nagarajan et~al\mbox{.}(2024)]%
        {10576030}
\bibfield{author}{\bibinfo{person}{Senthil~Murugan Nagarajan}, \bibinfo{person}{Ganesh~Gopal Devarajan}, \bibinfo{person}{M.~Suresh Thangakrishnan}, \bibinfo{person}{T.~V. Ramana}, \bibinfo{person}{Ali~Kashif Bashir}, {and} \bibinfo{person}{Ahmad~Ali AlZubi}.} \bibinfo{year}{2024}\natexlab{}.
\newblock \showarticletitle{Artificial Intelligence-Based Zero Trust Security Approach for Consumer Industry}.
\newblock \bibinfo{journal}{\emph{IEEE Transactions on Consumer Electronics}} \bibinfo{volume}{70}, \bibinfo{number}{3} (\bibinfo{year}{2024}), \bibinfo{pages}{5411--5418}.
\newblock


\bibitem[Fang et~al\mbox{.}(2024)]%
        {10413289}
\bibfield{author}{\bibinfo{person}{He Fang}, \bibinfo{person}{Yongxu Zhu}, \bibinfo{person}{Yan Zhang}, {and} \bibinfo{person}{Xianbin Wang}.} \bibinfo{year}{2024}\natexlab{}.
\newblock \showarticletitle{Decentralized Edge Collaboration for Seamless Handover Authentication in Zero-Trust IoV}.
\newblock \bibinfo{journal}{\emph{IEEE Transactions on Wireless Communications}} \bibinfo{volume}{23}, \bibinfo{number}{8} (\bibinfo{year}{2024}), \bibinfo{pages}{8760--8772}.
\newblock


\bibitem[Abdelnabi et~al\mbox{.}(2025)]%
        {comm2}
\bibfield{author}{\bibinfo{person}{Sahar Abdelnabi}, \bibinfo{person}{Amr Gomaa}, \bibinfo{person}{Eugene Bagdasarian}, \bibinfo{person}{Per~Ola Kristensson}, {and} \bibinfo{person}{Reza Shokri}.} \bibinfo{year}{2025}\natexlab{}.
\newblock \showarticletitle{Firewalls to Secure Dynamic {LLM} Agentic Networks}.
\newblock \bibinfo{journal}{\emph{ArXiv preprint: ArXiv:2502.01822}} (\bibinfo{year}{2025}).
\newblock


\bibitem[Gehman et~al\mbox{.}(2020)]%
        {gehman-etal-2020-realtoxicityprompts}
\bibfield{author}{\bibinfo{person}{Samuel Gehman}, \bibinfo{person}{Suchin Gururangan}, \bibinfo{person}{Maarten Sap}, \bibinfo{person}{Yejin Choi}, {and} \bibinfo{person}{Noah~A. Smith}.} \bibinfo{year}{2020}\natexlab{}.
\newblock \showarticletitle{{R}eal{T}oxicity{P}rompts: Evaluating Neural Toxic Degeneration in Language Models}. In \bibinfo{booktitle}{\emph{Proc. EMNLP}}. \bibinfo{pages}{3356--3369}.
\newblock


\bibitem[Sun et~al\mbox{.}(2024)]%
        {llmad}
\bibfield{author}{\bibinfo{person}{Yuan Sun}, \bibinfo{person}{Navid Salami~Pargoo}, \bibinfo{person}{Peter Jin}, {and} \bibinfo{person}{Jorge Ortiz}.} \bibinfo{year}{2024}\natexlab{}.
\newblock \showarticletitle{Optimizing Autonomous Driving for Safety: A Human-Centric Approach with {LLM}-Enhanced {RLHF}}. In \bibinfo{booktitle}{\emph{Proc. UbiComp}}. \bibinfo{pages}{76–80}.
\newblock


\bibitem[Nahar et~al\mbox{.}(2024)]%
        {10589640}
\bibfield{author}{\bibinfo{person}{Nurun Nahar}, \bibinfo{person}{Karl Andersson}, \bibinfo{person}{Olov Schelén}, {and} \bibinfo{person}{Saguna Saguna}.} \bibinfo{year}{2024}\natexlab{}.
\newblock \showarticletitle{A Survey on Zero Trust Architecture: Applications and Challenges of 6G Networks}.
\newblock \bibinfo{journal}{\emph{IEEE Access}}  \bibinfo{volume}{12} (\bibinfo{year}{2024}), \bibinfo{pages}{94753--94764}.
\newblock


\bibitem[Zhou et~al\mbox{.}(2025)]%
        {10916520}
\bibfield{author}{\bibinfo{person}{Fanqin Zhou}, \bibinfo{person}{Lei Zhang}, \bibinfo{person}{Zhixiang Yang}, {and} \bibinfo{person}{Lei Feng}.} \bibinfo{year}{2025}\natexlab{}.
\newblock \showarticletitle{Radio Frequency-Enhanced Multi-Factor {I}o{T} Device Authentication via Swarm Learning}.
\newblock \bibinfo{journal}{\emph{IEEE Transactions on Network Science and Engineering}} \bibinfo{volume}{12}, \bibinfo{number}{4} (\bibinfo{year}{2025}), \bibinfo{pages}{2487--2499}.
\newblock


\bibitem[Bouchiha et~al\mbox{.}(2024)]%
        {10633559}
\bibfield{author}{\bibinfo{person}{Mouhamed~Amine Bouchiha}, \bibinfo{person}{Quentin Telnoff}, \bibinfo{person}{Souhail Bakkali}, \bibinfo{person}{Ronan Champagnat}, \bibinfo{person}{Mourad Rabah}, \bibinfo{person}{Mickaël Coustaty}, {and} \bibinfo{person}{Yacine Ghamri-Doudane}.} \bibinfo{year}{2024}\natexlab{}.
\newblock \showarticletitle{{LLMC}hain: Blockchain-Based Reputation System for Sharing and Evaluating Large Language Models}. In \bibinfo{booktitle}{\emph{Proc. COMPSAC}}. \bibinfo{pages}{439--448}.
\newblock


\bibitem[Mao et~al\mbox{.}(2025)]%
        {agentsafe}
\bibfield{author}{\bibinfo{person}{Junyuan Mao} {et~al\mbox{.}}} \bibinfo{year}{2025}\natexlab{}.
\newblock \showarticletitle{AgentSafe: Safeguarding Large Language Model-based Multi-agent Systems via Hierarchical Data Management}.
\newblock \bibinfo{journal}{\emph{ArXiv preprint: ArXiv:2503.04392}} (\bibinfo{year}{2025}).
\newblock


\bibitem[Rezazadeh et~al\mbox{.}(2025)]%
        {accesscontrol1}
\bibfield{author}{\bibinfo{person}{Alireza Rezazadeh}, \bibinfo{person}{Zichao Li}, \bibinfo{person}{Ange Lou}, \bibinfo{person}{Yuying Zhao}, \bibinfo{person}{Wei Wei}, {and} \bibinfo{person}{Yujia Bao}.} \bibinfo{year}{2025}\natexlab{}.
\newblock \showarticletitle{Collaborative Memory: Multi-User Memory Sharing in LLM Agents with Dynamic Access Control}.
\newblock \bibinfo{journal}{\emph{ArXiv preprint: ArXiv:2505.18279}} (\bibinfo{year}{2025}).
\newblock


\bibitem[Kwon et~al\mbox{.}(2023)]%
        {pageattention}
\bibfield{author}{\bibinfo{person}{Woosuk Kwon} {et~al\mbox{.}}} \bibinfo{year}{2023}\natexlab{}.
\newblock \showarticletitle{Efficient Memory Management for Large Language Model Serving with PagedAttention}. In \bibinfo{booktitle}{\emph{Proc. SOSP}}. \bibinfo{pages}{1--16}.
\newblock


\bibitem[Prabhu et~al\mbox{.}(2025)]%
        {vattention}
\bibfield{author}{\bibinfo{person}{Ramya Prabhu}, \bibinfo{person}{Ajay Nayak}, \bibinfo{person}{Jayashree Mohan}, \bibinfo{person}{Ramachandran Ramjee}, {and} \bibinfo{person}{Ashish Panwar}.} \bibinfo{year}{2025}\natexlab{}.
\newblock \showarticletitle{vAttention: Dynamic Memory Management for Serving LLMs without PagedAttention}. In \bibinfo{booktitle}{\emph{Proc. ASPLOS}}. \bibinfo{pages}{1--18}.
\newblock


\bibitem[Henderson et~al\mbox{.}(2023)]%
        {ephemeral}
\bibfield{author}{\bibinfo{person}{Peter Henderson}, \bibinfo{person}{Eric Mitchell}, \bibinfo{person}{Christopher Manning}, \bibinfo{person}{Dan Jurafsky}, {and} \bibinfo{person}{Chelsea Finn}.} \bibinfo{year}{2023}\natexlab{}.
\newblock \showarticletitle{Self-Destructing Models: Increasing the Costs of Harmful Dual Uses of Foundation Models}. In \bibinfo{booktitle}{\emph{Proc. AIES}}. \bibinfo{pages}{287--296}.
\newblock


\bibitem[Fu et~al\mbox{.}(2023)]%
        {serveless1}
\bibfield{author}{\bibinfo{person}{Yao Fu}, \bibinfo{person}{Leyang Xue}, \bibinfo{person}{Yeqi Huang}, {and} \bibinfo{person}{Andrei-Octavian Brabete}.} \bibinfo{year}{2023}\natexlab{}.
\newblock \showarticletitle{Serverless{LLM}: low-latency serverless inference for large language models}. In \bibinfo{booktitle}{\emph{Proc. OSDI}}. \bibinfo{pages}{135--153}.
\newblock


\bibitem[Kinkelin et~al\mbox{.}(2020)]%
        {9110311}
\bibfield{author}{\bibinfo{person}{Holger Kinkelin}, \bibinfo{person}{Richard von Seck}, \bibinfo{person}{Christoph Rudolf}, {and} \bibinfo{person}{Georg Carle}.} \bibinfo{year}{2020}\natexlab{}.
\newblock \showarticletitle{Hardening X.509 Certificate Issuance using Distributed Ledger Technology}. In \bibinfo{booktitle}{\emph{Proc. NOMS}}. \bibinfo{pages}{1--6}.
\newblock


\bibitem[Arias-Cabarcos et~al\mbox{.}(2019)]%
        {10.1145/3336117}
\bibfield{author}{\bibinfo{person}{Patricia Arias-Cabarcos}, \bibinfo{person}{Christian Krupitzer}, {and} \bibinfo{person}{Christian Becker}.} \bibinfo{year}{2019}\natexlab{}.
\newblock \showarticletitle{A Survey on Adaptive Authentication}.
\newblock \bibinfo{journal}{\emph{ACM Computing Survey}} \bibinfo{volume}{52}, \bibinfo{number}{4} (\bibinfo{year}{2019}), \bibinfo{pages}{1--30}.
\newblock


\bibitem[Shaina~Raza(2025)]%
        {reputation3}
\bibfield{author}{\bibinfo{person}{Manoj Karkee Christos~Emmanouilidis Shaina~Raza, Ranjan~Sapkota}.} \bibinfo{year}{2025}\natexlab{}.
\newblock \showarticletitle{TRiSM for Agentic {AI}: A Review of Trust, Risk, and Security Management in {LLM}-based Agentic Multi-Agent Systems}.
\newblock \bibinfo{journal}{\emph{ArXiv preprint: ArXiv:2506.04133}} (\bibinfo{year}{2025}).
\newblock


\bibitem[Geng et~al\mbox{.}(2021)]%
        {9500841}
\bibfield{author}{\bibinfo{person}{Ziye Geng}, \bibinfo{person}{Yunhua He}, \bibinfo{person}{Chao Wang}, \bibinfo{person}{Gang Xu}, \bibinfo{person}{Ke Xiao}, {and} \bibinfo{person}{Shui Yu}.} \bibinfo{year}{2021}\natexlab{}.
\newblock \showarticletitle{A Blockchain based Privacy-Preserving Reputation Scheme for Cloud Service}. In \bibinfo{booktitle}{\emph{Proc. ICC}}. \bibinfo{pages}{1--6}.
\newblock


\bibitem[Ghafoorian et~al\mbox{.}(2019)]%
        {8466653}
\bibfield{author}{\bibinfo{person}{Mahdi Ghafoorian}, \bibinfo{person}{Dariush Abbasinezhad-Mood}, {and} \bibinfo{person}{Hassan Shakeri}.} \bibinfo{year}{2019}\natexlab{}.
\newblock \showarticletitle{A Thorough Trust and Reputation Based RBAC Model for Secure Data Storage in the Cloud}.
\newblock \bibinfo{journal}{\emph{IEEE Transactions on Parallel and Distributed Systems}} \bibinfo{volume}{30}, \bibinfo{number}{4} (\bibinfo{year}{2019}), \bibinfo{pages}{778--788}.
\newblock


\bibitem[Kayes et~al\mbox{.}(2020)]%
        {s20092464}
\bibfield{author}{\bibinfo{person}{A. Kayes} {et~al\mbox{.}}} \bibinfo{year}{2020}\natexlab{}.
\newblock \showarticletitle{A Survey of Context-Aware Access Control Mechanisms for Cloud and Fog Networks: Taxonomy and Open Research Issues}.
\newblock \bibinfo{journal}{\emph{Sensors}} \bibinfo{volume}{20}, \bibinfo{number}{9} (\bibinfo{year}{2020}), \bibinfo{pages}{1--34}.
\newblock


\bibitem[Hu et~al\mbox{.}(2024)]%
        {memory1}
\bibfield{author}{\bibinfo{person}{Cunchen Hu} {et~al\mbox{.}}} \bibinfo{year}{2024}\natexlab{}.
\newblock \showarticletitle{MemServe: Context Caching for Disaggregated {LLM} Serving with Elastic Memory Pool}.
\newblock \bibinfo{journal}{\emph{ArXiv preprint: ArXiv:2406.17565}} (\bibinfo{year}{2024}).
\newblock


\bibitem[CUD({[n.\,d.]})]%
        {CUDA}
 \bibinfo{year}{[n.\,d.]}\natexlab{}.
\newblock \bibinfo{title}{{CUDA API} libraries. 2025}.
\newblock
\newblock
\urldef\tempurl%
\url{https://docs.nvidia.com/cuda/cuda-driver-api/group\_\_CUDA\_\_VA.html}
\showURL{%
\tempurl}


\bibitem[Kim et~al\mbox{.}(2024)]%
        {10827351}
\bibfield{author}{\bibinfo{person}{Minjae Kim}, \bibinfo{person}{Taehyeong Kwon}, \bibinfo{person}{Kibeom Shim}, {and} \bibinfo{person}{Beonghoon Kim}.} \bibinfo{year}{2024}\natexlab{}.
\newblock \showarticletitle{Protection of {LLM} Environment Using Prompt Security}. In \bibinfo{booktitle}{\emph{Proc. ICTC}}. \bibinfo{pages}{1715--1719}.
\newblock


\bibitem[Zhang et~al\mbox{.}(2025)]%
        {JailGuard}
\bibfield{author}{\bibinfo{person}{Xiaoyu Zhang} {et~al\mbox{.}}} \bibinfo{year}{2025}\natexlab{}.
\newblock \showarticletitle{JailGuard: A Universal Detection Framework for Prompt-based Attacks on {LLM} Systems}.
\newblock \bibinfo{journal}{\emph{ACM Transactions on Software Engineering and Methodology}} (\bibinfo{date}{Mar.} \bibinfo{year}{2025}).
\newblock


\bibitem[Muliarevych(2024)]%
        {10755823}
\bibfield{author}{\bibinfo{person}{Oleksandr Muliarevych}.} \bibinfo{year}{2024}\natexlab{}.
\newblock \showarticletitle{Enhancing System Security: {LLM}-Driven Defense Against Prompt Injection Vulnerabilities}. In \bibinfo{booktitle}{\emph{Proc. TCSET}}. \bibinfo{pages}{420--423}.
\newblock


\bibitem[Li et~al\mbox{.}(2025)]%
        {SecurityLingua}
\bibfield{author}{\bibinfo{person}{Yucheng Li}, \bibinfo{person}{Surin Ahn}, \bibinfo{person}{Huiqiang Jiang}, \bibinfo{person}{Amir~H. Abdi}, \bibinfo{person}{Yuqing Yang}, {and} \bibinfo{person}{Lili Qiu}.} \bibinfo{year}{2025}\natexlab{}.
\newblock \showarticletitle{SecurityLingua: Efficient Defense of {LLM} Jailbreak Attacks via Security-Aware Prompt Compression}.
\newblock \bibinfo{journal}{\emph{ArXiv preprint: ArXiv:2506.12707}} (\bibinfo{year}{2025}).
\newblock


\bibitem[Zhang et~al\mbox{.}(2024)]%
        {psysafe}
\bibfield{author}{\bibinfo{person}{Zaibin Zhang} {et~al\mbox{.}}} \bibinfo{year}{2024}\natexlab{}.
\newblock \showarticletitle{PsySafe: A Comprehensive Framework for Psychological-based Attack, Defense, and Evaluation of Multi-agent System Safety}. In \bibinfo{booktitle}{\emph{Proc. ACL}}. \bibinfo{pages}{15202–15231}.
\newblock


\bibitem[Yu et~al\mbox{.}(2025)]%
        {netsafe}
\bibfield{author}{\bibinfo{person}{Miao Yu} {et~al\mbox{.}}} \bibinfo{year}{2025}\natexlab{}.
\newblock \showarticletitle{NetSafe: Exploring the Topological Safety of Multi-agent Networks}.
\newblock \bibinfo{journal}{\emph{ArXiv preprint: ArXiv:2410.15686}} (\bibinfo{year}{2025}).
\newblock


\bibitem[Wang et~al\mbox{.}(2025)]%
        {safeguard}
\bibfield{author}{\bibinfo{person}{Shilong Wang} {et~al\mbox{.}}} \bibinfo{year}{2025}\natexlab{}.
\newblock \showarticletitle{G-Safeguard: A Topology-Guided Security Lens and Treatment on LLM-based Multi-agent Systems}. In \bibinfo{booktitle}{\emph{Proc. ACL}}. \bibinfo{pages}{7261–7276}.
\newblock


\bibitem[Zhou et~al\mbox{.}(2025)]%
        {guardian}
\bibfield{author}{\bibinfo{person}{Jialong Zhou} {et~al\mbox{.}}} \bibinfo{year}{2025}\natexlab{}.
\newblock \showarticletitle{GUARDIAN: Safeguarding LLM Multi-Agent Collaborations with Temporal Graph Modeling}.
\newblock \bibinfo{journal}{\emph{ArXiv preprint: ArXiv:2505.19234}} (\bibinfo{year}{2025}).
\newblock


\bibitem[He et~al\mbox{.}(2025)]%
        {SentinelAgent}
\bibfield{author}{\bibinfo{person}{Xu He}, \bibinfo{person}{Di Wu}, \bibinfo{person}{Yan Zhai}, {and} \bibinfo{person}{Kun Sun}.} \bibinfo{year}{2025}\natexlab{}.
\newblock \showarticletitle{SentinelAgent: Graph-based Anomaly Detection in Multi-Agent Systems}.
\newblock \bibinfo{journal}{\emph{ArXiv preprint: ArXiv:2505.24201}} (\bibinfo{year}{2025}).
\newblock


\bibitem[Liu et~al\mbox{.}(2019)]%
        {8664132}
\bibfield{author}{\bibinfo{person}{Yinqiu Liu}, \bibinfo{person}{Kun Wang}, \bibinfo{person}{Yun Lin}, {and} \bibinfo{person}{Wenyao Xu}.} \bibinfo{year}{2019}\natexlab{}.
\newblock \showarticletitle{LightChain: A Lightweight Blockchain System for Industrial Internet of Things}.
\newblock \bibinfo{journal}{\emph{IEEE Transactions on Industrial Informatics}} \bibinfo{volume}{15}, \bibinfo{number}{6} (\bibinfo{year}{2019}), \bibinfo{pages}{3571--3581}.
\newblock


\bibitem[Huang et~al\mbox{.}(2025)]%
        {10908689}
\bibfield{author}{\bibinfo{person}{Huawei Huang} {et~al\mbox{.}}} \bibinfo{year}{2025}\natexlab{}.
\newblock \showarticletitle{BlockEmulator: An Emulator Enabling to Test Blockchain Sharding Protocols}.
\newblock \bibinfo{journal}{\emph{IEEE Transactions on Services Computing}} \bibinfo{volume}{18}, \bibinfo{number}{2} (\bibinfo{year}{2025}), \bibinfo{pages}{690--703}.
\newblock


\bibitem[Lei et~al\mbox{.}(2020)]%
        {9061111}
\bibfield{author}{\bibinfo{person}{Kai Lei}, \bibinfo{person}{Maoyu Du}, \bibinfo{person}{Jiyue Huang}, {and} \bibinfo{person}{Tong Jin}.} \bibinfo{year}{2020}\natexlab{}.
\newblock \showarticletitle{Groupchain: Towards a Scalable Public Blockchain in Fog Computing of {I}o{T} Services Computing}.
\newblock \bibinfo{journal}{\emph{IEEE Transactions on Services Computing}} \bibinfo{volume}{13}, \bibinfo{number}{2} (\bibinfo{year}{2020}), \bibinfo{pages}{252--262}.
\newblock


\bibitem[Leng et~al\mbox{.}(2022)]%
        {9271868}
\bibfield{author}{\bibinfo{person}{Jiewu Leng}, \bibinfo{person}{Man Zhou}, \bibinfo{person}{J.~Leon Zhao}, \bibinfo{person}{Yongfeng Huang}, {and} \bibinfo{person}{Yiyang Bian}.} \bibinfo{year}{2022}\natexlab{}.
\newblock \showarticletitle{Blockchain Security: A Survey of Techniques and Research Directions}.
\newblock \bibinfo{journal}{\emph{IEEE Transactions on Services Computing}} \bibinfo{volume}{15}, \bibinfo{number}{4} (\bibinfo{year}{2022}), \bibinfo{pages}{2490--2510}.
\newblock


\bibitem[Luo et~al\mbox{.}(2025)]%
        {blockchain2}
\bibfield{author}{\bibinfo{person}{Haoxiang Luo} {et~al\mbox{.}}} \bibinfo{year}{2025}\natexlab{}.
\newblock \showarticletitle{A Trustworthy Multi-LLM Network: Challenges, Solutions, and A Use Case}.
\newblock \bibinfo{journal}{\emph{ArXiv preprint: ArXiv:2505.03196}} (\bibinfo{year}{2025}).
\newblock


\bibitem[Karanjai and Shi(2024)]%
        {10634448}
\bibfield{author}{\bibinfo{person}{Rabimba Karanjai} {and} \bibinfo{person}{Weidong Shi}.} \bibinfo{year}{2024}\natexlab{}.
\newblock \showarticletitle{Trusted LLM Inference on the Edge with Smart Contracts}. In \bibinfo{booktitle}{\emph{Proc. ICBC}}. \bibinfo{pages}{1--7}.
\newblock


\bibitem[Mo et~al\mbox{.}(2025)]%
        {blockchain5}
\bibfield{author}{\bibinfo{person}{Zefeng Mo}, \bibinfo{person}{Zhihao Hou}, \bibinfo{person}{Ruilin Lai}, \bibinfo{person}{Xiaoyuan Wu}, \bibinfo{person}{Junjie Zhou}, {and} \bibinfo{person}{Gansen Zhao}.} \bibinfo{year}{2025}\natexlab{}.
\newblock \showarticletitle{A Blockchain-Based Framework for Crowdsourcing Evaluation of Large Language Models}. In \bibinfo{booktitle}{\emph{Blockchain and Web3.0 Technology Innovation and Application}}. \bibinfo{pages}{62--71}.
\newblock


\bibitem[Chen et~al\mbox{.}(2024)]%
        {blockchain3}
\bibfield{author}{\bibinfo{person}{Bei Chen}, \bibinfo{person}{Gaolei Li}, \bibinfo{person}{Xi Lin}, \bibinfo{person}{Zheng Wang}, {and} \bibinfo{person}{Jianhua Li}.} \bibinfo{year}{2024}\natexlab{}.
\newblock \showarticletitle{BlockAgents: Towards Byzantine-Robust {LLM}-Based Multi-Agent Coordination via Blockchain}. In \bibinfo{booktitle}{\emph{Proc. ACM-TURC}}. \bibinfo{pages}{187--192}.
\newblock


\bibitem[Xu et~al\mbox{.}(2025)]%
        {11011099}
\bibfield{author}{\bibinfo{person}{Peiming Xu}, \bibinfo{person}{Huan Xu}, \bibinfo{person}{Maoqiang Chen}, \bibinfo{person}{Zhihong Liang}, {and} \bibinfo{person}{Wenqian Xu}.} \bibinfo{year}{2025}\natexlab{}.
\newblock \showarticletitle{Privacy-Preserving Large Language Model in Terms of Secure Computing: A Survey}. In \bibinfo{booktitle}{\emph{Proc. ASENS}}. \bibinfo{pages}{286--294}.
\newblock


\bibitem[Qu et~al\mbox{.}(2025)]%
        {zkp}
\bibfield{author}{\bibinfo{person}{Wenjie Qu}, \bibinfo{person}{Yijun Sun}, \bibinfo{person}{Xuanming Liu}, \bibinfo{person}{Tao Lu}, \bibinfo{person}{Yanpei Guo}, \bibinfo{person}{Kai Chen}, {and} \bibinfo{person}{Jiaheng Zhang}.} \bibinfo{year}{2025}\natexlab{}.
\newblock \showarticletitle{zk{GPT}: An Efficient Non-interactive Zero-knowledge Proof Framework for {LLM} Inference}. In \bibinfo{booktitle}{\emph{Proc. USENIX Security}}. \bibinfo{pages}{1--19}.
\newblock


\bibitem[Liu et~al\mbox{.}(2025)]%
        {WLLM}
\bibfield{author}{\bibinfo{person}{Boyi Liu} {et~al\mbox{.}}} \bibinfo{year}{2025}\natexlab{}.
\newblock \showarticletitle{WiLLM: an Open Framework for LLM Services over Wireless Systems}.
\newblock \bibinfo{journal}{\emph{ArXiv preprint: ArXiv:2506.19030}} (\bibinfo{year}{2025}).
\newblock


\bibitem[Liu et~al\mbox{.}(2024)]%
        {10.1145/3666025.3699404}
\bibfield{author}{\bibinfo{person}{Boyi Liu}, \bibinfo{person}{Jingwen Tong}, {and} \bibinfo{person}{Jun Zhang}.} \bibinfo{year}{2024}\natexlab{}.
\newblock \showarticletitle{Poster Abstract: LLM-Slice: Dedicated Wireless Network Slicing for Large Language Models}. In \bibinfo{booktitle}{\emph{Proc. SenSys}}. \bibinfo{pages}{853–854}.
\newblock


\bibitem[Brown et~al\mbox{.}(2020)]%
        {chatGPT}
\bibfield{author}{\bibinfo{person}{Tom~B. Brown} {et~al\mbox{.}}} \bibinfo{year}{2020}\natexlab{}.
\newblock \showarticletitle{Language Models are Few-Shot Learners}.
\newblock \bibinfo{journal}{\emph{ArXiv preprint: ArXiv:2005.14165}} (\bibinfo{year}{2020}).
\newblock


\bibitem[Pan et~al\mbox{.}(2025)]%
        {fail}
\bibfield{author}{\bibinfo{person}{Melissa~Z Pan} {et~al\mbox{.}}} \bibinfo{year}{2025}\natexlab{}.
\newblock \showarticletitle{Why Do Multiagent Systems Fail?}. In \bibinfo{booktitle}{\emph{Proc. ICLR}}. \bibinfo{pages}{1--40}.
\newblock


\bibitem[Huang et~al\mbox{.}(2024)]%
        {Step}
\bibfield{author}{\bibinfo{person}{Xiaowei Huang} {et~al\mbox{.}}} \bibinfo{year}{2024}\natexlab{}.
\newblock \showarticletitle{A survey of safety and trustworthiness of large language models through the lens of verification and validation}.
\newblock \bibinfo{journal}{\emph{Artificial Intelligence Review}} \bibinfo{volume}{57}, \bibinfo{number}{175} (\bibinfo{year}{2024}), \bibinfo{pages}{1--53}.
\newblock


\bibitem[Elhaminia et~al\mbox{.}(2023)]%
        {10024295}
\bibfield{author}{\bibinfo{person}{Behnaz Elhaminia} {et~al\mbox{.}}} \bibinfo{year}{2023}\natexlab{}.
\newblock \showarticletitle{Toxicity Prediction in Pelvic Radiotherapy Using Multiple Instance Learning and Cascaded Attention Layers}.
\newblock \bibinfo{journal}{\emph{IEEE Journal of Biomedical and Health Informatics}} \bibinfo{volume}{27}, \bibinfo{number}{4} (\bibinfo{year}{2023}), \bibinfo{pages}{1958--1966}.
\newblock


\bibitem[Zhang et~al\mbox{.}(2025)]%
        {agentFM}
\bibfield{author}{\bibinfo{person}{Lingzhe Zhang}, \bibinfo{person}{Yunpeng Zhai}, \bibinfo{person}{Tong Jia}, \bibinfo{person}{Xiaosong Huang}, \bibinfo{person}{Chiming Duan}, {and} \bibinfo{person}{Ying Li}.} \bibinfo{year}{2025}\natexlab{}.
\newblock \showarticletitle{Agent{FM}: Role-Aware Failure Management for Distributed Databases with {LLM}-Driven Multi-Agents}. In \bibinfo{booktitle}{\emph{Proc. FSE-IVR}}. \bibinfo{pages}{1--5}.
\newblock


\bibitem[Chang and Geng(2025)]%
        {saga}
\bibfield{author}{\bibinfo{person}{Edward~Y. Chang} {and} \bibinfo{person}{Longling Geng}.} \bibinfo{year}{2025}\natexlab{}.
\newblock \showarticletitle{Saga{LLM}: Context Management, Validation, and Transaction Guarantees for Multi-Agent {LLM} Planning}.
\newblock \bibinfo{journal}{\emph{ArXiv preprint: ArXiv:2505.20824}} (\bibinfo{year}{2025}).
\newblock


\bibitem[Tiwari and Farag(2025)]%
        {10947002}
\bibfield{author}{\bibinfo{person}{Abhinav Tiwari} {and} \bibinfo{person}{Hany E.~Z. Farag}.} \bibinfo{year}{2025}\natexlab{}.
\newblock \showarticletitle{Responsible AI Framework for Autonomous Vehicles: Addressing Bias and Fairness Risks}.
\newblock \bibinfo{journal}{\emph{IEEE Access}}  \bibinfo{volume}{13} (\bibinfo{year}{2025}), \bibinfo{pages}{58800--58822}.
\newblock


\bibitem[Kherraf et~al\mbox{.}(2019)]%
        {8624371}
\bibfield{author}{\bibinfo{person}{Nouha Kherraf}, \bibinfo{person}{Hyame~Assem Alameddine}, \bibinfo{person}{Sanaa Sharafeddine}, \bibinfo{person}{Chadi~M. Assi}, {and} \bibinfo{person}{Ali Ghrayeb}.} \bibinfo{year}{2019}\natexlab{}.
\newblock \showarticletitle{Optimized Provisioning of Edge Computing Resources With Heterogeneous Workload in IoT Networks}.
\newblock \bibinfo{journal}{\emph{IEEE Transactions on Network and Service Management}} \bibinfo{volume}{16}, \bibinfo{number}{2} (\bibinfo{year}{2019}), \bibinfo{pages}{459--474}.
\newblock


\end{thebibliography}


\end{document}